\documentclass[%
reprint,
superscriptaddress,
amsmath,
amssymb,
aps,
longbibliography
]{revtex4-2}

\usepackage{graphicx}%
\usepackage{dcolumn}%
\usepackage{bm}%
\usepackage{hyperref}%
\usepackage{amssymb}
\usepackage{amsmath}
\usepackage{siunitx}
\usepackage{mathtools}
\usepackage{xr}
\usepackage{float}
\usepackage[usenames,dvipsnames]{color}
\usepackage[export]{adjustbox}
\usepackage{gensymb}
\usepackage{soul,xcolor}
\usepackage{multirow}
\usepackage{tikz}
\usepackage{dcolumn}
\usepackage{longtable}
\usepackage{makecell}
\usepackage[position=top,font=large,caption=false]{subfig}
\usepackage{placeins}
\usepackage{gensymb}

\DeclareSIUnit\gauss{G}
\DeclareSIUnit\sig{\mbox{$\sigma$}}
\DeclareSIUnit\debye{D}

\newcommand{\expval}[1]{\langle#1\rangle}

\newcommand{\D}{\mathrm{d}}

\newcommand{\bracket}[3]{\langle {#1} | {#2} | {#3} \rangle}

\newcommand{\aver}[1]{\ensuremath{\langle {#1} \rangle}}

\newcommand{\bbar}[1]{\overline{\overline{#1}}}
\renewcommand{\vec}[1]{\mbox{\boldmath $#1$}}
\newcommand{\drt}[0]{\ensuremath{\delta \aver{r^2}}}
\newcommand{\drf}[0]{\ensuremath{\delta \aver{r^4}}}
\newcommand{\drtsq}{[\drt^2]}
\newcommand{\mr}[1]{\expval{r^{#1}}}
\newcommand{\dmr}[1]{\delta \mr{#1}}
\newcommand{\ambit}{\textsc{amb}{\footnotesize
		i}\textsc{t}}
\newcommand{\Fy}{Fy($\Delta r$)}

\newcommand{\beginsupplement}{%
	\setcounter{equation}{0}
	\renewcommand{\theequation}{S\arabic{equation}}%
	\setcounter{table}{0}
	\renewcommand{\thetable}{S\arabic{table}}%
	\setcounter{figure}{0}
	\renewcommand{\thefigure}{S\arabic{figure}}%
}

\begin{document}
	
	\title{Evidence of Two-Source King Plot Nonlinearity in Spectroscopic Search for New Boson}
	
	\author{Joonseok Hur}
	\thanks{These authors contributed equally to this work.}
	\affiliation{Department of Physics and Research Laboratory of Electronics, Massachusetts Institute of Technology, Cambridge, Massachusetts 02139, USA}
	
	\author{Diana P. L. Aude Craik}
	\thanks{These authors contributed equally to this work.}
	\affiliation{Department of Physics and Research Laboratory of Electronics, Massachusetts Institute of Technology, Cambridge, Massachusetts 02139, USA}
	
	\author{Ian Counts}
	\thanks{These authors contributed equally to this work.}\textbf{}
	\affiliation{Department of Physics and Research Laboratory of Electronics, Massachusetts Institute of Technology, Cambridge, Massachusetts 02139, USA}
	
	\author{Eugene Knyazev}
	\affiliation{Department of Physics and Research Laboratory of Electronics, Massachusetts Institute of Technology, Cambridge, Massachusetts 02139, USA}
	
	\author{Luke Caldwell}
	\affiliation{JILA, NIST and University of Colorado, Boulder, Colorado 80309, USA}
	
	\author{Calvin Leung}
	\affiliation{Department of Physics and Research Laboratory of Electronics, Massachusetts Institute of Technology, Cambridge, Massachusetts 02139, USA}
	
	\author{Swadha Pandey}
	\affiliation{Department of Physics and Research Laboratory of Electronics, Massachusetts Institute of Technology, Cambridge, Massachusetts 02139, USA}
	
	\author{Julian C. Berengut}
	\affiliation{School of Physics, University of New South Wales, Sydney, New South Wales 2052, Australia}
	
	\author{Amy Geddes}
	\affiliation{School of Physics, University of New South Wales, Sydney, New South Wales 2052, Australia}
	
	\author{Witold Nazarewicz}
	\affiliation{Facility for Rare Isotope Beams and Department of Physics and Astronomy, Michigan State University, East Lansing, Michigan 48824, USA}
	
	\author{Paul-Gerhard Reinhard}
	\affiliation{Institut f{\"u}r Theoretische Physik, Universit{\"a}t Erlangen, Erlangen, Germany.}
	
	\author{Akio Kawasaki}
	\affiliation{National Metrology Institute of Japan (NMIJ), National Institute of Advanced Industrial Science and Technology (AIST),
		1-1-1 Umezono, Tsukuba, Ibaraki 305-8563, Japan}
	
	\author{Honggi Jeon}
	\affiliation{Department of Physics and Astronomy, Seoul National University, Seoul 151-747, Korea}
	
	\author{Wonho Jhe}
	\affiliation{Department of Physics and Astronomy, Seoul National University, Seoul 151-747, Korea}
	
	\author{Vladan Vuleti\'c}
	\email{vuletic@mit.edu}
	\affiliation{Department of Physics and Research Laboratory of Electronics, Massachusetts Institute of Technology, Cambridge, Massachusetts 02139, USA}
	
	\begin{abstract}
		Optical precision spectroscopy of isotope shifts can be used to test for new forces beyond the Standard Model, and to determine basic properties of atomic nuclei. We measure isotope shifts on the highly forbidden ${}^2S_{1/2} \rightarrow {}^2F_{7/2}$ octupole transition of trapped $^{168,170,172,174,176}$Yb ions. When combined with previous measurements in Yb$^+$ and very recent measurements in Yb, the data reveal a King plot nonlinearity of up to 240$\sigma$.
		The trends exhibited by experimental data are explained by nuclear density functional theory calculations with the Fayans functional. We also find, with 4.3$\sigma$ confidence, that there is a second distinct source of nonlinearity, and discuss its possible origin.
	\end{abstract}
	
	\maketitle
	
	Despite ample evidence for the existence of dark matter \cite{GallaxyRotation, GallaxyCollision,GravLensing,CMB} and concerted experimental searches for candidate particles \cite{PDG, ALPs, EDM, BSMwAMO}, its origin and composition remain unknown. Isotope-shift (IS) spectroscopy has been recently proposed as a tabletop method to search for dark matter candidates in the intermediate mass range $\lesssim 100$MeV$/c^2$ \cite{Delaunay2017,Berengut2018}. In particular, IS spectroscopy can be used to search for a hypothetical new boson, $\phi$, that mediates interactions between quarks and leptons. An observable consequence is an additional isotope shift that arises from the effective interaction between neutrons and electrons.
	Such a shift could be detected as a deviation from linearity in a King plot \cite{King1984} that compares the normalized isotope shifts for two different transitions. If at least three isotope shifts in each transition are measured, a deviation from linearity can be detected. The nonlinearity can also be caused by higher-order nuclear effects \cite{Flambaum2018,Allehabi2021,Mikami2017,Tanaka2019,Reinhard2020,Allehabi2020,Muller2021}.
	
	In our previous work, we reported evidence, at the $3\sigma$ level, for a nonlinearity in a King plot that compared two optical quadrupole transitions (${}^2S_{1/2} \rightarrow {}^2D_{3/2}, {}^2D_{5/2}$) in a trapped Yb$^+$ ion \cite{counts2020}.
	The measurement was performed for five even isotopes, one more than required, and we also proposed a new method to assign the nonlinearity to different possible physical origins based on the observed nonlinearity pattern. At the reported measurement accuracy of $\sim 300$\,Hz on two relatively similar electronic excited states, the source of the nonlinearity could not be discriminated, and was consistent both with a new boson and with Standard-Model (SM) nuclear shifts. IS spectroscopy in Ca$^+$, which has lighter nuclei and therefore lower sensitivity to both new physics and nuclear effects than Yb$^+$ \cite{Berengut2018}, showed no King nonlinearity at the 20\,Hz level \cite{Solaro2020}. At the time of completion of the present work, large King nonlinearities were also reported when comparing transitions in neutral Yb \cite{ono2021,Figueroa2021} with the quadrupole transitions in Yb$^+$.

\begin{figure}
		\centering
		\includegraphics[width=\columnwidth]{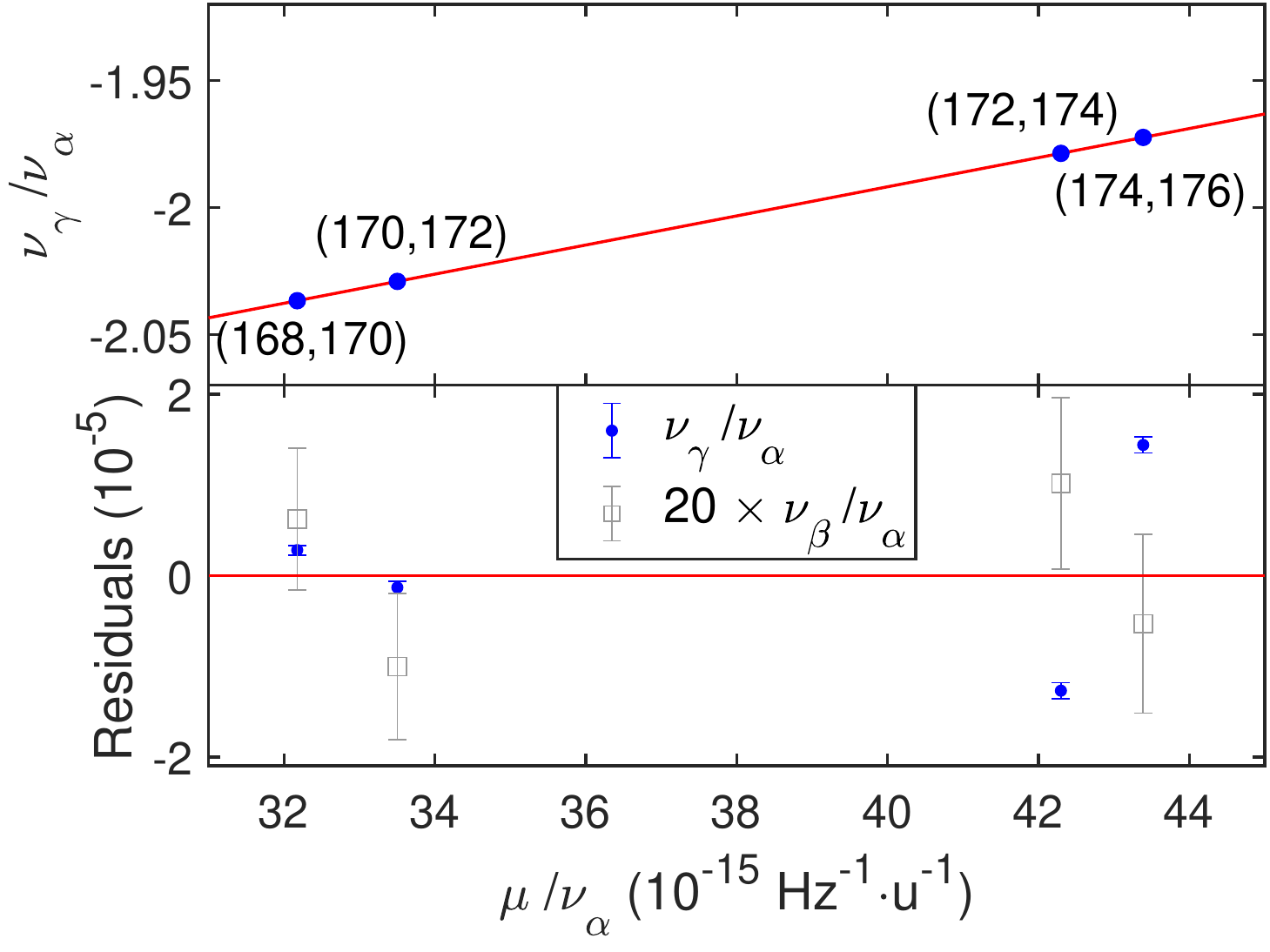}
		\caption{Frequency-normalized King plot (top)
			and residuals (bottom, blue) for the $\gamma$ (${}^2S_{1/2} \rightarrow {}^2F_{7/2}$) transition and reference transition $\alpha$ (${}^2S_{1/2} \rightarrow {}^2D_{5/2}$) for even-neighbor pairs ($A'=A+2$) of Yb$^+$ isotopes.  A deviation from linearity (red line) by 41 standard deviations $\sigma$ is observed. For reference, residuals for the $\beta$ (${}^2S_{1/2} \rightarrow {}^2D_{3/2}$) transition \cite{counts2020}, magnified 20-fold, are also plotted in gray. The error bars indicate $2 \sigma$ uncertainties; for correlations between the errors, see SMat.
		} 
		\label{fig:king_plot}
	\end{figure}

	In this Letter, we report IS laser spectroscopy for the highly forbidden octupole transition ${}^2S_{1/2} \rightarrow {}^2F_{7/2}$ in Yb$^+$. The electron configuration in the $F$ state is very different from the previously measured $D$ states \cite{counts2020}, which increased the size of the observed King nonlinearity 20-fold (see Fig.~\ref{fig:king_plot}).
	At a measurement resolution of $\sim 500$\,Hz, we observe a King plot nonlinearity with 41 standard deviations $\sigma$.
	Including the recent data for neutral Yb \cite{ono2021,Figueroa2021} into our analysis, the significance of the nonlinearity rises to 240$\sigma$, and analyzing the patterns \cite{counts2020} we show that the measurements 
	can be consistently explained by microscopic calculations carried out within nuclear density functional theory (DFT), which provides agreement with ground-state properties of complex deformed Yb isotopes \cite{Reinhard2020,Reinhard2021}. Combining all measured transitions in Yb$^+$ and Yb, we further find evidence, at the  $4.3\sigma$ level, of a second, smaller source of nonlinearity, and discuss implications for limits on a new boson. Finally, we also extract nuclear data that can be used to further fine-tune nuclear energy density functionals.

	\begingroup
	\squeezetable
	\begin{table}
		\caption{\label{tab:IsotopeShift}
			Isotope shifts $\nu^{AA'}_{\gamma,\alpha} =\nu_{\gamma,\alpha}^{A} - \nu_{\gamma,\alpha}^{A'}$ measured for the $\gamma: {}^2S_{1/2} \rightarrow {}^2F_{7/2}$ (this work) and $\alpha: {}^2S_{1/2} \rightarrow {}^2D_{5/2}$ \cite{counts2020} transitions for pairs $(A,A')$ of stable Yb$^+$ even isotopes. Inverse-mass differences $\mu^{AA'}=1/m^{A} - 1/m^{A'}$ calculated from \cite{Nesterenko2020,AME2016_1,AME2016_2,Rana2012} with the Yb ionization energy set to 6.254\,eV are also listed. Numbers in parentheses indicate 1$\sigma$ statistical uncertainties.
		}
		\begin{ruledtabular}
			\begin{tabular}{cccD{.}{.}{2.11}}
				$(A, A')$ & $\nu^{AA'}_{\gamma}$ [MHz] %
				& $\nu^{AA'}_{\alpha}$ [MHz] & 
				\multicolumn{1}{c}{$\mu^{AA'}$ $\left[ 10^{-6} \mathrm{u}^{-1} \right]$}	  \\
				\hline
				(168,170) & -4 438.160 30(50) & 2 179.098 93(21) & 70.113 619 5(36) \\
				(170,172) & -4 149.190 38(45) & 2 044.854 78(34) & 68.506 890 49(63)  \\
				(172,174) & -3 132.321 60(50)  & 1 583.068 42(36) & 66.958 651 95(64) \\
				(174,176) & -2 976.391 60(48) & 1 509.055 29(28) & 65.474 078 21(65) \\
				(168,172) & -8 587.352 00(47) &                  &  \\
				(170,174) & -7 281.511 88(45) & 3 627.922 95(50) &  \\
				(172,176) & -6 108.712 93(44) &                  &
			\end{tabular}
		\end{ruledtabular}
	\end{table}
	\endgroup

	Our IS measurements are performed on individual cold trapped ${}^{A}\text{Yb}^+$ ions with zero nuclear spin ($A\in \{ 168, 170, 172, 174, 176 \}$). To make an IS measurement on the octupole transition ${}^2S_{1/2} \rightarrow {}^2F_{7/2}$ near 467\,nm that we label $\gamma$, we first load a single ion of one isotope $A$ into the trap, Doppler cool it to $\sim500$\,$\mu$K, and measure the excitation probability when scanning the frequency of our probe laser, a frequency-doubled Ti:Sapphire laser which is locked to an ultralow-thermal-expansion (ULE) cavity with linewidth $\kappa_c/(2\pi)=30$\,kHz. We measure two transitions between Zeeman sublevels that are symmetrically detuned from the zero-field transition $\nu_\gamma$, and determine the center frequency $\nu_\gamma^{A}$ as the mean (see Supplemental Material (SMat) \cite{SM}).
	A second isotope $A'$ is then loaded into the trap and its center frequency $\nu_\gamma^{A'}$ is measured. We alternate several times between the two isotopes, achieving an accuracy of $\sim500$\,Hz in our measurement of the IS $\nu^{AA'}_{\gamma} \equiv \nu_\gamma^{A}- \nu_\gamma^{A'}$, limited mainly by the long-term stability of the ULE cavity.
	Our measured ISs $\nu^{AA'}_{\gamma}$ are given in Table~\ref{tab:IsotopeShift}. Table~\ref{tab:AbsoluteFreq467} lists the absolute transition frequencies derived from our measured IS in combination with the absolute transition frequency for $^{172}\text{Yb}^+$ ~\cite{Furst2020,MehlstaublerPrivComm}. 
	
	\begin{table}
		\caption{\label{tab:AbsoluteFreq467}Absolute frequencies of the $\gamma: {}^2S_{1/2} \rightarrow {}^2F_{7/2}$ transition extracted from our IS measurements and the absolute frequency measurement in Ref.~\cite{Furst2020,MehlstaublerPrivComm}} 
		\begin{ruledtabular}
			\begin{tabular}{@{}clr@{}}
				Isotope & Absolute frequency [THz] & Ref. \\
				\hline
				168 & 642.108 197 799 37(37) & [this work] \\ 
				170 & 642.112 635 960 21(32) & [this work] \\ 
				172 & 642.116 785 150 879 5(24) & \cite{Furst2020,MehlstaublerPrivComm} \\
				174 & 642.119 917 472 25(33) & [this work] \\ 
				176 & 642.122 893 863 83(36) & [this work]
			\end{tabular}
		\end{ruledtabular}
	\end{table}

	To a very good approximation, the IS can be factored into an electronic component, which is transition dependent (labeled by a greek letter subscript) but does not depend on the isotope, and a nuclear contribution, which depends on the isotopes (labeled by $AA'$) but not on the electronic transition \cite{King1984,Mikami2017,Delaunay2017,counts2020}:
	\begin{equation}
		\begin{aligned}
			\nu^{AA'}_{\gamma} = F_\gamma \drt^{AA'} + K_\gamma \mu^{AA'} + G^{(4)}_\gamma \drf^{AA'} + \\  + G^{(2)}_\gamma \drtsq^{AA'}  + \upsilon_{ne} D_\gamma a^{AA'} + \cdots
			\label{eq:IsotopeShift}
		\end{aligned}
	\end{equation}
	Here $\dmr{n}^{AA'} \equiv \mr{n}^A - \mr{n}^{A'}$ is the difference in the $n$-th nuclear charge moment between isotopes $A$ and $A'$, $\mu^{AA'} \equiv 1/m^A - 1/m^{A'}$ is the inverse-mass difference, and $\drtsq^{AA'} \equiv (\drt^{AA''})^2 - (\drt^{A'A''})^2$, with $A''$ denoting a reference isotope (we use $A'' = 172$). The quantity $\upsilon_{ne}=(-1)^{s+1} y_n y_e /(4 \pi \hbar c)$ is the product of the coupling constants of the new boson to the neutron $y_n$ and electron $y_e$, resulting in a Yukawa-like potential given by $V_{ne}(r) = \hbar c \, \upsilon_{ne} \exp(-r/\lambdabar_c)/r$ for a boson with spin $s$, mass $m_\phi$, and reduced Compton wavelength $\lambdabar_c=\hbar/(m_\phi c)$ \cite{Mikami2017,Delaunay2017,counts2020}. $a^{AA'} = A-A'$ is the neutron-number difference between the two isotopes. The coefficients $F$, $K$, $G^{(4)}$, $G^{(2)}$, and $D$ are transition-dependent quantities that quantify the field shift, the mass shift, the fourth-moment shift, the quadratic field shift (QFS), and the sensitivity to the new boson, respectively.
	
	To eliminate the large field shift $F$ (associated with the size change of the nucleus $\drt$, of order $\sim 4$\,GHz), and mass shift $K$ (of order $\sim 0.2$\,GHz) contributions, one can use a second set of isotope shifts measured on a different reference transition $\tau$ to generate a King plot \cite{King1984}. In its frequency-normalized version \cite{counts2020}, the relationship studied can be written as
	\begin{equation}
		\begin{aligned}
			\overline{\nu}^{AA'}_\gamma &= f_{\gamma \tau} + K_{\gamma \tau} \overline{\mu}^{AA'} + \\& G_{\gamma \tau}^{(4)} \overline{ \delta \aver{r^4}}^{AA'} + G_{\gamma \tau}^{(2)} \overline{\drtsq}^{AA'} + \upsilon_{ne} D_{\gamma \tau} \overline{a}^{AA'}
			\label{eq:KingPlot}
		\end{aligned}
	\end{equation}
	where the notation $\overline{x}^{AA'} \equiv x^{AA'}/{\nu^{AA'}_{\tau}}$ indicates frequency-normalized terms. We define $z_{\gamma \tau} \equiv Z_{\gamma}/Z_{\tau}$ as the ratio of coefficients for transitions $\gamma$ and $\tau$, and $Z_{\gamma \tau} \equiv Z_\gamma (1- f_{\gamma \tau}/z_{\gamma \tau})$ for $Z \in \{F, K, G^{(2)}, G^{(4)}, D\}$. The first two terms in Eq.~(\ref{eq:KingPlot}) represent the linear relation between $\overline{\nu}_{\gamma}$ and $\overline{\mu}$ in the King plot, while the remaining terms possibly violate the linearity.
	
\begin{figure}
		\includegraphics[width=\columnwidth]{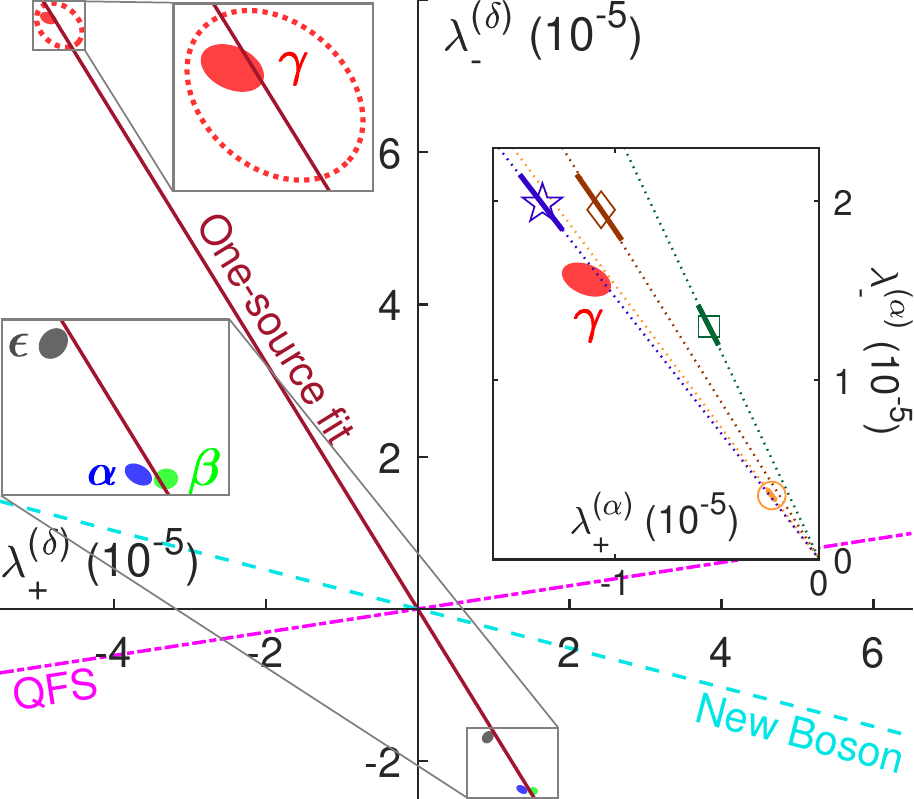}
		\caption{Decomposition of the measured nonlinearity (solid ellipses, 95\% confidence interval) onto the ($\lambda_+$, $\lambda_-$) basis for the transitions $\alpha: {}^2S_{1/2} \rightarrow {}^2D_{5/2}$ in Yb$^+$ (blue) \cite{counts2020}, $\beta: {}^2S_{1/2} \rightarrow {}^2D_{3/2}$ in Yb$^+$ (green) \cite{counts2020}, $\epsilon: {}^1S_{0} \rightarrow {}^1D_{2}$ (dark grey) in Yb \cite{Figueroa2021}, and $\gamma: {}^2S_{1/2} \rightarrow {}^2F_{7/2}$ in Yb$^+$ (red, this work). The corresponding frequency-normalized King plot is generated with the reference transition $\delta: {}^1S_{0} \rightarrow {}^3P_{0}$ in Yb \cite{ono2021} ($\lambda^{(\delta)}_\pm$) that has been measured with the highest frequency accuracy.
			The red dotted ellipse indicates a previous preliminary measurement for the $\gamma$ transition \cite{CountsThesis}.
			The dashed lines indicate the ratio $\lambda_+/\lambda_-$ that would arise solely from a new boson (light blue dashed) or the QFS (pink dash-dotted).
			The brown solid line is a single-source fit to all four transitions $\alpha, \beta, \gamma, \epsilon$, yielding evidence for a second nonlinearity source with 4.3$\sigma$ significance ($\chi^2 = 25.4$).
			The largest inset shows the nonlinearity in a King plot with $\alpha$ as the reference transition ($\lambda^{(\alpha)}_\pm$). Open symbols indicate the nonlinearity due to $\drf^{AA'}$ from  nuclear DFT calculations 
			with SV-min  (square), RD-min (diamond), UNEDF1  (circle), and {\Fy} (star) energy density functionals. Short bold lines indicate the uncertainty in electronic-structure calculations (see SMat).
		}
		\label{fig:lambda_map}
	\end{figure}

	Figure~\ref{fig:king_plot} shows a frequency-normalized King plot using the previously measured transition $\alpha: {}^2S_{1/2} \rightarrow {}^2D_{5/2}$ near 411\,nm \cite{counts2020} as the reference transition $\tau$. The residuals from the linear fit reveal a nonlinearity at the $10^{-5}$ level, corresponding to $41\sigma$. The nonlinearity is 20 times larger than the nonlinearity we observed previously \cite{counts2020} comparing the two quadrupole transitions, $\alpha$  and $\beta: {}^2S_{1/2} \rightarrow {}^2D_{3/2}$, that have a more similar electronic structure. The recent measurements in neutral Yb \cite{ono2021,Figueroa2021}, when combined with our $\alpha$ or $\beta$ transition data, confirm a nonlinearity of a similar size (see also Fig.~\ref{fig:lambda_map}).
	
	Having unambiguously established a King nonlinearity, we can gain information about the sources of nonlinearity by analyzing the deviation patterns \cite{counts2020}. With four isotope-shift data points, we can rewrite Eq.~(\ref{eq:KingPlot}) in terms of four-dimensional vectors as follows:
	
	\begin{equation}
		\vec{\overline{\nu}}_{\gamma} = f_{\gamma \tau}\mathbf{1} + K_{\gamma \tau}\vec{\overline{\mu}} + (\lambda_+ \vec{\Lambda}_+ + \lambda_- \vec{\Lambda}_-)
		\label{eq:vectorKingPlotEq}
	\end{equation}
	where the vector space inhabited by the vectors $\vec{z} \equiv (z_1,z_2,z_3,z_4)$ with $z_k \equiv z^{A,A+2}$ ($A=166+2k$ for $k=1,2,3,4$, $z \in \{\overline{\mu}, \overline{\nu}_\gamma\}$) is spanned by the basis $(\vec{1},\vec{\overline{\mu}},\vec{\Lambda_+}, \vec{\Lambda_-})$.
	
	The first two vectors, $\vec{1} \equiv (1,1,1,1)$ and $\vec{\overline{\mu}}$, define a plane of King linearity (i.e. the component of $\vec{\overline{\nu}}_{\gamma}$ in this plane does not give rise to King nonlinearities), while the unit vectors $\vec{\Lambda}_+$ and $\vec{\Lambda}_-$, defined as $\vec{\Lambda}_{+} \propto (\overline{\mu}_{3} - \overline{\mu}_{2}, \overline{\mu}_{1} - \overline{\mu}_{4}, \overline{\mu}_{4} - \overline{\mu}_{1}, \overline{\mu}_{2} - \overline{\mu}_{3})$ and 
	$ \vec{\Lambda}_{-} \propto (\overline{\mu}_{4} - \overline{\mu}_{2}, \overline{\mu}_{1} - \overline{\mu}_{3}, \overline{\mu}_{2} - \overline{\mu}_{4}, \overline{\mu}_{3} - \overline{\mu}_{1})$, span the out-of-plane space of vectors that produce a King nonlinearity (see SMat). Any vector with nonzero residuals from the linear King plot fit hence has components in the space spanned by $(\vec{\Lambda}_+, \vec{\Lambda}_-)$, and can be expressed as in terms of its scalar components $\lambda_+$ and $\lambda_-$ along $\vec{\Lambda}_+$ and $\vec{\Lambda}_-$, respectively. ($\vec{\Lambda}_+$ and $\vec{\Lambda}_-$ correspond approximately to the zigzag $+-+-$ and curved $+--+$ patterns of residuals introduced in Ref.~\cite{counts2020}.) Both SM and new-boson effects produce nonlinearities with a defined $\lambda_+/\lambda_-$ ratio, given by the associated nuclear factors $x^{AA'}$, and are characterized by lines along definite directions in the $\lambda_{\pm}$-plane (see Fig.~\ref{fig:lambda_map}).
	
	Figure~\ref{fig:lambda_map} displays the measured nonlinearity in the $\lambda_{\pm}$ plane for the $\gamma$ transition, as well as for the previously measured $\alpha$ and $\beta$ transitions in Yb$^+$ \cite{counts2020}, and the recently measured $\epsilon: {}^1S_{0} \rightarrow {}^1D_{2}$ transition in Yb \cite{Figueroa2021}. For the reference transition $\tau$ in Eq.~(\ref{eq:KingPlot}), we choose in Fig.~\ref{fig:lambda_map} the transition $\delta: {}^1S_{0} \rightarrow {}^3P_{0}$ in Yb that has been very recently measured with the highest frequency accuracy \cite{ono2021}. All measured transitions $\alpha, \beta, \gamma, \epsilon, \delta$ are consistent with each other in that they lie nearly along the same direction in the $\lambda_{\pm}$ plane, indicating that the nonlinearity originates from a common dominant source for all transitions. This direction corresponds neither to a new boson $a^{AA'}$ nor to the QFS $\drtsq^{AA'}$. 
	
	To interpret the IS measurements, we performed quantified nuclear calculations of $\aver{r^2}$ and $\aver{r^4}$
	using nuclear density functional theory (DFT) with realistic energy density functionals (EDFs).
	The nuclear charge radial moments were obtained directly from calculated charge densities as discussed in Refs.~\cite{Reinhard2020,Reinhard2021}.
	To explore a
	possible span of predictions, we consider four different EDFs:
	Skyrme functionals  SV-min and UNEDF1, extended Skyrme functional RD-min, and the Fayans functional {\Fy}.
	The calculated $\drf$ are multiplied by $G^{(4)}_{\gamma\alpha}$ from atomic structure calculations to predict the nonlinearity for $G^{(4)}_{\gamma\alpha} \drf$.
	For details on the calculations, see Refs.~\cite{Reinhard2021c,Klupfel2009} and SMat.

	The predicted values of  $\aver{r^2}$ and $\aver{r^4}$ are impacted by several effects \cite{Brown1984,Otten1989,Reinhard2020,Reinhard2021}, including: the surface thickness of nuclear density that shows a pronounced particle-number dependence due to shell effects; the relativistic corrections that contain contributions from the intrinsic nucleon form factors; and nuclear deformation and pairing effects, which also give rise to the fragmentation \cite{Reinhard2021} of the single-particle spin-orbit strength that affects spin-orbit contributions to charge moments. Our DFT calculations take all these effects into account. In this respect, a King plot nonlinearity may be rooted in several nuclear structure effects impacting $\aver{r^2}$ and $\aver{r^4}$, not just one as discussed in Ref.~\cite{Allehabi2021}.
	As shown in the large inset to Fig.~\ref{fig:lambda_map}, our DFT results agree well with the observed direction in the $\lambda_\pm$ plane (see SMat for details).

	\begin{figure}
		\centering
		\includegraphics[width=\columnwidth]{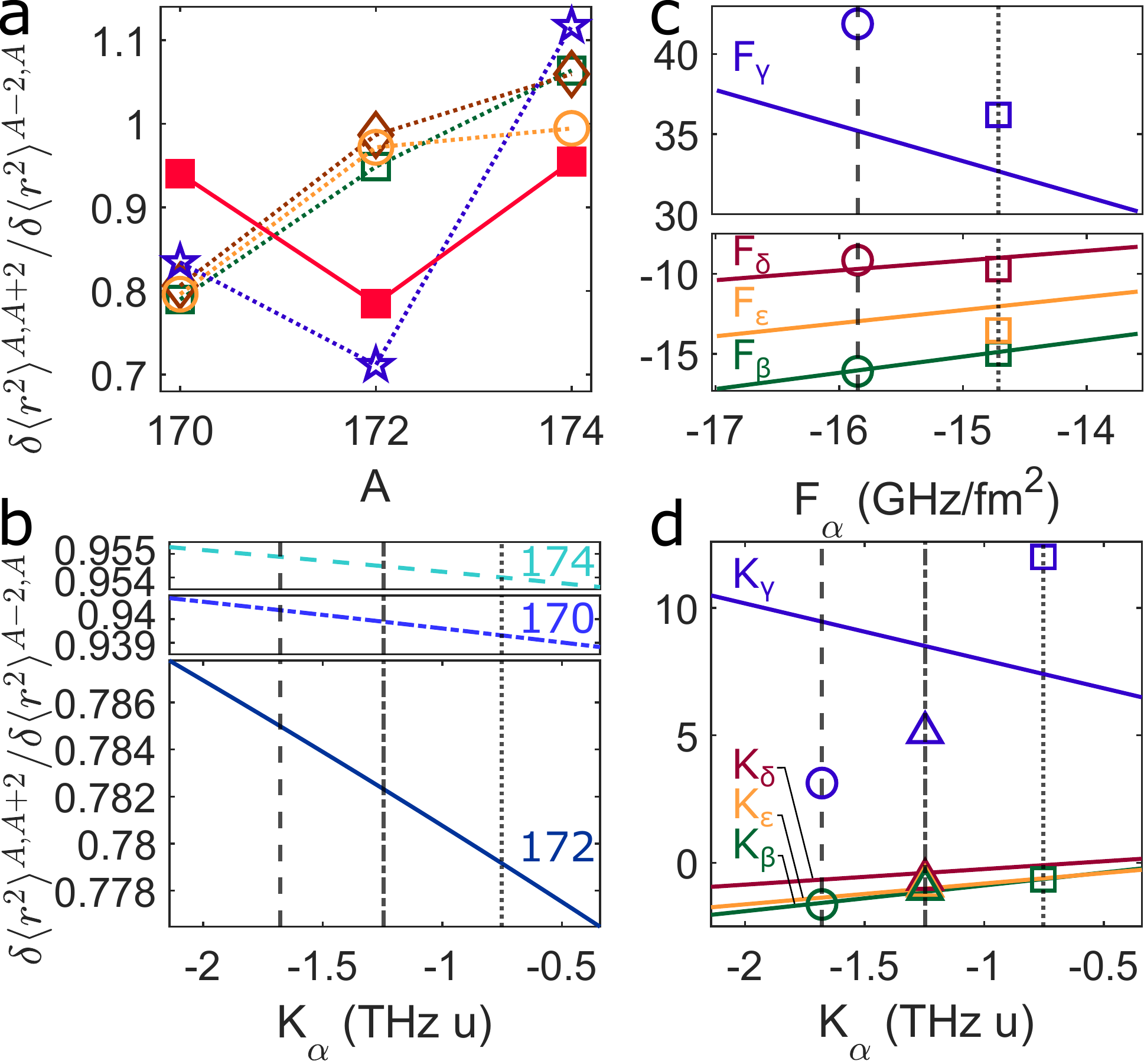}
		\caption{(a)  Comparison plot of derived values for the ratio of the mean-square nuclear radius differences between $(A,A+2)$ isotope pairs. 
			Open symbols mark the values derived from 
			nuclear calculations using SV-min, RD-min, UNEDF1, and {\Fy}  energy density functionals 
			(see Fig.~\ref{fig:lambda_map} for symbol assignments). The red filled square symbols are values derived from measured ISs on the 411\,nm transition in combination with mass shifts from configuration interaction (CI) \cite{Jonsson1996,Porsev2009,Fawcett1991,Biemont1998} calculations. (b) Plot of derived values for the ratio of the mean square nuclear radius between sequential isotope pairs as a function of $K_\alpha$, showing very weak dependence on $K_\alpha$. (c, d) Derived values of $F_\beta, F_\gamma, F_\delta, F_\epsilon$ ($K_\beta, K_\gamma, K_\delta, K_\epsilon$) as a function of $F_\alpha$ ($K_\alpha$), using the experimentally-determined ratios $F_{\kappa\alpha}$ ($K_{\kappa\alpha}$) for $\kappa={\beta,\gamma,\delta,\epsilon}$.
			In (b), (c), and (d), dashed (dotted) vertical lines and round (square) markers indicate values from CI calculations using GRASP2018 \cite{FroeseFischer2018} (\ambit{} \cite{Kahl2019}). Dash-dotted lines and open triangle markers correspond to CI and many-body perturbation theory (CI+MBPT) \cite{Dzuba1996} calculations using \ambit{}.} 
		\label{fig:dr2ratio}
	\end{figure}
	
	\begin{figure}
		\includegraphics[width=\columnwidth]{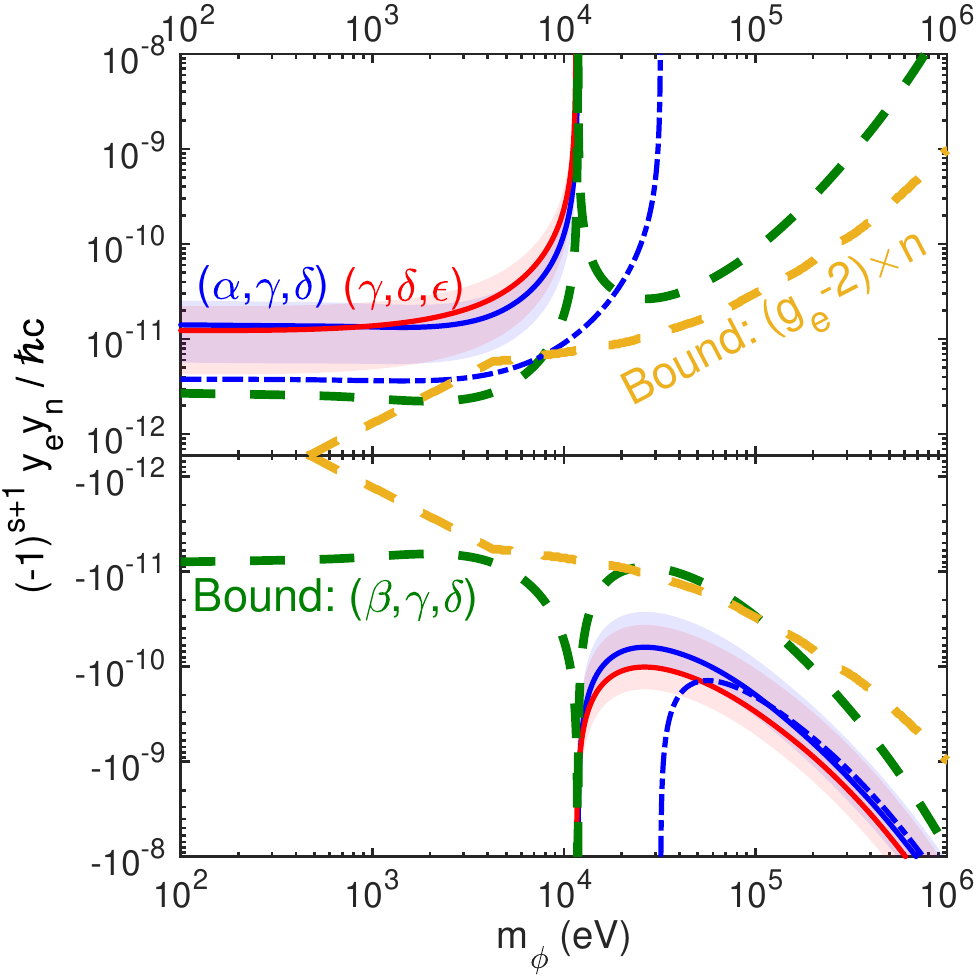}
		\caption{Product of coupling constants $y_e y_n$ of a new boson with spin $s$ versus boson mass $m_\phi$, derived from generalized-King-plot analyses \cite{Mikami2017,Berengut2020} of groups of three transitions $(\alpha,\gamma,\delta)$ (blue), $(\gamma,\delta,\epsilon)$ (red), and $(\beta,\gamma,\delta)$ (green), assuming that the observed second nonlinearity is dominated by a new boson. Dashed lines indicate the lower bounds of $y_ey_n$'s excluded magnitude. 
			Solid lines and shaded areas are center values and confidence intervals for configuration-interaction calculations using \ambit{}. [we show the $\approx 95\%$ confidence interval (see SMat) that arises from the statistical uncertainty in the measured ISs. The systematic uncertainty due to the atomic structure calculations is larger; the dash-dotted line shows the center value of $y_ey_n$ for the $(\alpha,\gamma,\delta)$ transition combination using GRASP2018 calculation results, for comparison.] 
			The yellow line indicates the bound derived from electron $g_e-2$ measurements \cite{Hanneke2008,Aoyama2012,Bouchendira2011,Davoudiasl2014} in combination with with neutron scattering measurements \cite{Barbieri1975,Leeb1992,Pokotilovski2006,Nesvizhevsky2008} from Ref.~\cite{Berengut2018}.}
		\label{fig:yeyn_vs_m}
	\end{figure}
	
	We can also directly compare the calculated changes in the nuclear size $\drt$ to the measured values. In order to be insensitive to the electronic factor $F$ in Eq.~(\ref{eq:IsotopeShift}), which can currently only be calculated with a typical uncertainty of $\lesssim$\,30\%, we plot in Fig.~\ref{fig:dr2ratio}(a) the ratios $\drt^{A,A+2}/\drt^{A-2,A}$ that can be determined from the experimental data with much higher accuracy. The nuclear calculations agree with the IS data to within 20\%.
	The ratios obtained from nuclear theory show monotonically increasing trends for the three EDFs SV-min, RD-min, and UNEDF1. Only {\Fy}  produces a trend that is consistent with data. This is yet another demonstration that the  Fayans functional is better adapted to local nonmonotonic trends in charge radius data, see also Refs.~\cite{Hammen2018,Gorges2019,Miller2019}.  We note
	that {\Fy} also provides a better description of nuclear quadrupole deformations as compared to other EDFs, see SMat for details. This  demonstrates that high-precision data on nuclear radii deliver important information for discrimination and further development of nuclear models.

	Our data also provide strong tests for electronic-structure calculations, as shown in Fig.~\ref{fig:dr2ratio}(c,d): The field (mass) shift coefficient $F_{\tau}$ ($K_{\tau}$) on one transition $\tau$ determines the coefficients on all other transitions $\kappa$ via the experimentally determined value of $F_{\kappa\tau}$ ($K_{\kappa\tau}$) (see SMat for details).

	While all transitions $\alpha, \beta, \gamma, \epsilon$ lie near a line through the origin in Fig.~\ref{fig:lambda_map}, there is a deviation from that line for all four transitions (plus the reference transition $\delta$) with 4.3$\sigma$ significance. (In contrast, the generalized King plot proposed in previous studies \cite{Mikami2017,Berengut2020} provides a test only for three transitions, giving significance less than 4$\sigma$ for any choices of three transitions, see SMat). This second nonlinearity is too large to be explained by the QFS, which is expected to be the next largest source of nonlinearity within the SM (see SMat). In Fig.~\ref{fig:yeyn_vs_m}, we show the strength of the coupling constant $y_e y_n$ for a new boson vs boson mass under the assumption that the new boson is the sole source of the second nonlinearity. Different combinations of measured transitions give similar values or bounds for the coupling strength $y_e y_n$ that is near or slightly exceeds the best other laboratory bounds given by the combination of $g-2$ measurements on the electron and neutron scattering experiments \cite{Hanneke2008,Aoyama2012,Bouchendira2011,Davoudiasl2014,Barbieri1975,Leeb1992,Pokotilovski2006,Nesvizhevsky2008,Berengut2018}.
	
	In the future, it should be possible to reduce the experimental uncertainties by up to four orders of magnitude to sub-Hz levels, as has been demonstrated with simultaneously trapped Sr$^+$ ions \cite{Manovitz2019}. In combination with improved electronic and nuclear calculations, it should then be possible to determine unambiguously if some part of the observed nonlinearity cannot be explained by physics within the SM. Besides better measurements on (more) transitions, it may also become possible to perform further measurements on unstable isotopes, which would allow the direct extraction (and elimination) of additional nuclear effects.
	
	\begin{acknowledgments}
		This work was supported by the NSF CUA and the U.S.\ Department of Energy, Office of Science, Office of Nuclear Physics under award numbers DE-SC0013365 and DE-SC0018083 (NUCLEI SciDAC-4 collaboration). This project has received funding from the European Union's Horizon 2020 research and innovation programme under the Marie Sklodowska-Curie grant agreement No 795121. J. C. B. is supported by the Australian Research Council (DP190100974). C. L. was supported by the U. S. Department of Defense (DoD) through the National Defense Science \& Engineering Graduate Fellowship (NDSEG) Program.
	\end{acknowledgments}

	\providecommand{\noopsort}[1]{}\providecommand{\singleletter}[1]{#1}%
	%

	\clearpage

	\beginsupplement

		\begin{center}
			\textbf{\large Supplemental Material}
		\end{center}

		\section{Experimental Details}
		\label{sec:experimentalDetails}
		
\begin{figure}
			\includegraphics[width=\columnwidth]{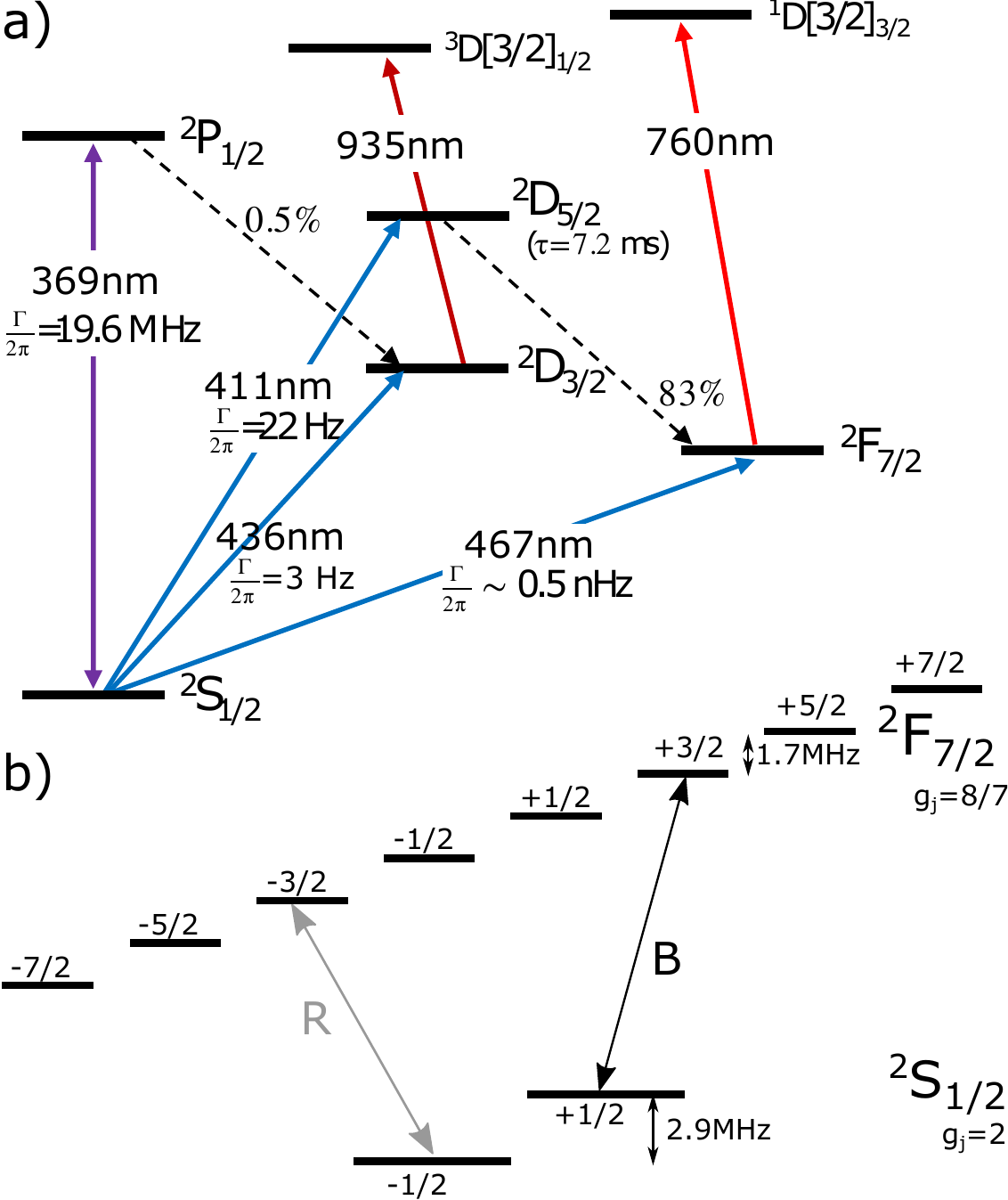}
			\caption{
				(a) Partial level diagram for the Yb$^+$ ion. In this work, we measure the 467-nm transition and use our previous measurements of the 411-nm and 436-nm transitions \cite{counts2020} to produce the King plots.
				(b) Zeeman levels of the ground $^2S_{1/2}$ and excited $^2F_{7/2}$ states of the 467-nm transition. We use a static magnetic field $B_0 = 1.05$\,G to split the Zeeman levels.
			}
			\label{fig:Yblevels}
		\end{figure}

\begin{figure}
			\includegraphics[width=\columnwidth]{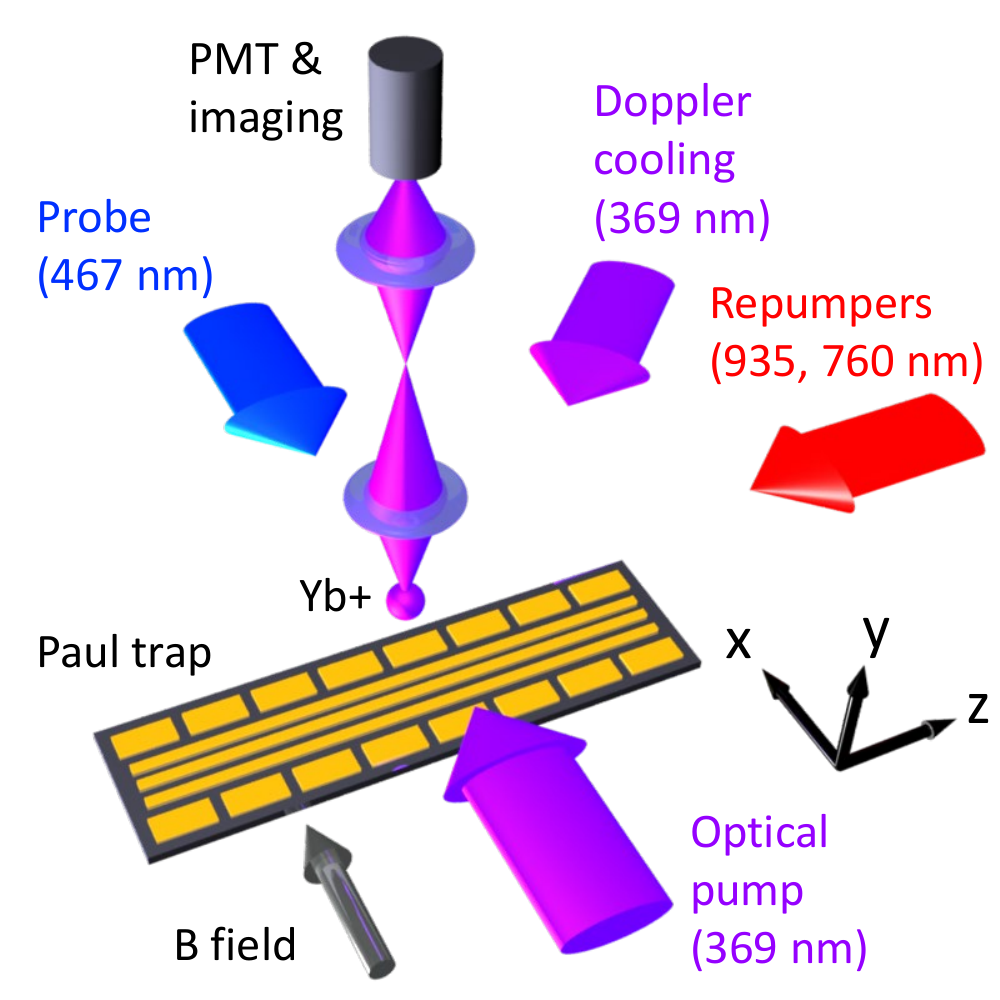}
			\caption{
				Schematic drawing of the experimental setup. A single ytterbium ion is trapped 135~$\mu$m away from the surface of a microfabricated planar Paul trap housed in an ultrahigh vacuum chamber. The propagation directions of the laser beams used for cooling, repumping, optical pumping, and probing the ion are indicated by labeled arrows. Fluorescence from the ion is collected using either a photo-multiplier tube (PMT) or a camera.
				The probe laser beam is linearly polarized along the trap axis (the $z$ direction in this figure).}
			\label{fig:ExperimentalSetup}
		\end{figure}

		Fig.~\ref{fig:Yblevels} shows the relevant transitions of Yb$^+$, with the narrow transitions used for isotope shift measurements depicted in blue. In this work, we probe the 467-nm transition and combine our measurements with our previous data on isotope shifts on the 411\,nm and 435\,nm transitions from Ref.~\cite{counts2020}, as well as data from transitions in neutral Yb from Refs.~\cite{ono2021, Figueroa2021}. We Doppler-cool the trapped ion (see Fig.~\ref{fig:ExperimentalSetup}) to $\sim500 \mu$K using 369-nm light which, aided by repumpers at 935\,nm and 760\,nm, drives a cycling transition between the $^2S_{1/2}$ and $^2P_{1/2}$ levels. Here we also report for the first time the frequencies (for all stable even isotopes) of the 760\,nm-transition which is used to repump the ion from the ${}^2F_{7/2}$ state (see Table~\ref{tab:Yb+_trap_freqs} and Sec.~\ref{sec:760freq}). The frequencies of the cooling transition at 370\,nm and the repumping transition for the ${}^2D_{3/2}$ state at  935\,nm for $^{168}$Yb$^+$ can be found in Table~\ref{tab:Yb+_trap_freqs} as well.
		
		To produce 467\,nm probe light, we frequency-double a Ti:Sapphire laser at 934\,nm with an M Squared ECD-X external cavity doubler. As described in detail in Ref.~\cite{counts2020}, we divert some of the light before the doubling cavity and pass it through an electro-optic modulator (EOM) to produce a sideband several gigahertz away from the carrier. We frequency-stabilize this sideband to a ultralow-thermal-expansion (ULE) high-finesse cavity using the Pound-Drever-Hall (PDH) protocol. Coarse frequency tuning of the probe light is then achieved simply by scanning the sideband frequency. Fine-tuning of the probe frequency is accomplished with an acousto-optic modulator (AOM) for the frequency-doubled light. Our probe beam power is \SI{160}{\milli\watt} and an achromatic lens is used to focus the beam to a waist ($1/e^2$-intensity radius) of 15\,$\mu$m at the ion.
		
		To determine the center of the 467\,nm transition, we drive two transitions, labeled $R$ and $B$ in Fig.~\ref{fig:Yblevels}(b), between symmetrically red and blue-detuned Zeeman components of the $^{2}S_{1/2}$ and $^{2}F_{7/2}$ states, and average their center frequencies. To minimize the effect of magnetic-field drifts, we interleave the scans of $R$ and $B$ (i.e. we record one data point on the frequency scan of $R$, then shift the frequency of the probe laser and take a data point on the frequency scan of $B$, then take another point on the scan of $R$ and so on). A 0.5\,s pause time is used after shifting the frequency between the $R$ and $B$ transitions to allow the laser to settle.
		
		\begin{figure}
			\centering
			\includegraphics[width=\columnwidth]{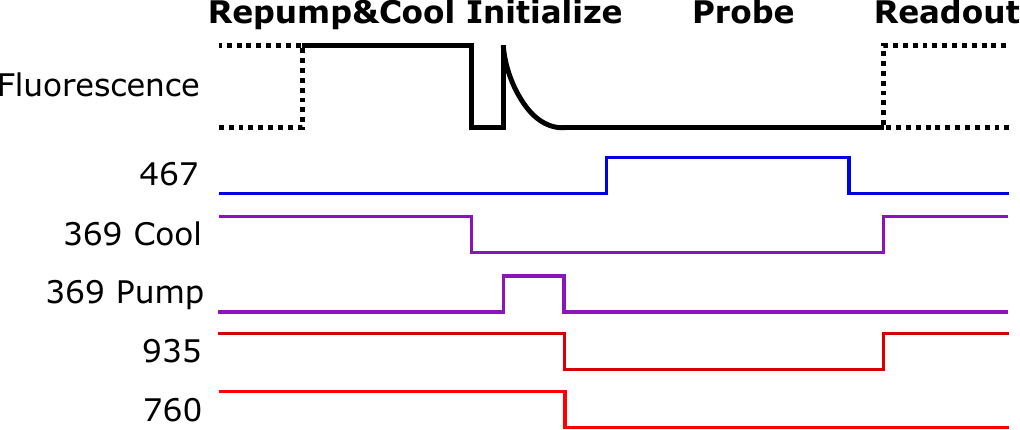}
			\caption{Time sequence of the experiment with wavelengths of lasers indicated.}
			\label{fig:pulse_sequence}
		\end{figure}
		
\begin{figure}
			\includegraphics[width=\columnwidth]{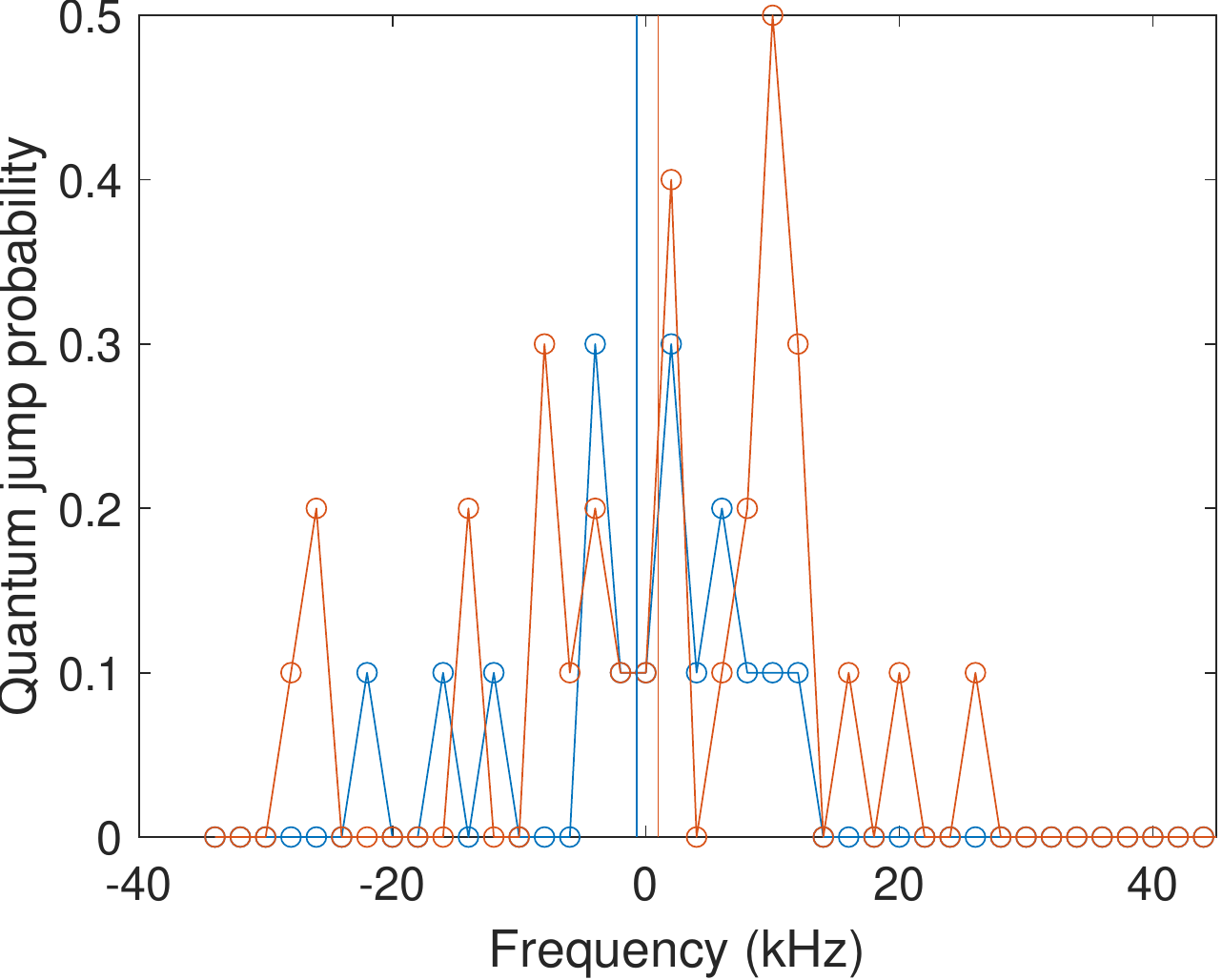}
			\caption{
				Example of simultaneous frequency scans of the probe transition on the $R$ and $B$ Zeeman components. Here the horizontal axis has been offset so that the Zeeman splitting is not shown. Vertical lines correspond to the statistical mean value for the frequency for each peak. 
			}
			\label{fig:QuantumJumpProbability}
		\end{figure}

		Fig.~\ref{fig:pulse_sequence} shows the 500\,ms-long laser pulse sequence used to drive the $467\,$nm transition, and
		Fig.~\ref{fig:ExperimentalSetup} depicts the polarization and propagation direction of the laser beams. The sequence begins by cooling the ion with 369\,nm light and optically pumping it with a circularly-polarized $369\,$nm-beam to one of the two $m_s=\pm \frac{1}{2}$ levels of the $^{2}S_{1/2}$ ground state. We record fluorescence emitted during cooling to confirm that the ion has been correctly initialized to the ground state and has not been shelved to one of the long-lived $D$ or $F$ states. (If the ion goes dark during this time, the corresponding period of the sequence is ignored in the data.) A probe laser pulse is then applied for 390\,ms, followed by readout of fluorescence by electron shelving \cite{counts2020,Taylor1997}. During readout, the 369\,nm-cooling light is again applied. When the transition has occurred, the ion will be in the $F_{7/2}$ state and will not fluoresce when illuminated by the cooling light. In this case, we consider the ion to have performed a quantum jump. However, if the ion did not make the transition and remained in the ground state, it will emit fluorescence on the cycling transition when driven by the cooling light. For each point on a frequency scan of the probe laser, we repeat this sequence until 10 successful periods of the sequence (i.e., the periods staring with the ion in the ground state) are observed, and determine what fraction of attempts resulted in a quantum jump. Fig.~\ref{fig:QuantumJumpProbability} shows the quantum jump probability versus probe laser frequency for one frequency scan.

	\begin{figure}
		\includegraphics[width=\columnwidth]{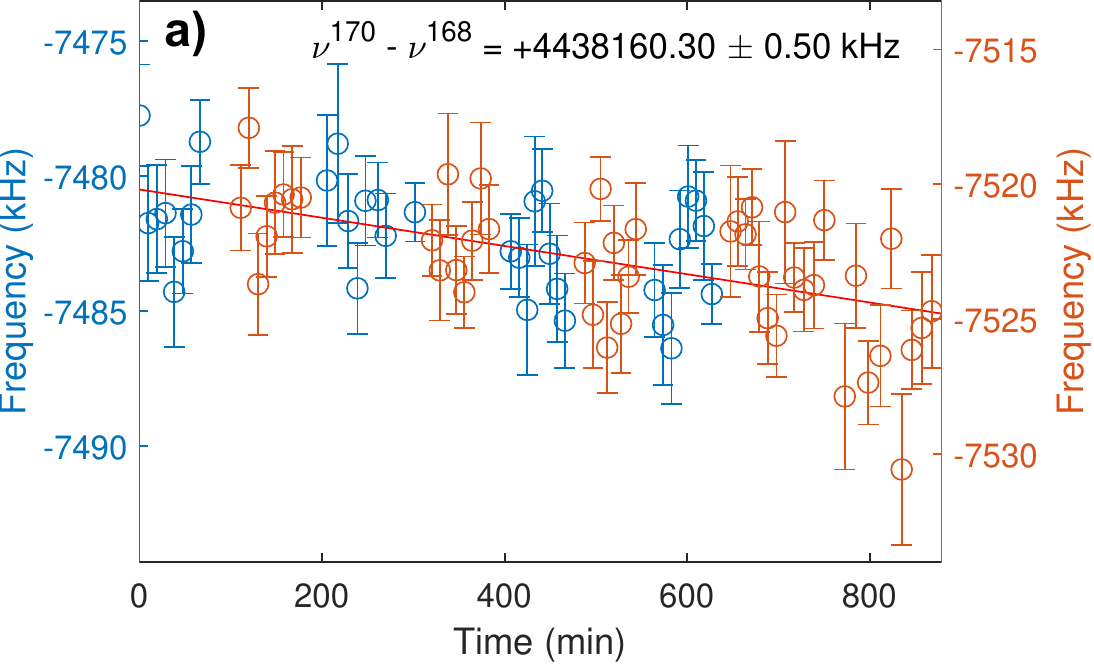}
		\includegraphics[width=0.85\columnwidth]{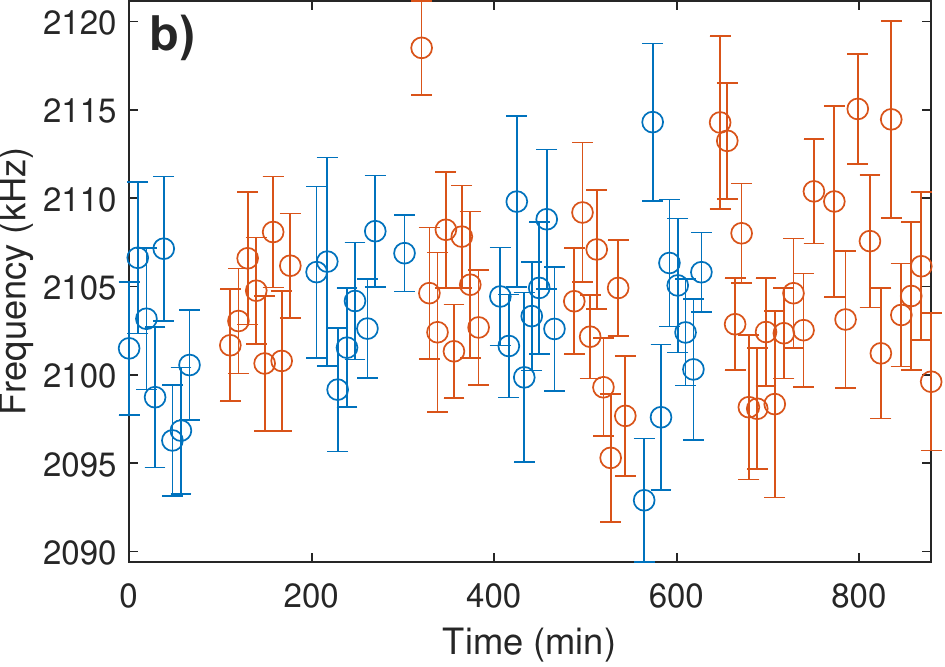}
		\caption{\label{fig:CommonDiffDrift}
			(a) Example of the common frequency drift of Zeeman peaks $R$ and $B$ for the IS measurement. Linear fit follows the drift of the ULE cavity frequency reference. Orange and blue color correspond to ${}^{168}\text{Yb}^+$ and ${}^{170}\text{Yb}^+$, respectively.
			(b) Example of the frequency difference between the Zeeman peaks $R$ and $B$. The Zeeman splitting is affected by variations of the external magnetic field, but not by the drifts of the ULE cavity.
		}
	\end{figure}

		We measure isotope shifts between pairs of isotopes by loading individual ions of each isotope in turn into the trap. (We can selectively photoionize different isotopes by tuning the frequency of our $399$\,nm photoionization laser.) For a given trapped isotope, we take at least 7 simultaneous frequency scans of the $R$ and $B$ Zeeman transitions before switching to the other isotope. This process is repeated at least four times throughout the course of a day of data taking (i.e., four data segments for each isotope). We calculate the common frequency drift for each pair of scans $R$ and $B$ to determine the center frequency, see Fig.~\ref{fig:CommonDiffDrift}(a). We then fit this data using a least-squares fit with varying offset [see the Supplemental Material of Ref.~\cite{counts2020}, Eq.~(S1)]. As described in detail in Sec.~\ref{sec:crosscheckMeasurements}, we measured all seven possible combinations of nearest-even-neighbor $(A,A+2)$ and next-to-nearest-even-neighbor isotope pairs $(A,A+4)$ in order to cross-check our measurement results for systematic errors, and to improve our precision.
		
		\subsection{Repumping from the ${}^2D_{3/2}$ and ${}^2F_{7/2}$-states}
		\label{sec:760freq}
		
		The state ${}^2F_{7/2}$ state is repumped by a 760\,nm laser beam that drives the $^{}2F_{7/2} \rightarrow {}^1D[3/2]_{3/2}$ transition \cite{Huntemann2012, Ransford2020, Jau2015, Sugiyama2000, Mulholland2019, Edmunds2021} with a $\lesssim 10$-ms time constant for a $\sim 7$\,mW beam focused to a waist of 100\,$\mu$m (consistent with Refs.~\cite{Huntemann2012,Ransford2020}). The absolute frequency of the 760\,nm beam is measured by and actively stabilized to a Fizeau wavemeter (HighFinesse/$\textrm{\AA}$ngstrom WS/7). The frequencies for this transition for all the isotopes, as well as the frequencies for the ${}^2D_{3/2} \rightarrow {}^3D[3/2]_{1/2}$ repumping transition at 935\,nm, are shown in Table~\ref{tab:Yb+_trap_freqs}.

		\begin{table*}
			\centering
			\caption{Measured values of absolute frequencies $\nu^{A}$ (upper table) and isotope shifts $\nu^{AA'} = \nu^{A} - \nu^{A'}$ (lower table) for the ${}^2S_{1/2} \rightarrow {}^2P_{1/2}$ (369\,nm) cooling transition, and the ${}^2D_{3/2} \rightarrow {}^3D[3/2]_{1/2}$ (935\,nm) and ${}^2D_{3/2} \rightarrow {}^1D[3/2]_{3/2}$ (760\,nm) repumping transitions. 100\,MHz, 60\,MHz, and 50\,MHz uncertainties in measured absolute frequencies of the 369\,nm, 760\,nm, and 935\,nm transitions, respectively, are specified by the manufacturer of the wavemeter (HighFinesse/$\textrm{\AA}$ngstrom WS/7). The differences in the transition frequencies are measured with better precision; 60\,MHz, 20\,MHz, and 20\,MHz are given as upper bounds of the uncertainties due to the drift of the wavemeter.}
			\label{tab:Yb+_trap_freqs}
			\begin{ruledtabular}
				\begin{tabular}{c|cc|cc|cl}
					\multirow{3}{*}{Isotope $A$} & \multicolumn{6}{c}{Transition frequency [THz]} \\
					& \multicolumn{2}{c}{369\,nm transition} & \multicolumn{2}{c}{935\,nm transition} & \multicolumn{2}{c}{760\,nm transition} \\
					& This work & Reference & This work & Reference & This work & \multicolumn{1}{c}{Reference} \\
					\hline
					168 & 811.29611(10) & & 320.562190(50) & & 394.432865(60) &  \\
					170 & 811.29439(10) & 811.29440(13) \cite{McLoughlin2011} & 320.565910(50) & 320.56593(7) \cite{McLoughlin2011} & 394.429590(60) &  \\
					172 & 811.29274(10) & 811.29284(13) \cite{McLoughlin2011} & 320.569390(50) & 320.56941(7) \cite{McLoughlin2011} & 394.426550(60)  & 394.4266\footnotemark[1] \cite{Jau2015} \\
					174 & 811.29146(10) & 811.29154(13) \cite{McLoughlin2011} & 320.572010(50) & 320.57201(7) \cite{McLoughlin2011} & 394.424145(60)  & 394.424\footnotemark[1] \cite{Sugiyama2000} \\
					& & & & & & 394.423900\footnotemark[1] \cite{Ransford2020} \\
					176 & 811.29025(10) & 811.29031(13) \cite{McLoughlin2011} & 320.574515(50) & 320.57449(7) \cite{McLoughlin2011} & 394.421885(60)  &  \\
				\end{tabular}
			\end{ruledtabular}
			\begin{ruledtabular}
				\begin{tabular}{c|ccc}
					Isotope pair $(A,A')$ & \multicolumn{3}{c}{Isotope shift [MHz]} \\
					& 369\,nm & 935\,nm & 760\,nm \\
					\hline
					(168,170) & 1 720(60) & -3 720(20) & 3 275(20) \\
					(170,172) & 1 650(60) & -3 480(20) & 3 040(20) \\
					(172,174) & 1 280(60) & -2 620(20) & 2 405(20) \\
					(174,176) & 1 210(60) & -2 505(20) & 2 260(20) \\
				\end{tabular}
			\end{ruledtabular}
			\footnotetext[1]{Uncertainty not specified.}
		\end{table*}
		
		\section{Data Analysis}
		\subsection{Determining the transition center frequency}
		As described in Sec.~\ref{sec:experimentalDetails}, for each isotope, we determine the $^{2}S_{1/2}\rightarrow^{2}F_{7/2}$ transition center frequency by driving two symmetrically detuned transitions between Zeeman levels of the ground and excited states, transitions $R$ and $B$ [see Fig.~\ref{fig:Yblevels}(b)], and averaging their center frequencies. We scan the probe laser frequency and plot the quantum jump probability versus frequency, see Fig.~\ref{fig:QuantumJumpProbability}. To determine the center frequencies of $R$ and $B$ from our data, we take the statistical mean of our data points. This allows us to determine the center frequency without assuming a known lineshape fit function. However, this method of determining the center is susceptible to a small amount of frequency pulling if our scan range is not centered at the transition resonance frequency (we discuss this effect in detail in Sec.~\ref{sec:frequencyPulling} and determine that it is significantly smaller than our leading error sources). An analysis where we fit lineshapes Gaussian function with background offset to find the transition centers gives deviations that are smaller than our statistical error bars.

		\subsection{Inverse-mass difference $\mu^{AA'}$}
		
		Assuming that the errors of the measured masses of all five isotopes of interest are uncorrelated, we calculate the inverse-mass differences $\mu^{AA'} = 1/m^A - 1/m^{A'}$ (where $m^A$ is the mass of the $^{A}$Yb$^{+}$ ion) and the correlations between different $\mu^{AA'}$ for different isotope pairs $(A,A')$. The values for $m^A$, $\mu^{AA'}$, and their correlations, are listed in Table~\ref{tab:Yb+mass}, Table~I in the main text, and Table~\ref{tab:mu_corr}, respectively.
		
		The uncertainties in measured atomic masses of Yb isotopes appear as $x$-errors in frequency-normalized King plots [see Eq.~(2) in the main text]. The effect of mass uncertainties is largely suppressed due to the small slopes in King plots (given by $K_{\kappa\chi}$, the two-transition mass shift coefficients). The maximum uncertainty in measured Yb masses $m^A$ is currently $10^{-7}$\,u (limited by the $^{168}$Yb isotope), translating into an uncertainty of $3.6\times10^{-12}$\,u$^{-1}$ in $\mu^{AA'}$ (see Table~\ref{tab:Yb+mass} and Table~I in the main text). This mass uncertainty leads to a  $22$\,Hz uncertainty in $K_{\kappa\chi}\mu^{AA'}$, for the maximum value of $K_{\kappa\chi} = 6002$\,GHz$\cdot$u (see Table~\ref{tab:el_factors_2}), which is smaller than the IS uncertainty in this work. As the precision of IS measurements increases to $O(1\,\text{Hz})$ \cite{Furst2020, ono2021} and further to $O(1\,\text{mHz})$ \cite{Manovitz2019}, the atomic masses should be measured with uncertainty below $O(10^{-8}\,\text{u})$ and $O(10^{-11}\,\text{u})$, respectively, to avoid uncertainties in King plots being dominated by the mass uncertainties.
		
		\begin{table}
			\caption{Masses of isotopes from Ref.~\cite{Nesterenko2020} for the $^{168}$Yb$^+$ ion, and from Refs.~\cite{AME2016_1, AME2016_2, Rana2012} for the remaining isotopes. The ionization energy of 6.254~eV for a neutral Yb atom \cite{NIST_ASD, Aymar1980} is used to calculate the ion mass from the neutral-atom mass.}
			\centering
			\begin{ruledtabular}
				\begin{tabular}{cc}
					Isotope $A$ & $m^A$ [u]\\
					\hline
					168 & 167.93389132(10)$\,\,\,$ \\ 
					170 & 169.934767246(11) \\ 
					172 & 171.936386659(15) \\
					174 & 173.938867548(12) \\
					176 & 175.942574709(16)
				\end{tabular}
			\end{ruledtabular}
			\label{tab:Yb+mass}
		\end{table}
		
		\begin{table}
			\caption{Correlation coefficients between inverse-mass differences $\mu^{AA'}$ for different nearest-even neighboring isotope pairs.}
			\centering
			\begin{ruledtabular}
				\begin{tabular}{c|cccc}
					Isotope pair ($A$,$A'$) & (168,170) & (170,172) & (172,174) & (174,176) \\
					\hline
					(168,170) & & -0.4430 & 0.1879 & -0.0906 \\ 
					(170,172) &  & & -0.4241 & 0.2045 \\
					(172,174) &  &  & & -0.4822 \\
					(174,176) &  &  &  & \\
				\end{tabular}
			\end{ruledtabular}
			\label{tab:mu_corr}
		\end{table}
		
		\subsection{Cross-checks and improved isotope shifts}
		\label{sec:crosscheckMeasurements}
		
		To check for systematic errors and improve our uncertainties, we perform additional redundant measurements of the IS by measuring next-next-even neighbor ISs. Then each measured quantity is a linear combination of other quantities (e.g., $\nu^{170,174} = \nu^{170,172} + \nu^{172,174})$. By combining the measured values, each of the quantities can be better estimated with a reduced uncertainty, at the slight complication of correlations between different quantities. In this work, the measured ISs for nearest-even neighboring isotope pairs $(166 + 2i,166 + 2i + 2)$, $i = 1, \cdots, 4$, and next-nearest-even neighboring isotope pairs $(166 + 2i,166 + 2i + 4)$, $i = 1, \cdots, 3$, producing improved values of the nearest-even neighboring isotope pairs' ISs from the following relation.
		
		\begin{equation}
			\underbrace{
				\left[
				\begin{array}{c}
					\nu^{168,170}_{\alpha} \\
					\nu^{170,172}_{\alpha} \\
					\nu^{172,174}_{\alpha} \\
					\nu^{174,176}_{\alpha} \\
					\nu^{168,172}_{\alpha} \\
					\nu^{170,174}_{\alpha} \\
					\nu^{172,176}_{\alpha}
				\end{array}
				\right]
			}_{\vec{y}}
			= \underbrace{
				\left[
				\begin{array}{cccc}
					1 & 0 & 0 & 0  \\
					0 & 1 & 0 & 0  \\
					0 & 0 & 1 & 0  \\
					0 & 0 & 0 & 1  \\
					1 & 1 & 0 & 0  \\
					0 & 1 & 1 & 0  \\
					0 & 0 & 1 & 1
				\end{array}
				\right]
			}_{X}
			\underbrace{
				\left[
				\begin{array}{c}
					\nu^{168,170}_{\alpha} \\
					\nu^{170,172}_{\alpha} \\
					\nu^{172,174}_{\alpha} \\
					\nu^{174,176}_{\alpha}
				\end{array}
				\right]
			}_{\vec{\beta}}
			\label{eq:Redundance}
		\end{equation}
		Finding the best estimates of the four isotope shifts $\vec{\hat{\beta}}$ from the observations $\vec{y}$ becomes a typical least square problem $\vec{y} = X\vec{\beta}$. The improved IS values $\vec{\hat{\beta}}$ are obtained via a weighted least squares fit with the weights given by inverse-squared measurement uncertainties. The values, errors, and correlations of $\vec{\hat{\beta}}$ are listed in Table~\ref{tab:improved_IS}.
		
		\begin{table*}
			\caption{Improved values and errors of ISs $\nu^{AA'} = \nu^{A} - \nu^{A'}$ between nearest-neighboring even isotope pairs (diagonal elements; in kHz) from the redundant measurements listed in Table~I in the main text. Correlation coefficients between $\nu^{AA'}$ for different isotope pairs are given as off-diagonal elements.}
			\centering
			\begin{ruledtabular}
				\begin{tabular}{lc|cccc}
					Transition & Isotope pair $(A,A')$ & (168,170) & (170,172) & (172,174) & (174,176) \\
					\hline
					\multirow{4}{*}{\makecell[c]{$\gamma$: 467\,nm\\(this work)}} & (168,170) & -4 438 160.85(38) & -0.4430 & 0.1879 & -0.0906 \\ 
					& (170,172) &  & -4 149 190.66(32) & -0.4241 & 0.2045 \\
					& (172,174) &  &  & -3 132 321.38(33) & -0.4822 \\
					& (174,176) &  &  &  & -2 976 391.58(37) \\
					\hline
					\multirow{4}{*}{$\alpha$: 411\,nm \cite{counts2020}} & (168,170) & 2 179 098.93(21) &  &  & \\ 
					& (170,172) &  & 2 044 854.73(30) & -0.3286 &  \\
					& (172,174) &  &  & 1 583 068.35(31) &  \\
					& (174,176) &  &  &  & 1 509 055.29(28) \\
					\hline
					\multirow{4}{*}{$\beta$: 436\,nm \cite{counts2020}} & (168,170) & 2 212 391.85(37) &  &  & \\ 
					& (170,172) &  & 2 076 421.04(28) & -0.4235 &  \\
					& (172,174) &  &  &  1 609 181.29(20) & \\
					& (174,176) &  &  &  & 1 534 144.06(24) \\
					\hline
					\multirow{4}{*}{$\delta$: 578\,nm \cite{ono2021}} & (168,170) & 1 358 484.4763(23) &  &  & \\ 
					& (170,172) &  & 1 275 772.0060(30) & -0.7546 &  \\
					& (172,174) &  &  &  992 714.5867(23) & \\
					& (174,176) &  &  &  & 946 921.7751(30) \\
					\hline
					\multirow{4}{*}{$\epsilon$: 361\,nm \cite{Figueroa2021}} & (168,170) & 1 781 784.73(55) & -0.2210 &  & \\ 
					& (170,172) &  & 1 672 021.40(29) &  &  \\
					& (172,174) &  &  &  1 294 454.41(21) & -0.3885 \\
					& (174,176) &  &  &  & 1 233 942.14(25) \\
				\end{tabular}
			\end{ruledtabular}
			\label{tab:improved_IS}
		\end{table*}
		
\begin{figure}
			\centering
			\includegraphics[width=\columnwidth]{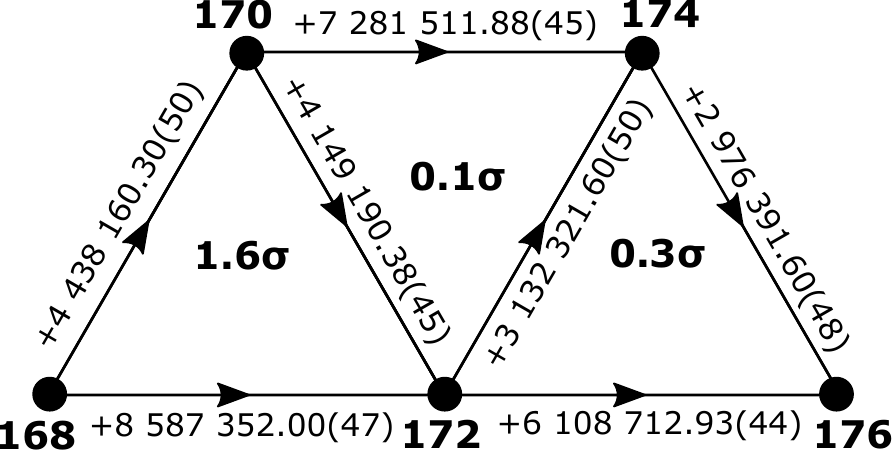}
			\caption{Measured values of ISs (values along edges in kHz) for different pairs of Yb$^+$ isotopes (vertices labeled with the mass numbers $A$ of isotopes $^A$Yb$^+$) and consistency of values forming shortest loops ($\sigma$-significance in the center of each triangle). The IS value $\nu^{AA'} = \nu^{A} - \nu^{A'}$ is shown for an edge directed from vertex $A$ to vertex $A'$. The measured values agree overall with 0.86$\sigma$ significance.}
			\label{fig:isotope_trangle}
		\end{figure}

		The self-consistency of three measured IS values $\nu^{AA'}$, $\nu^{A'A''}$ and $\nu^{AA''}$ that involves three isotopes $A$, $A'$, and $A''$ can be tested by, e.g., checking if the difference of $\nu^{AA'} + \nu^{A'A''}$ and $\nu^{AA''}$ is within the combined measurement uncertainty $\sqrt{(\Delta\nu^{AA'})^2 + (\Delta\nu^{A'A''})^2 + (\Delta\nu^{AA''})^2}$ (see Fig.~\ref{fig:isotope_trangle}). Eq.~(\ref{eq:Redundance}) serves as a test of the self-consistency of our measurements; we find agreement of our data with the linear relation given by Eq.~(\ref{eq:Redundance}) within $0.86\sigma$.
		
		\subsection{Fitting points in King plots}
		\label{sec:fitting}
		
		Fitting our data requires a procedure that both accounts for the fact that our data points have error bars along the $x$ axis of the King Plot (as well as along the $y$ axis), and for the existence of correlations between the data points. These correlations arise mainly because we make redundant measurements of the isotope shifts and use them to reduce the uncertainties in the pairwise shifts. For data with uncertainties only along the $y$ axis, correlations can be straightforwardly accounted for by using a generalized least squares (GLS) fitting procedure, which takes as an input the covariance matrix of the data. To adapt the procedure to account for errors along $x$, we translate the $x$-errors into $y$-errors via the slope of the fit line, and then perform an iterative GLS fit to our data.
		
		For the two-dimensional (2D) King plot (Fig.~1 in the main text), the effect of $x$-errors and correlations on the fit result is not significant if the King plot is frequency-normalized (see Eq.~(2) in the main text) \cite{counts2020}. This is is because the $x$-errors, when propagated to the $y$ direction, are significantly smaller than the $y$-errors. This is true in general for heavy atomic species where the field shift (FS) is significantly larger than the mass shift (MS).
		
		Table~\ref{tab:el_factors_2} compares the results of the fit with and without $x$-errors and correlations; we see that the two agree. This provides one of the main motivations to use the frequency-normalized King plot instead of the more conventional inverse-mass-normalized King plot, as one can easily obtain reliable fitting results via standard GLS fitting procedures, which have analytic solutions.
		
		\subsection{Analysis of the nonlinearity pattern}
		\label{sec:nonlin_analysis}
		We use the following vector notation for isotope-pair-dependent parameters:
		\begin{equation}\label{eq:vector_notation}
			\vec{x} = (x^{A_1 A'_1}, x^{A_2 A'_2}, x^{A_3 A'_3}, x^{A_4 A'_4})
		\end{equation}
		where $A_k A'_k$ are the isotopes in the $k$-th pair.
		This notation provides an alternative view of King plot: if the King plot is linear, then the vector with components $\overline{\nu}^{AA'}_{\kappa} = \nu^{AA'}_{\kappa}/ \nu^{AA'}_{\tau}$ resides in the plane that two \textit{King vectors} $\vec{1}$ and $\vec{\overline{\mu}}$ define, with $f_{\kappa\tau}$ and $K_{\kappa\tau}$ as the coefficient of the vectors, respectively. 
		
		Since the vectors are four-dimensional, one can define two vectors $\vec{\Lambda}_+$ and $\vec{\Lambda}_-$ (that we call \textit{nonlinearity vectors}) that span the space orthogonal to the \textit{King plane}. When measured normalized ISs $\overline{\nu}_{\kappa}^{AA'}$ do not exactly lie in the King plane, the out-of-plane component can be decomposed along the nonlinearity vectors with components $\lambda_+$ and $\lambda_-$. In other words, the King plane and nonlinearity plane corresponds to the best fit and the remaining residuals of the ordinary-least-square fit in the King plot, respectively. There is an infinite number of ways to define nonlinearity vectors, and we suggest the following unit vectors:
		\begin{equation}\label{eq:nonlinearity_vectors}
			\begin{aligned}
				\vec{\Lambda}_+ &\propto (\overline{\mu}_{3} - \overline{\mu}_{2}, \overline{\mu}_{1} - \overline{\mu}_{4}, \overline{\mu}_{4} - \overline{\mu}_{1}, \overline{\mu}_{2} - \overline{\mu}_{3}) \\
				\vec{\Lambda}_- &\propto (\overline{\mu}_{4} - \overline{\mu}_{2}, \overline{\mu}_{1} - \overline{\mu}_{3}, \overline{\mu}_{2} - \overline{\mu}_{4}, \overline{\mu}_{3} - \overline{\mu}_{1})
			\end{aligned}
		\end{equation}
		where $\overline{\mu}_k \equiv \overline{\mu}^{A_kA'_k}$. The proposed nonlinearity vectors have several advantages: They have a fairly simple, linear form while being orthogonal to the King vectors. This simplifies error propagation in the measured quantities $\mu^{AA'}$, $\nu_{\tau}^{AA'}$, and $\nu_{\kappa}^{AA'}$ to $\vec{\Lambda}_\pm$ and $\lambda_\pm$. Furthermore, the $\vec{\Lambda}_+$ and $\vec{\Lambda}_-$ vectors represent zigzag (+\,--\,+\,--) and curved (+\,--\,--\,+) patterns of nonlinearity if $m_1$ to $m_4$ are in increasing order (i.e., $m_k < m_{k+1}$), replacing the role of $\vec{\zeta}_\pm = (1,-1,\pm1,\mp1)$ in our previous work \cite{counts2020}.
		
		A drawback of the above basis is that in general $\vec{\Lambda}_+$ and $\vec{\Lambda}_-$ are not orthogonal to each other. One can alternatively, for instance, keep $\vec{\Lambda}_+$ and define $\vec{\Lambda}_-$ as the vector that is orthogonal to the two King vectors and $\vec{\Lambda}_+$. With this kind of choices, however, the propagation of the uncertainty is less straightforward. Interestingly, the values of $\overline{\mu}^{AA'}$ for Yb are such that the nonlinearity vectors in Eq.~(\ref{eq:nonlinearity_vectors}) are  very close to being perpendicular to each other ($\vec{\Lambda}_+ \cdot \vec{\Lambda}_- = 0.0014$).
		
		The points representing the measured ISs in the $(\lambda_+, \lambda_-)$ plane of expansion coefficients for the $\alpha$: ${}^2S_{1/2} \rightarrow {}^2D_{5/2}$ (411\,nm), $\beta$: ${}^2S_{1/2} \rightarrow {}^2D_{3/2}$ (436\,nm), and $\gamma$: ${}^2F_{7/2} \rightarrow {}^2D_{3/2}$ (467\,nm) transitions in Yb$^+$ ions; and the $\delta$: ${}^1S_0 \rightarrow {}^3P_0$ (578\,nm), and $\epsilon$: ${}^1S_0 \rightarrow {}^1D_2$ (361\,nm) transitions in neutral Yb atoms, using $\vec{\Lambda}_\pm$ as given by Eq.~(\ref{eq:nonlinearity_vectors}), for two different choices of the transition for the normalization (\textit{reference transition}) are shown in Fig.~2 in the main text. The $\lambda_\pm$ plane referenced to the $\delta$ transition ($\lambda^{(\delta)}_\pm$) is introduced as the main graph because the ISs for the $\delta$ transition have been measured with a precision that is much higher than the other transitions, so that the uncertainties in the $\lambda_\pm$ values (shown as ellipses) for different transitions are not correlated to each other.

		\subsection{Three-dimensional (3D) King plot}
		\label{sec:3Dking}
		The 3D King plot is a special case of the generalized King plot introduced in Ref.~\cite{Mikami2017}. If there is one source of isotope shifts in addition to the MS and FS, denoted as $X_\alpha x^{AA'}$, then the isotope shift is given as
		\begin{equation}\label{eq:3Dking_IS}
			\nu_{\alpha}^{AA'} = K_\alpha \mu^{AA'} + F_\alpha \drt^{AA'} + X_\alpha x^{AA'} + Y_\alpha y^{AA'}
		\end{equation}
		where $Y_\alpha y^{AA'}$ is a small fourth contribution to the IS shifts (i.e., $X_\alpha x^{AA'} \gg Y_\alpha y^{AA'}$) that we want to test for. If the ISs of three transitions $\nu_\alpha$, $\nu_\beta$, and $\nu_\gamma$ are measured, then the unknown quantities $\drt^{AA'}$ and $x^{AA'}$ can be eliminated from the expression by solving Eq.~(\ref{eq:3Dking_IS}) for the three transitions.
		
		\begin{equation}
			\begin{bmatrix} \nu_{\alpha}^{AA'} \\ \nu_{\beta}^{AA'} \\ \nu_{\gamma}^{AA'} \end{bmatrix}
			- \begin{bmatrix} Y_\alpha \\ Y_\beta \\ Y_\gamma \end{bmatrix} y^{AA'}
			=
			\underbrace{
				\begin{bmatrix}
					K_\alpha & F_\alpha & X_\alpha \\ 
					K_\beta & F_\beta & X_\beta \\ 
					K_\gamma & F_\gamma & X_\gamma
				\end{bmatrix}
			}_{T}
			\begin{bmatrix} \mu^{AA'} \\ \drt^{AA'} \\ x^{AA'} \end{bmatrix}
		\end{equation}
		\begin{equation}\label{eq:3Dking_mu}
			\mu^{AA'} = \sum_{\chi = \alpha, \beta, \gamma} \left(T^{-1}\right)_{1\chi} \left(\nu_{\chi}^{AA'} - Y_\chi y^{AA'}\right)
		\end{equation}
		By rearranging Eq.~(\ref{eq:3Dking_mu}), we obtain the expression for inverse-mass-normalized 3D King plot as follows:
		\begin{equation}\label{eq:3Dking}
			\bbar{\nu}_{\gamma}^{AA'} = K_{\gamma\beta\alpha} + f_{\gamma\beta\alpha} \bbar{\nu}_{\alpha}^{AA'} + f_{\gamma\alpha\beta} \bbar{\nu}_{\beta}^{AA'} +
			Y_{\gamma\beta\alpha} \bbar{y}^{AA'}
		\end{equation}
		where $\bbar{z}^{AA'} \equiv z^{AA'}/\mu^{AA'}$ ($z \in \{ \nu_\alpha, \nu_\beta, \nu_\gamma, y \}$) are inverse-mass-normalized quantities \cite{counts2020},
		
		\begin{align}
			f_{\gamma\beta\alpha} &= \frac{\frac{F_\gamma}{F_\beta} - \frac{X_\gamma}{X_\beta}}{\frac{F_\alpha}{F_\beta} - \frac{X_\alpha}{X_\beta}} = \frac{X_{\gamma\beta}}{X_{\alpha\beta}} \text{ and} \\
			f_{\gamma\alpha\beta} &= \frac{\frac{F_\gamma}{F_\alpha} - \frac{X_\gamma}{X_\alpha}}{\frac{F_\beta}{F_\alpha} - \frac{X_\beta}{X_\alpha}} = \frac{X_{\gamma\alpha}}{X_{\beta\alpha}}
		\end{align}
		are the slopes of the plane in 3D the King plot along the axes corresponding to transitions $\alpha$ and $\beta$, respectively,
		
		\begin{equation}\label{eq:3Dking_Zgammaalphabeta}
			\begin{aligned}
				Z_{\gamma\beta\alpha} &= Z_\gamma - f_{\gamma\beta\alpha} Z_\alpha -     f_{\gamma\alpha\beta} Z_\beta \\
				&= Z_{\gamma\alpha} - \frac{X_{\gamma\alpha}}{X_{\beta\alpha}} Z_{\beta\alpha} \\
				&= Z_{\beta\alpha}(z_{\gamma\alpha\beta} - f_{\gamma\alpha\beta})
			\end{aligned}
		\end{equation}
		where $Z = K, Y$ are the $z$-intercept of the plane and the electronic factor associated with the nonlinearity source $y^{AA'}$, respectively, and $z_{\gamma\alpha\beta} \equiv z_{\gamma\alpha}/z_{\beta\alpha}$ (see the main text for the definition of $z_{\kappa\chi}$ and $Z_{\kappa\chi}$ for $\chi \neq \kappa$). 
		There is only one source of nonlinearity in the 2D King plot (namely $X_\alpha x^{AA'}$) if all the points in 3D King plot lie in a plane, barring the case that $\vec{y}$ can be decomposed to $\vec{\mu}$, $\vec{\drt}$, and $\vec{x}$ (see Sec.~\ref{sec:nonlin_analysis} for the vector notation), or $Y_\alpha$, $Y_\beta$, and $Y_\gamma$ cancel out in $Y_{\gamma\beta\alpha}$. Therefore, fitting the points in the 3D King plot serves as a test if there are one or two contributions to the IS besides the MS and the FS.
		
		We fit 3D King plots in a similar way as 2D King plots (see Sec.~\ref{sec:fitting}). Uncertainties and correlations in the $x$ and $y$ values are propagated to the $z$ direction, and iterative GLS fits are performed.
		
\begin{figure}
			\includegraphics[width=\columnwidth]{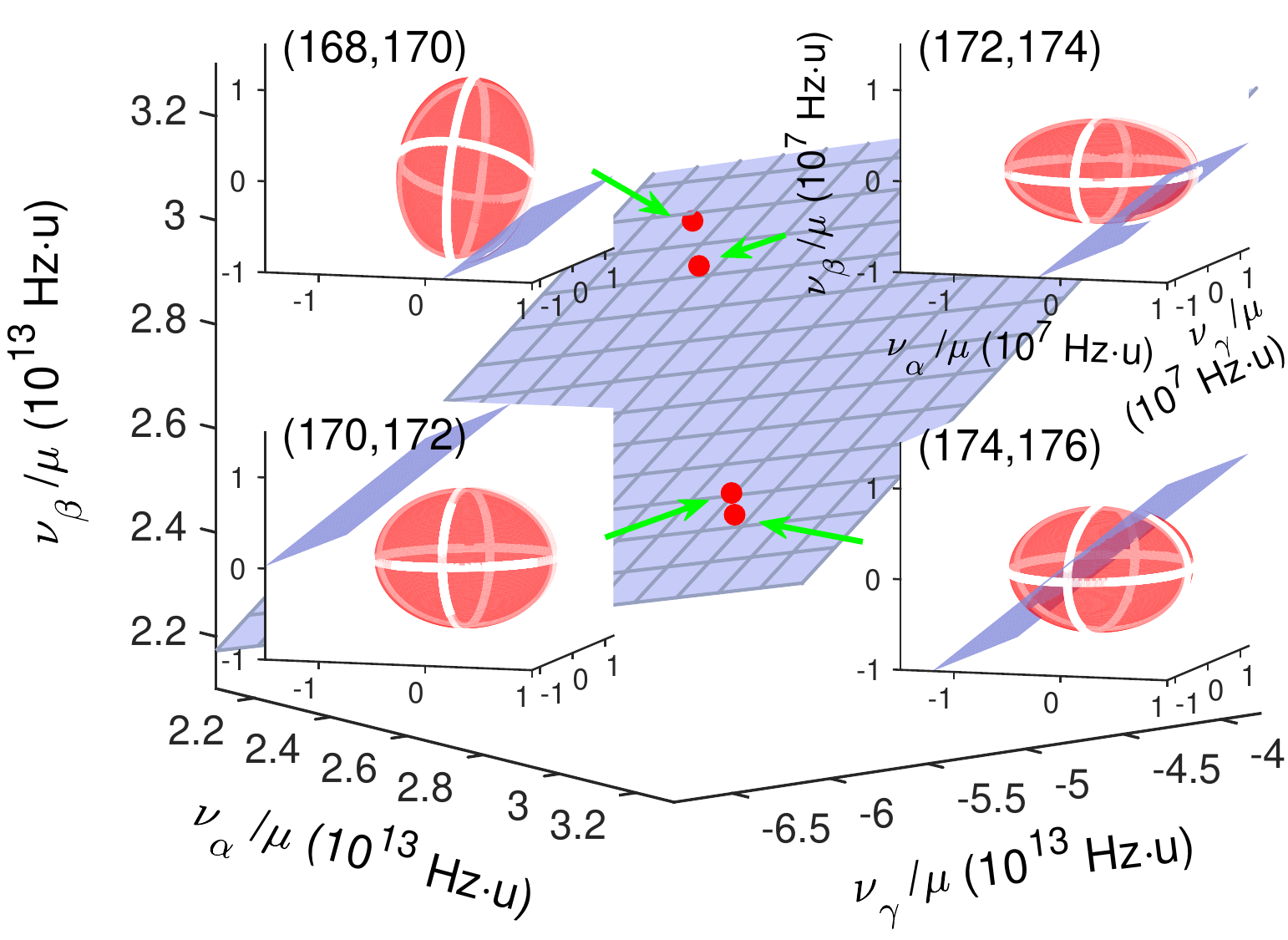}
			\caption{Plane fitted to a 3D inverse-mass-normalized King plot constructed from isotope shifts measured on the $\alpha = 411$\,nm \cite{counts2020}, $\beta = 436$\,nm \cite{counts2020} and $\gamma = 467$\,nm (this work) transitions for nearest-neighbor pairs of even Yb$^+$ isotopes, as described by Eq.~(\ref{eq:3Dking}). Insets display magnified view of each point to show deviation from the fitted plane. The origin of the inset axes has been set to the center of each point. The red ellipsoids depict 1$\sigma$ confidence intervals of the data. The fit to the plane gives $3.2\sigma$ significance of nonlinearity (see Table~\ref{tab:el_factors_3}). Each point in the King plot is correlated with other points (see Sec.~\ref{sec:fitting}).}
			\label{fig:3Dking}
		\end{figure}

		An example 3D King plot and the linear fit are shown in Fig.~\ref{fig:3Dking} for the $(\alpha,\gamma,\beta)$ transitions.
		
		\subsubsection{\label{sec:3Dking_equiv_lambda}Equivalence between 3D King-plot linearity and single-source fit in $\lambda_{\pm}$ plane}
		
		\begin{figure}
			\centering
			\includegraphics{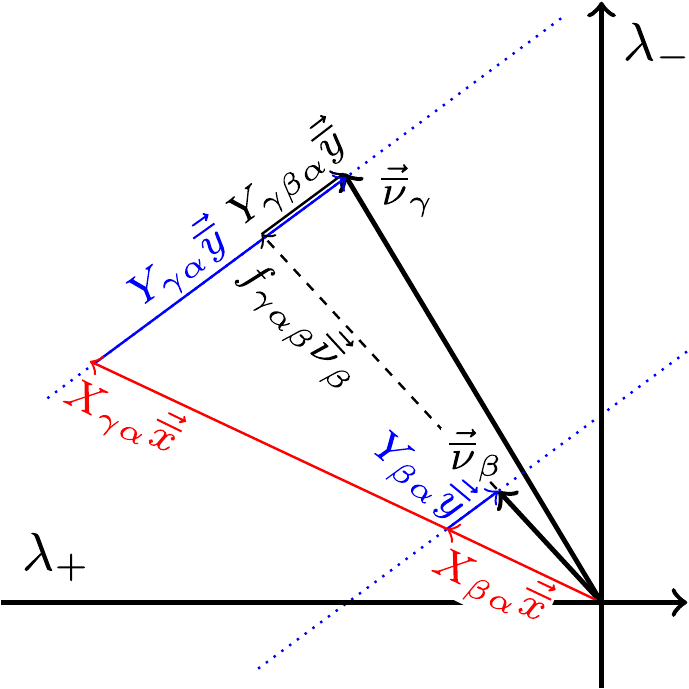}
			\caption{\label{fig:lambda_3Dking}
				Two-source-nonlinearity analysis in the $(\lambda_+,\lambda_-)$ map with reference transition $\alpha$. Thick black arrows indicate the measured $\nu_\beta$ and $\nu_\gamma$. The nonlinearity from $x^{AA'}$ ($y^{AA'}$) is coded with red (blue) color. The blue dotted line shows the direction of $\lambda_\pm$ due to $y^{AA'}$. The 3D King plot corresponds to stretching the nonlinearity from $\nu_\beta$ (dashed black arrow; $f_{\gamma\beta\alpha}\vec{\overline{\nu}}_\beta$) and moving along $y^{AA'}$'s direction (thin black arrow; $Y_{\gamma\beta\alpha}\vec{\overline{y}}$) to form a triangle with nonlinearity for $\nu_\gamma$.}
		\end{figure}
		
		Consider a frequency-normalized 3D King plot for $(\alpha,\beta,\gamma)$ transitions [equivalent to Eq.~(\ref{eq:3Dking})]:
		\begin{equation}
			\vec{\overline{\nu}}_{\gamma} = K_{\gamma\beta\alpha} \vec{\overline{\mu}} + f_{\gamma\beta\alpha} \vec{1} + f_{\gamma\alpha\beta} \vec{\overline{\nu}}_{\beta} +
			Y_{\gamma\beta\alpha} \vec{\overline{y}}
		\end{equation}
		[see Eq.~(\ref{eq:vector_notation}) for the vector notation]. From Fig.~\ref{fig:lambda_3Dking}, one can easily see that linear fit in 3D King plot corresponds to finding the values of $f_{\gamma\alpha\beta}$ and $Y_{\gamma\alpha\beta}$ to form a triangle along nonlinearity patterns for $\nu_\beta^{AA'}$, $\nu_\gamma^{AA'}$, and $y^{AA'}$. 
		
		Therefore, if the 3D King plot is linear, the area of the corresponding triangle vanishes. It implies $\vec{\overline{\nu}}_\beta$ and $\vec{\overline{\nu}}_\gamma$ have to be parallel to each other. Thus a test of whether two data points lie along a line through the origin in the $\lambda_\pm$ plane can be used to probe for the existence of a second nonlinearity source $y^{AA'}$  (see Fig.~2 in the main text).
		
		It is straightforward to see that the ratio $f_{\eta\chi\kappa} = G^{(4)}_{\eta\chi}/G^{(4)}_{\kappa\chi}$ (ratio of red arrows' lengths) determines the $\lambda_-/\lambda_+$ ratio for $y^{AA'}$ (dotted lines in Fig.~\ref{fig:lambda_3Dking}) and vice versa, independent of the $\lambda_-/\lambda_+$ ratio of the dominant source of nonlinearity $x^{AA'}$ (i.e., the direction of the red arrow). The former one is equivalent to fitting the 3D King plot with a known nonlinearity pattern from nuclear parameters $y^{AA'}$ (see Eqs.~\ref{eq:3Dking_boson} and \ref{eq:3Dking_QFS}). The latter suggests that if $f_{\eta\chi\kappa}$ can be calculated precisely in the future, the $\lambda_-/\lambda_+$ ratio of the second nonlinearity source can be deduced and compared with $\lambda_-/\lambda_+$ from Quadratic field shift (QFS), a new boson, or any other proposed sources.

		\subsection{\label{sec:3Dking_boson}New-boson range}
		
		As the dominant source of nonlinearity observed in 2D King plot is expected to be from $G^{(4)}\drf$ (see Fig.~3 in the main text), we can eliminate the dominant source by drawing a 3D King plot. If we assume that the nonlinearity remaining in the 3D King plot is originating primarily from the new boson, we can obtain the value of $\upsilon_{ne}D_{\eta\kappa\chi}$ by fitting the King plot using the relation
		\begin{equation}\label{eq:3Dking_boson}
			\bbar{\nu}_{\eta}^{AA'} = K_{\eta\kappa\chi} + f_{\eta\kappa\chi} \bbar{\nu}_{\chi}^{AA'} + f_{\eta\chi\kappa} \bbar{\nu}_{\kappa}^{AA'} + \upsilon_{ne} D_{\eta\kappa\chi} \bbar{a}^{AA'}.
		\end{equation}
		We obtain a perfect fit as there are four fitting parameters $K_{\eta\kappa\chi}$, $f_{\eta\kappa\chi}$, $f_{\eta\chi\kappa}$, and $\upsilon_{ne} D_{\eta\kappa\chi}$ for four isotope pairs $(A,A')$ (see Table~\ref{tab:el_factors_3}). On the other hand, the calculated $D_{\eta\kappa\chi}$ at light new-boson mass $m_\phi$ has statistical uncertainty from the fitted value of $f_{\eta\chi\kappa}$ (see Sec.~\ref{sec:el_factors_3}). Therefore, the value of new-boson-coupling product
		\begin{equation}\label{eq:yeyn}
			y_ey_n = (-1)^{s+1} 4\pi \hbar c \frac{(\upsilon_{ne}D_{\eta\kappa\chi})_{\textrm{fit}}}{(D_{\eta\kappa\chi})_{\textrm{cal}}}
		\end{equation}
		is given by a ratio of fitted parameter to calculated parameter. Here we use a simple way to treat uncertainties in the numerator and the denominator. We consider the 95\% confidence interval of each value and conservatively obtain the range for $y_e y_n$ from the intervals.
		
		The values of $y_e y_n$ as a function of the new-boson mass $m_\phi$ obtained using Eq.~(\ref{eq:yeyn}) for some choices of $(\chi,\kappa,\eta)$ from five transitions $\alpha$ to $\epsilon$ are shown in Fig.~4 in the main text. There the fitted $\upsilon_{ne} D_{\eta\chi\kappa}$ have much bigger fractional uncertainties than the calculated $D_{\eta\kappa\chi}$ for the $(\chi,\kappa,\eta) = (\alpha,\gamma,\delta)$, $(\beta,\gamma,\delta)$, and $(\gamma,\delta,\epsilon)$ transitions (see Table~\ref{tab:el_factors_3}), except in the regions where $D_{\eta\kappa\chi}$ is close to zero, and we have no sensitivity to a new boson for the given transition (i.e., $y_e y_n$ diverges).

		\subsection{\label{sec:3Dking_QFS}QFS range}
		
		Similar to the new-boson bound, the experimental range of the quadratic field shift $\drtsq$ can be obtained by assuming that it is the dominant source of the observed nonlinearity in 3D King plot,
		\begin{equation}\label{eq:3Dking_QFS}
			\bbar{\nu}_{\eta}^{AA'} = K_{\eta\kappa\chi} + f_{\eta\kappa\chi} \bbar{\nu}_{\chi}^{AA'} + f_{\eta\chi\kappa} \bbar{\nu}_{\kappa}^{AA'} + G^{(2)}_{\eta\kappa\chi} \bbar{\drtsq}^{AA'}
		\end{equation}
		We believe that the calculation for the $(\alpha,\gamma,\beta)$ transitions is the most reliable as the transitions are obtained simultaneously from a single run of the CI calculation (from GRASP2018; see Sec.~\ref{sec:ASC}), providing maximum consistency between the calculations for the different transitions. The fitted value of $G^{(2)}_{\beta\gamma\alpha}$ has different sign and bigger magnitude than the two-transition factor $G^{(2)}_{\beta\alpha}$ (see Tables~\ref{tab:el_factors_2} and \ref{tab:el_factors_3}). However, we expect the three-transition factor to be significantly smaller than the two-transition factor for $G^{(2)}$ (see Sec.~\ref{sec:el_factors_3}). This implies that the observed nonlinearity might not be mainly from QFS, although future measurements of the $\alpha$ and $\beta$ transitions  with the better precision might result in smaller fitted $G^{(2)}_{\beta\gamma\alpha}$. %
		
		\begin{figure}
			\centering
			\includegraphics[width=\columnwidth]{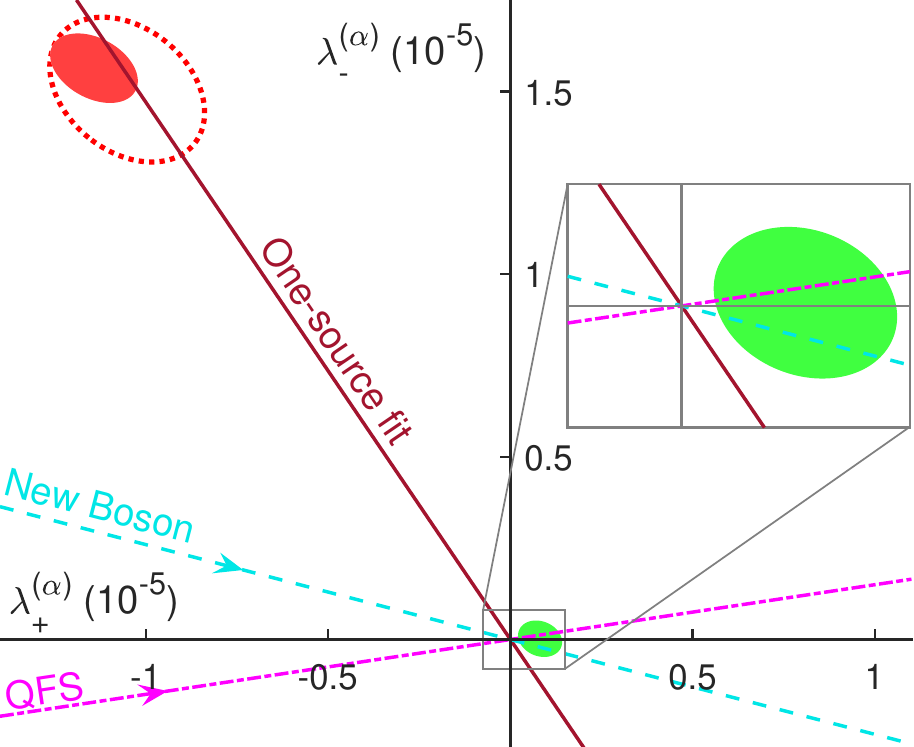}
			\caption{\label{fig:lambda_map_411_436_467}
				$(\lambda_+,\lambda_+)$  values of measured ISs for $\beta$ (green) and $\gamma$ (red) transitions normalized by ISs for $\alpha$ transition. See Fig.~2 in the main text for details.
			}
		\end{figure}

		\section{Estimation of systematic effects and errors}
		
		Our measurement error is determined directly from the scatter of our data points. Most of the systematic effects pertaining to measurements of transition frequencies in atoms are common-mode between the isotopes, with only a small differential component that affects our measured ISs. Drifts in experimental parameters can lead to uncertainties on these differential shifts, and these are the main source of our measurement error. Many of the systematic uncertainties affecting our experiment are the same as for our previous measurement of the ISs of the quadrupole transitions \cite{counts2020} and are discussed in detail in the Supplemental Material of that work. While we summarize all effects in Table~\ref{tab:systematic-shifts}, here we discuss primarily systematic effects which differ from the previous work, due to the transition or modifications in the experimental setup. 
		
		\begin{table*}
			\caption{Estimated contributions to systematic shift on the $\gamma$ transition. The systematic uncertainty is dominated by the laser-induced AC Stark shift.}
			\centering
			\begin{ruledtabular}
				\begin{tabular}{@{}lrr@{}}
					& Estimated magnitude of absolute shift (Hz) & Estimated differential shift (Hz) \\
					\hline
					Laser-induced Stark shift & \num{7+-1e3}& \num{0+-1e2} \\
					Linear Zeeman shift & \num{0+-3e2} & \num{0+-3e2} \\
					Absolute frequency stability probe laser & \num{0+-3e2} & \num{0+-3e2} \\
					Second-order Doppler shift & \num{5+-10e-1}& \num{2+-100e-2} \\
					Micromotional Stark shift & \num{5+-1e-1} & \num{2+-100e-3} \\
					Electric quadrupole shift & \num{3+-3e-2} & \num{0+-3e-2} \\
					Black-body shift & \num{68+-3e-3}& \num{0+-3e-3} \\
					
					Quadratic Zeeman shift & \num{1000+-2e-4} & \num{0+-2e-4} \\
					Gravitational red shift & \num{100+-1e-4}& \num{0+-1e-4} \\
				\end{tabular}
			\end{ruledtabular}
			\label{tab:systematic-shifts}
		\end{table*}

		\subsection{AC Stark shift}
		\label{sec:ACstark}
		
		\begin{figure}
			\centering
			\includegraphics[width=\columnwidth,trim=20 20 20 10,clip]{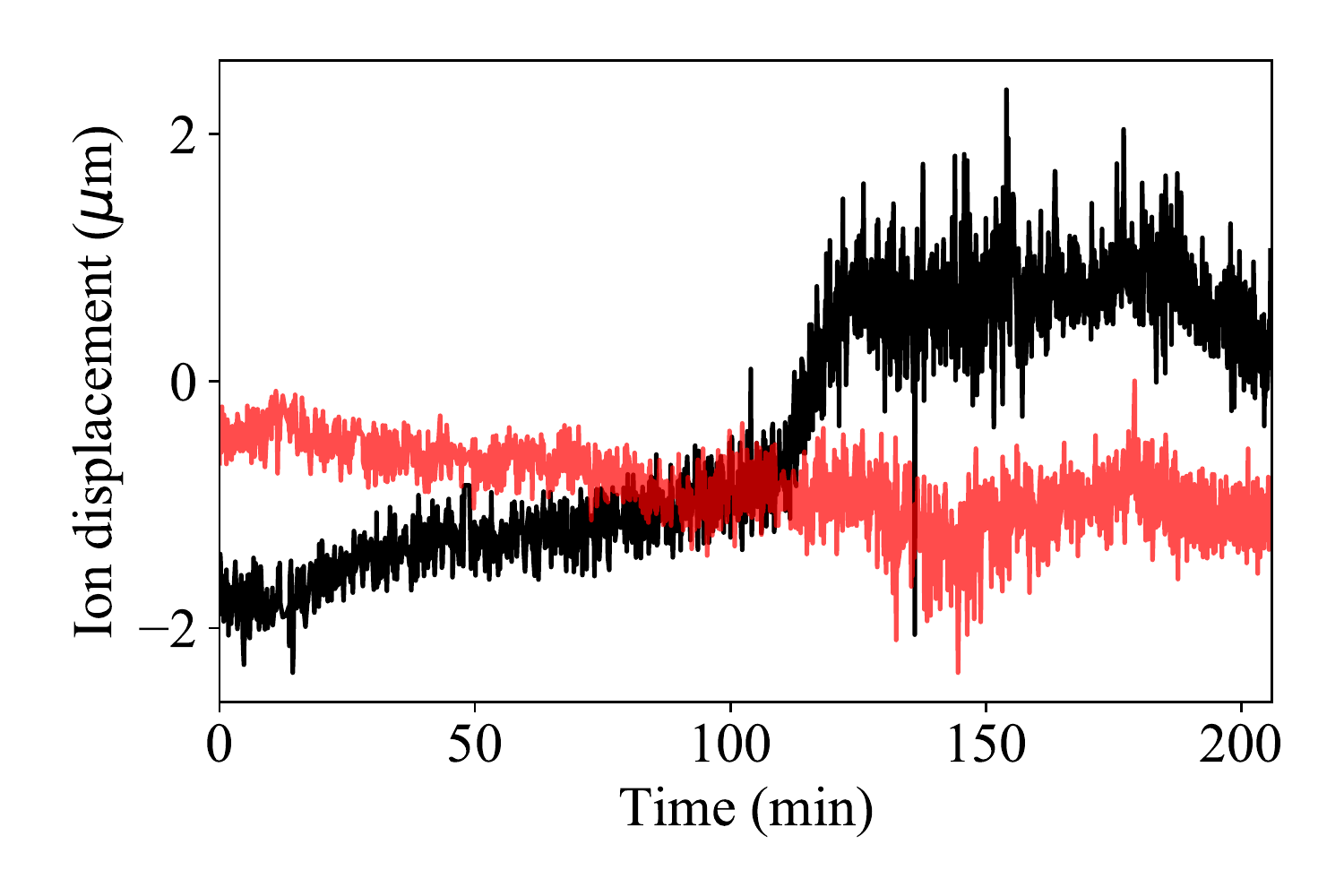}
			\caption{Position of the ion in the plane of the trap as a function of time. The $z$ direction is along the trap axis (black), and $x$ is the direction perpendicular to the trap axis and parallel to the plane of the trap (red). The ion is interrogated with the same laser pulse sequence used during the experiment, but the probe laser is detuned from resonance. Decreased ion displacement along $x$ reflects the tighter confinement along the radial trap direction.}
			\label{fig:ion_camera_motion}
		\end{figure}
		
		The \SI{467}{\nano\meter} probe light used to drive the octupole transition can also couple to other transitions in the atom, causing an intensity-dependent AC Stark shift of the transition frequency. This shift has been measured to be \SI{5.9+-0.8e-5}{\hertz\per\watt\meter\squared} \cite{Furst2020}. For our beam waist, power and polarization, this leads to a shift of \SI{\sim 7}{\kilo\hertz}. The shift is common-mode between different isotopes but can introduce an error in the IS measurement if there is a systematic variation of the probe laser intensity when tuned to different isotope transition frequencies or due to random fluctuations in intensity between isotopes which do not completely average out over our measurements. We stabilize the power of the probe beam such that power fluctuations at a monitor photodiode are kept below $0.5\%$, corresponding to an error of \SI{35}{\hertz}. We also note, however, that our UV laser beams can periodically charge the trap chip, shifting the minimum of the trapping potential and moving the ion relative to the center of the beam. Any such effect could lead to changes in the effective intensity seen by the ion. To determine the size of this effect, we monitored the ion fluorescence on a camera and tracked the ion's position while subjecting it to the sequence of laser pulses used in the experiment. We observed a drift $\lesssim 2$~$\mu$m in the ion position in the plane of the trap (see Fig.~\ref{fig:ion_camera_motion}), indicating that the ion could experience intensity fluctuations of around $1\%$, giving an error of order \SI{100}{\hertz}. We note, however, that our camera monitoring of the ion position has no sensitivity to drifts perpendicular to the plane of the trap. It is possible that drifts in this direction are a contributing source of the $\sim\SI{2}{\kilo\hertz}$ spread we observe in the data [see Fig.~\ref{fig:CommonDiffDrift}(a) for an example dataset].
		
		\subsection{Absolute frequency stability of probe laser}
		
		The frequency stability of our probe laser is currently limited by a residual amplitude modulation (RAM) of the PDH error signal used to lock the laser to a resonance of the ULE cavity.
		
		\subsubsection{RAM stabilization}
		
		RAM originating from the fiber EOM has the potential to introduce noise in the PDH frequency stabilization. Following Ref.~\cite{Zhang2014}, we employ a RAM stabilization scheme feeding back to the DC voltage input of the fiber EOM and the temperature control of the EOM crystal.
		Deviating from the scheme employed in Ref.~\cite{Zhang2014}, we feed the DC voltage signal into the temperature control servo.
		
		To obtain a measure of the RAM remaining in our system, we continuously monitored the off-resonant PDH error signal. With our stabilization system engaged, we found that the effect of RAM was suppressed to a level below our measurement resolution, bounding its contribution to the probe laser frequency instability to $\leq$ 300 Hz.

		\subsubsection{ULE-cavity transmission power stabilization}
		
		Light at infrared wavelengths is used to lock the Ti:Sapphire laser to the ULE cavity.
		Drifts in the power of the intracavity light can change the heating in the mirror coatings, and systematically shift the ULE cavity frequency. To counter this effect, we stabilize the power transmitted through the cavity with an AOM.
		
\begin{figure}[] 
			\centering
			\includegraphics[width=\columnwidth]{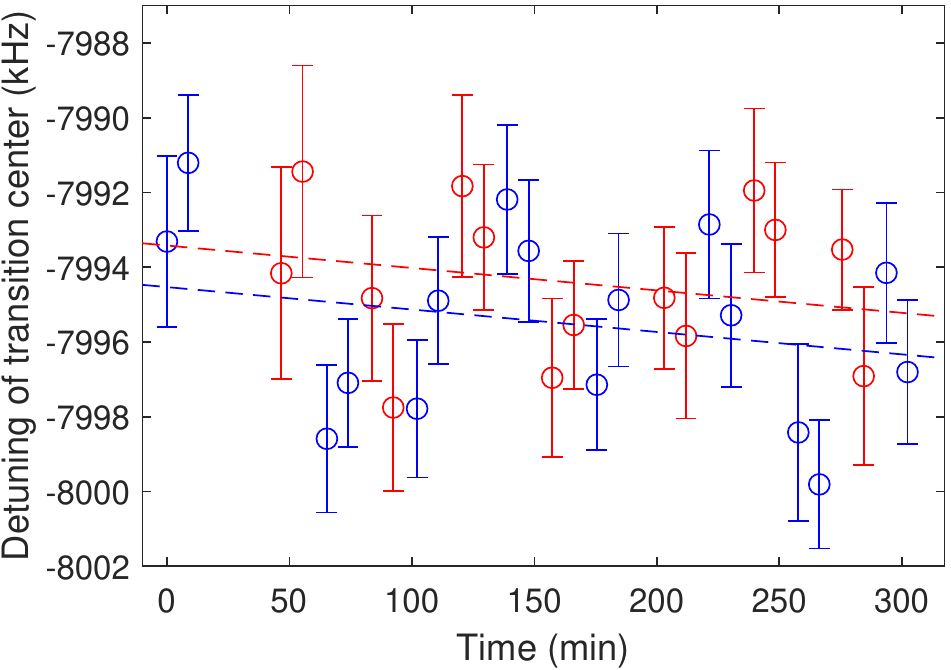}
			\caption{Measured transition center frequency plotted against time for two different values of the ULE cavity transmission power. Blue and red data points correspond to $42\,\mu W$ and $29\,\mu W$ of transmitted power, respectively. An offset was subtracted from the vertical axis, which is not shown here. The linear fit corresponds to the common drift of the reference cavity. The fitted frequency shift is 1.11(85)\,kHz.}
			\label{fig:ULEtrasmissionSpectScan}
		\end{figure}

		To quantify the effect of intracavity power fluctuations on the probe laser frequency, we performed a spectroscopy experiment on the 467\,nm transition of $^{174}$Yb$^+$. The cavity transmission power was varied between two values ($29$\,$\mu$W and $42$\,$\mu$W). For each value of power, we took two transition frequency scans, then switched to the other power. Each transition scan takes $\sim 8$ minutes, so that the total duration of the experiment was 5 hours. We plot the results in Fig.~\ref{fig:ULEtrasmissionSpectScan} and determine that the reference frequency drifts $-1.11(85)$\,kHz for 13\,$\mu$W increase in optical power (i.e., $-85(65)$\,Hz/$\mu$W).
		
		Residual frequency drifts in the probe laser can originate from variations in the set point of the servo loop for the cavity transmission power, at 24\,$\mu$W. Assuming temperature variations in the laboratory of $\pm 2^\circ$C, the drift in the control electronics can introduce a maximum error of 25\,Hz.

		\subsection{Linear Zeeman shift}
		
		In order to minimize the uncertainty on the measurement of the transition frequency introduced by drifts in the magnetic field between scans, we perform interleaved scans of transitions $B$ and $R$ (the two transitions we measure to determine the center, which are symmetrically blue and red detuned from the center) -- i.e. we measure one point on the scan of the red transition, then one point on the scan of the blue transition, then the next point on the red transition and so on. The time needed to measure a point on a given transition and switch to measuring the other transition is of order \SI{10}{s}.
		
		We can extract an estimate for our magnetic-field noise by evaluating the differential drift of the measured resonant frequencies of the $B$ and $R$ transitions. We find that the RMS differential drift to be of order \SI{5}{\kilo\hertz}, which implies a magnetic-field noise on the order of \SI{3}{\milli G}. This level of noise is expected due to a local subway station and is consistent with what we measure in other experiments. We find no significant correlation between this measure of magnetic field and the measured centers of the transitions, indicating that it is unlikely that our magnetic-field noise is contributing systematic shifts to our measurement.

		\subsection{Black-body shift}
		
		The black-body radiation shifts on the transition are well approximated by \cite{Safronova2012}
		\begin{equation}
			\Delta \nu_{\rm BBR} = -\frac{1}{2} \Delta \alpha_0 (\SI{831.9}{\volt\per\meter})^2\left(\frac{T}{\SI{300}{\kelvin}}\right)^4
		\end{equation}
		where $\Delta \alpha_0$ is the difference in scalar polarizability between the atomic states associated with the transition of interest, measured to be \SI{-1.3+-0.6e-40}{\joule\per\volt\squared\meter\squared} \cite{Huntemann2012}. This gives a shift of \SI{68}{\milli\hertz} at \SI{300}{\kelvin}. We conservatively estimate that the temperature of the chamber can drift by \SI{3}{\kelvin} during a measurement, yielding a change in $\Delta \nu_{\rm BBR} \sim \SI{3}{\milli\hertz}$.
		
		\subsection{Electric quadrupole shift}
		
		A frequency shift results from the interaction of the electric quadrupole moment of the two states with electric field gradients from the trap. The shift is of order
		\begin{equation}
			\Delta \nu_{\rm quad} \sim \frac{\Theta\cdot \nabla E}{h}.
		\end{equation}
		The quadrupole moment of the $^2F_{7/2}$ state has been measured at \num{-0.041+-0.005}\,$e a_0^2$ \cite{Huntemann2012}. Time-varying electric field
		gradients due to patch potentials on the chip trap can
		lead to a differential shift between isotopes. We observe a typical day-to-day variation of the DC micromotion compensation voltages applied to our trap electrodes of \SI{20}{\milli\volt}. Conservatively, we consider a maximum variation of \SI{200}{\milli\volt} during the course of a shift measurement
		data-taking run. From this, we infer that differential
		patch-potential gradients of order $\sim\SI{1}{\volt\per\milli\meter\squared}$
		could occur, which would lead to a differential quadrupole shift
		of $\sim \SI{30}{\milli\hertz}$.
		
		\subsection{Second-order Doppler shift and Stark shift due to micromotion}
		Both the second-order Doppler shift and the Stark shift due to micromotion contribute systematic uncertainties that are several orders of magnitude below the leading systematics on our experiment. For completeness, we update our estimate of these systematics here employing the same calculation described in the Supplemental Material of Ref.~\cite{counts2020}. To estimate the stray DC fields and micromotion amplitudes experienced by the ion, we use our measurement of the maximum excursion made by the ion from the trap center over the course of a day while exposed to the sequence of laser pulses used in the experiment (as described in Sec.~\ref{sec:ACstark}, we expect that our tightly focused probe laser beam and UV Doppler cooling beams may cause charging of the trap chip, leading to drifts in the ion position). We estimate a contribution to our error budget on the order of \SI{1}{\hertz} from the second-order Doppler shift and  \SI{0.1}{\hertz} from the Stark shift.
		
		\subsection{Frequency pulling of the measured transition center due to imperfect centering of the scan range}
		
		\label{sec:frequencyPulling}
		As shown in Fig.~\ref{fig:pulse_sequence}, the 760\,nm repumper light is turned off during the readout stage of our laser pulse sequence. This introduces a small probability of a false quantum jump reading due to rare events where the probe transition has not been successfully driven but the ion still falls into the $F_{7/2}$ state through other channels (according to Ref.~\cite{Olmschenk2007}; this likely occurs due to collisions with background gas and happens once every few hours).
		Because we determine the center of a frequency scan by taking the statistical mean of the points, if our scan range is not perfectly centered on the transition resonance frequency, this effect could slightly pull our transition center frequency. However, if we instead find the center by fitting the transition lineshape, we should be insensitive to this effect (since it would, on average, contribute a background that is symmetric around the transition center). To bound this potential error source (and any other potential pulling of the line due to imperfect centering of our scan range), we compare the results of our analysis with one where we fit the datapoints to a Gaussian function with background offset. We find that there is no difference between the two methods within our statistical error bars.
		
		\section{\label{sec:ASC}Atomic structure calculations and electronic factors}
		
		Atomic-structure calculations (ASCs) are performed using Dirac-Hartree-Fock (DHF) \cite{Grant1980, Dyall1989} and subsequent configuration interaction (CI) methods \cite{Jonsson1996, Porsev2009, Fawcett1991, Biemont1998} using two different calculation packages available: GRASP2018 \cite{FroeseFischer2018, counts2020} and \ambit{} \cite{Kahl2019, Berengut2020}.
		
		\subsection{Calculations using GRASP2018}
		We use the popular package GRASP2018 \cite{FroeseFischer2018} to solve for the electronic wavefunction associated with each atomic state. We perform two calculations with GRASP2018: one for the $^1S_0$ and $^3P_{0}$ states in neutral Yb for the 576\,nm clock transition, and another for the $^1S_{1/2}$, $^2D_{3/2}$, $^2D_{5/2}$, and $^2F_{7/2}$ states in singly-ionized Yb for the 435, 411, and 467\,nm clock transitions. In both calculations we use multi-configuration DHF calculations; first we obtain radial wavefunctions for orbitals in the $^{172}$Yb core (up to $5s^2 5p^6 4f^{14}$) followed by the valence orbitals ($6s$, $6p$, and $5d$). Then, we construct a basis for correlation orbitals. Finally, we perform a configuration interaction (CI) calculation to obtain mixing coefficients for the different configuration state functions (CSFs) in the expansion.
		
		For neutral Yb, correlation orbitals up to $10spdfg$ are constructed in the Thomas-Fermi approximation.
		To construct the CSFs for the $^1S_0$ and $^3P_{0}$ states, we begin with a multireference consisting of the $4f^{14} 6s^2$, $4f^{14} 6s 6p$, $4f^{13} 6s^2 5d$, and $4f^{14} 6s5d$ configurations. We allow for a single excitation originating from any of the valence orbitals or select core orbitals ($4spd$ and $5sp$); we find this produces sufficient agreement with experimentally-measured clock transition wavelengths.
		
		For the Yb$^+$ ion, correlation orbitals up to $8spdf$ are calculated via DHF. For $^2S_{1/2}$, $^2D_{5/2}$, $^2D_{3/2}$, and $^2F_{7/2}$ states, single and double excitations from $6s$, $6p$, and $5d$ shells and single excitations from $4f$ shell are allowed in $4f^{14}6s$, $4f^{14}6p$, $4f^{14}5d$, $4f^{13}6s^2$, $4f^{13}6s5d$, $4f^{13}6p^2$, and $4f^{13}5d^2$ to generate the CSFs. Single excitations from $4sp$ and $5sp$ core shells are also allowed for $4f^{14}6s$, $4f^{14}6p$, and $4f^{13}6s^2$ configurations. The total number of excitations is limited to two.
		
		From the calculated wavefunction for a state specified by its total angular momentum $J$ and parity $P$, the radial electron density $\rho(r)$\footnote{It is the one-dimensional density and normalized as follows: $\int \D r \rho(r) = N$.}  can be obtained from the expression $\rho(r) = \bracket{\Psi}{\sum_{i=1}^{N} \delta(r-|\mathbf{r}_i|)}{\Psi}$
		where $\Psi = \sum_{\nu} c_\nu \Phi(\gamma_\nu PJM_J)$ is the atomic state function with CSFs $\Phi(\gamma_\nu PJM_J)$ and associated mixing coefficients $c_\nu$, $\delta(r - |\mathbf{r}_i|)$ are one-dimensional Dirac-delta functions for $i$-th electron's position $\mathbf{r}_i$, and $N= Z-I$ is the number of electrons in an (ionized) atom \cite{Ekman2019}.
		REDF1, a program for extracting radial electron densities from GRASP2018 calculation results has been developed by modifying and merging the source codes for RHFS routine in GRASP2018 and RIS4 routine \cite{Ekman2019} since our previous work \cite{counts2020}. 
		The routine is available in Ref.~\cite{REDF}.
		Finally, the change in the electron density during the $\chi$ transition is given as $\rho_\chi(r) = \rho_\chi^{(e)}(r) - \rho_\chi^{(g)}(r)$ where $\rho_\chi^{(g)}(r)$ and $\rho_\chi^{(e)}(r)$ are the densities for the ground and excited states, respectively.

		\subsection{Calculations using AMBiT}
		The particle-hole CI calculations using \ambit{}~\cite{Berengut2018} are performed in the closed-core DHF potential ($V^{N-1}$ for Yb$^+$). The valence $6s$, $6p$, and $5d$ DHF orbitals are generated in this potential. Higher orbitals $nlj$ are constructed by multiplying the upper component of the $(n-1)lj$ orbital by the simple radial function $r$, and orthogonalizing with the lower orbitals~\cite{Bogdanovich1983}. The lower component is constructed from the upper component using the Dirac equation. The $5f$ orbital is specially created by multiplying the $5d$ orbital by $r$ and orthogonalizing to $4f$.
		
		For Yb$^+$, the CI calculation includes orbitals up to $8spdf$. Configurations were then generated by allowing single and double-electron excitations from the valence orbitals in the leading configurations $6s$, $5d$, $6p$, $4f^{-1}\ 6s^2$, $4f^{-1}\ 5d\ 6s$, $4f^{-1}\ 5d^2$, and $4f^{-1}\ 6p^2$. One additional excitation from the $4f$ shell was also allowed. In this way we captured most of the important configurations. CSFs were then created for each total angular momentum and parity $J^{\pi}$.
		
		The calculation for neutral Yb was very similar. The basis was extended to $12spdf$ and single and double-electron excitations were generated from the leading configurations $6s^2$, $6s\ 6p$, $6p^2$, and $6s\ 5d$, with additional single excitations from the $5s$ and $5p$ orbitals.
		
		\subsection{Single-transition electronic factors}
		\label{sec:el_factors_1}
		Single-transition electronic factors can be derived from the wavefunctions or transition frequencies calculated via ASCs. From the GRASP2018 output REDF1 routine, the change in electron density over space $\rho_{\chi}(\vec{r})$ during the transition $\chi$ can be extracted, and the procedures to obtain single-transition electronic factors $F_\chi$, $K_\chi$, $G^{(4)}_\chi$, and $D_\chi$ are elaborated in the Supplement Material of our previous paper \cite{counts2020}. We have changed the strategy to obtain $G^{(2)}_\chi$ to avoid numerical noise from repeated ASCs pointed out in Ref.~\cite{Allehabi2021}. It is assumed that the finite size of the nucleus caps the electronic wavefunction which would diverge at the origin if the nucleus were a point charge. This gives the relation
		\begin{equation}
			\rho_{\chi}(0; \mr{2}) = C_\chi \rho^P_\chi(r^2 = \mr{2})
		\end{equation}
		where $\rho_{\chi}(0)$\footnote{Here the electron density function is three-dimensional (i.e., $\int \D\vec{r} \rho^{(e,g)}(\vec{r}) = \text{(the number of electrons)}$ for ground or excited states in a given transition).} is the change in electronic density at the origin with the finite nuclear size $\mr{2}$ during the transition $\chi$, $\rho^P_{\chi}$ is the density for point-charge nucleus, and $C_\chi$ is a constant for the size of the nucleus. Then $G^{(2)}_\chi$ is given as
		\begin{equation}
			\begin{aligned}
				G^{(2)}_\chi &= \frac{1}{2}\frac{\partial^2 \nu_\chi}{(\partial \mr{2})^2}(\mr{2}^{A}) 
				= \frac{1}{2} \frac{\partial F_\chi}{\partial \mr{2}}(\mr{2}^{A}) \\
				&= \frac{c\alpha' Z}{96\pi^2} \frac{\partial \rho_\chi(0; \mr{2})}{\partial \mr{2}}(\mr{2}^{A}) \\
				&= C_\chi \frac{c\alpha' Z}{96\pi^2} \frac{\partial \rho^P_\chi}{\partial r^2}(\mr{2}^{A})
			\end{aligned}
		\end{equation}
		where $c$ is the speed of light, $\alpha' \approx 1/137$ is the fine structure constant, and $Z=70$ is the proton number of Yb, for a reference isotope $A$ (here we choose $A=172$). Therefore, a single atomic structure calculation with a point-charge nucleus is sufficient to obtain $G^{(2)}_\chi$. It is numerically observed that $C_\alpha = 1.04$, essentially unity, for transition $\alpha$: ${}^2S_{1/2} \rightarrow {}^2D_{1/2}$ (411\,nm) transition. A similar idea appears in Ref.~\cite{Flambaum2018} for the analytic estimation of the King plot nonlinearity.
		
		For \ambit{}, The ASCs are repeated for transition $\chi$ while varying nuclear parameters $z = \mu, \drt, a$,
		and the rates of the change in transition frequency $(\partial \nu_\chi)/(\partial z)$ are taken as the associated electronic factors $K_\chi$, $F_\chi$, and $D_\chi$, respectively. $G^{(2)}_\chi$ is given as the second derivative $\frac{1}{2}(\partial^2 \nu_\chi)/(\partial \drt)^2$.
		For $K_\chi$, the nuclear inverse mass $\mu$ is promoted to a finite field parameter by adding a relativistic mass shift operator to the Coulomb interaction~\cite{Berengut2003}.
		
		The values of the single-transition electronic factors for the five transitions $\alpha$ to $\epsilon$ in this paper are tabulated in Table~\ref{tab:el_factors_1} and shown in Fig.~\ref{fig:D1_vs_m}. 
		
		\subsection{Two-transition electronic factors}
		\label{sec:el_factors_2}
		
		Two-transition electronic factors $f_{\kappa\chi} = F_\kappa/F_\chi$ and $Z_{\kappa\chi} = Z_\kappa - f_{\kappa\chi}Z_\chi$ where $Z \in \{K, G^{(4)}, G^{(2)}, D\}$ are defined for (2D) King plots (see the main text), and are calculated from the single-transition factors from ASCs.
		
		The values of the two-transition electronic factors for all possible transition pairs out of the five transitions $\alpha$ to $\epsilon$ can be found in Table~\ref{tab:el_factors_2} and Fig.~\ref{fig:D2_vs_m}. 
		
		\subsection{Three-transition electronic factors}
		\label{sec:el_factors_3}
		
		Three-transition electronic factors $f_{\eta\chi\kappa} = G^{(4)}_{\eta\chi}/G^{(4)}_{\kappa\chi}$ and $Z_{\eta\kappa\chi} = Z_{\eta\chi} - f_{\eta\chi\kappa}Z_{\kappa\chi}$ where $Z \in \{K, G^{(2)}, D\}$ are defined for the 3D King plot (see Sec.~\ref{sec:3Dking}), assuming that the fourth-moment field shifts $G^{(4)}_{\chi, \kappa, \eta}\drf_{ji}$ are the dominant source of the nonlinearity in 2D King plot (see Fig.~2 in the main text). Their values are calculated from the two-transition factors.
		
		\begin{figure}
			\centering
			\includegraphics[width=\columnwidth]{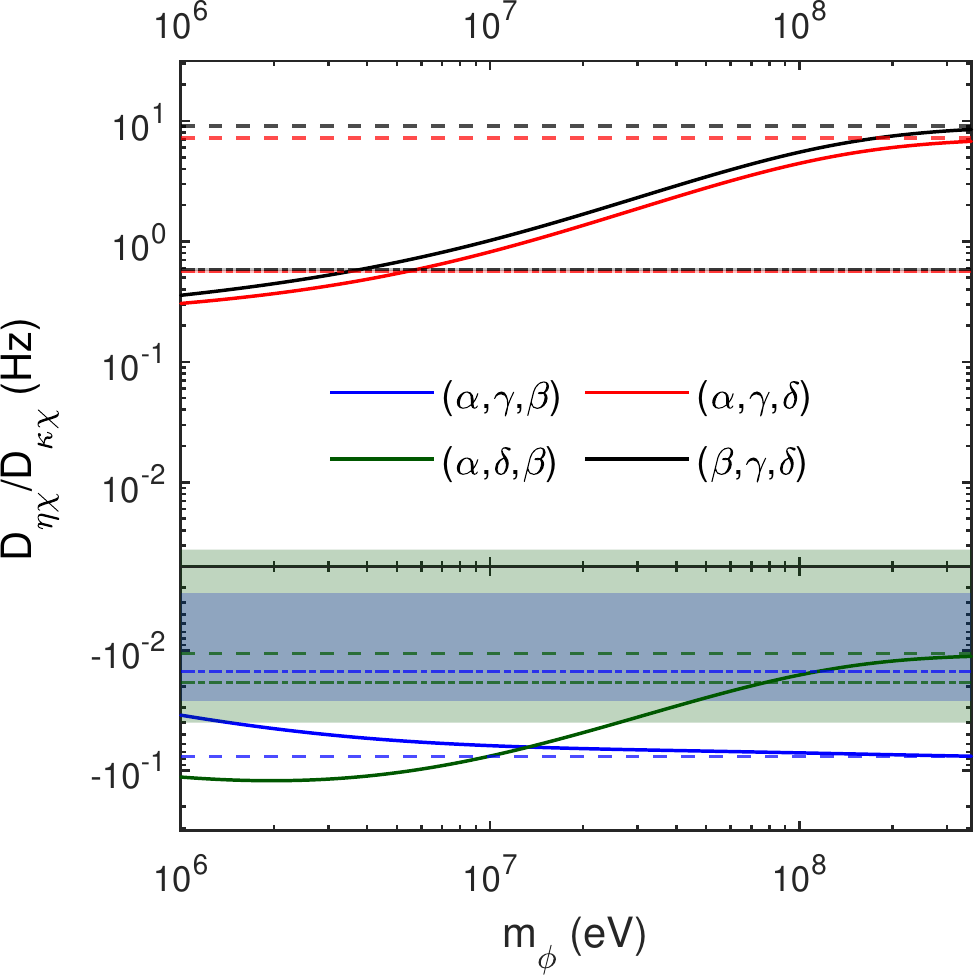}
			\caption{$d_{\eta\chi\kappa} = D_{\eta\chi}/D_{\kappa\chi}$ ratio derived from atomic structure calculations (ASCs) performed using GRASP2018 \cite{FroeseFischer2018} vs new boson mass $m_\phi$ for various transitions $(\chi,\kappa,\eta)$ (solid line) coded with different colors (see legend). Dashed lines indicate corresponding $f_{\eta\chi\kappa} = G^{(4)}_{\eta\chi}/G^{(4)}_{\kappa\chi}$ ratios derived from the ASCs. Dash-dotted lines and shaded area show $f_{\eta\chi\kappa}$ and their 1$\sigma$ uncertainties obtained from linear fit in the corresponding 3D King plots. (The shaded regions are not visible for $(\alpha,\gamma,\delta)$ and $(\beta,\gamma,\delta)$ transitions as the areas are too thin.) Theoretical and experimental values of  $f_{\eta\chi\kappa}$ can be found in Table~\ref{tab:el_factors_3}.}
			\label{fig:D2ratio_vs_m}
		\end{figure}
		
		We have the choice of using the calculated or the fitted $f_{\eta\chi\kappa}$ to obtain $Z_{\eta\kappa\chi} = Z_{\kappa\chi}(z_{\eta\chi\kappa}-f_{\eta\chi\kappa})$. Unfortunately, the calculated and fitted values of $f_{\eta\chi\kappa}$ are significantly different for the current accuracy of our ASCs (see Table~\ref{tab:el_factors_3}). For the electronic factors $Z$ that are expected to have a strong correlation to $G^{(4)}$ (i.e., $z_{\eta\chi\kappa}-f_{\eta\chi\kappa} \ll 1$) such as $G^{(2)}$ and $D$ at heavy new-boson mass $m_\phi \gtrsim 10^7$\,eV (corresponds to the nuclear size), using the calculated $f_{\eta\chi\kappa}$ would be better to ensure $z_{\eta\chi\kappa}-f_{\eta\chi\kappa} \ll 1$ and obtain the right order of magnitude (see Fig.~\ref{fig:D2ratio_vs_m}). The strong correlation is because all of the factors probe the properties of electronic wavefunction near the origin. For $K$ and $D$ at the lighter mass $m_\phi \lesssim 10^4$\,eV (corresponds to the Bohr radius), the correlation with $G^{(4)}$ is not expected in general as they encode the global shape of the wavefunction. Therefore, we determine that using experimental value of $f_{\eta\chi\kappa}$ is the better choice.
		
		Note that the situation for two-transition factors is similar, $Z_{\kappa\chi} = Z_\chi (z_{\kappa\chi}- f_{\kappa\chi})$, and here the calculated $f_{\kappa\chi}$ are used for all $Z$ as they agree sufficiently well with the fitted values (see Table~\ref{tab:el_factors_2}).
		
		The values of the electronic factors for all possible choices of three transitions out of the five transitions $\alpha$ to $\epsilon$ are listed in Table~\ref{tab:el_factors_3}, and plotted in Fig.~\ref{fig:D3_vs_m}.
		
		\subsection{Estimating mass shift coefficient $K_\kappa$ from reliable $K_\chi$ calculation and $K_{\kappa\alpha}$ from measured ISs}
		
		It is challenging to calculate mass shift coefficients $K_\chi$ for heavy atoms precisely \cite{Papoulia2016,Puchalski2010}. This turns out to be especially the case for the $\gamma$: ${}^2S_{1/2} \rightarrow {}^2F_{7/2}$ (467\,nm) transition; values from calculations with GRASP2018 and \ambit{} don't agree on the sign, and neither of them predicts $K_{\gamma\alpha}$ close enough the experimental value from the King plot (see Tables~\ref{tab:el_factors_1} and \ref{tab:el_factors_2}, and Fig.~3(d) in the main text).
		On the other hand, the calculated mass shift coefficient for the $\alpha$: ${}^2S_{1/2} \rightarrow {}^2D_{5/2}$ (411\,nm) transition and the $\beta$: ${}^2S_{1/2} \rightarrow {}^2D_{3/2}$ (436\,nm) transition are relatively reliable; values from GRASP2018 and \ambit{} agree to about factor of two, and the experimental value of $K_{\beta\alpha}$ agrees relatively well with the values from GRASP2018 and \ambit{}.
		This is presumably because the $\alpha$ and $\beta$ transitions have relatively simpler electronic configurations, in which a valence electron is excited to higher orbitals while the core configuration is maintained, while the $\gamma$ transition corresponds to the excitation of a core electron from the $4f$ shell to $6s$ valence orbital.
		In a case like this, where the value of $K_\chi$ is more reliable than $K_\kappa$, we can relate them via the experimentally accurately measured quantities $K_{\kappa\chi}$ and $f_{\kappa\chi}$,
		\begin{equation}
			K_\kappa = K_{\kappa\chi} + f_{\kappa\chi}K_{\chi},
		\end{equation}
		which serves as a benchmark for the calculated $K_\kappa$ (See Fig.~3(d) in the main text).

		\section{Nuclear calculations and nuclear charge moments}
		
		\begin{figure}[htb]
			\includegraphics[width=\columnwidth]{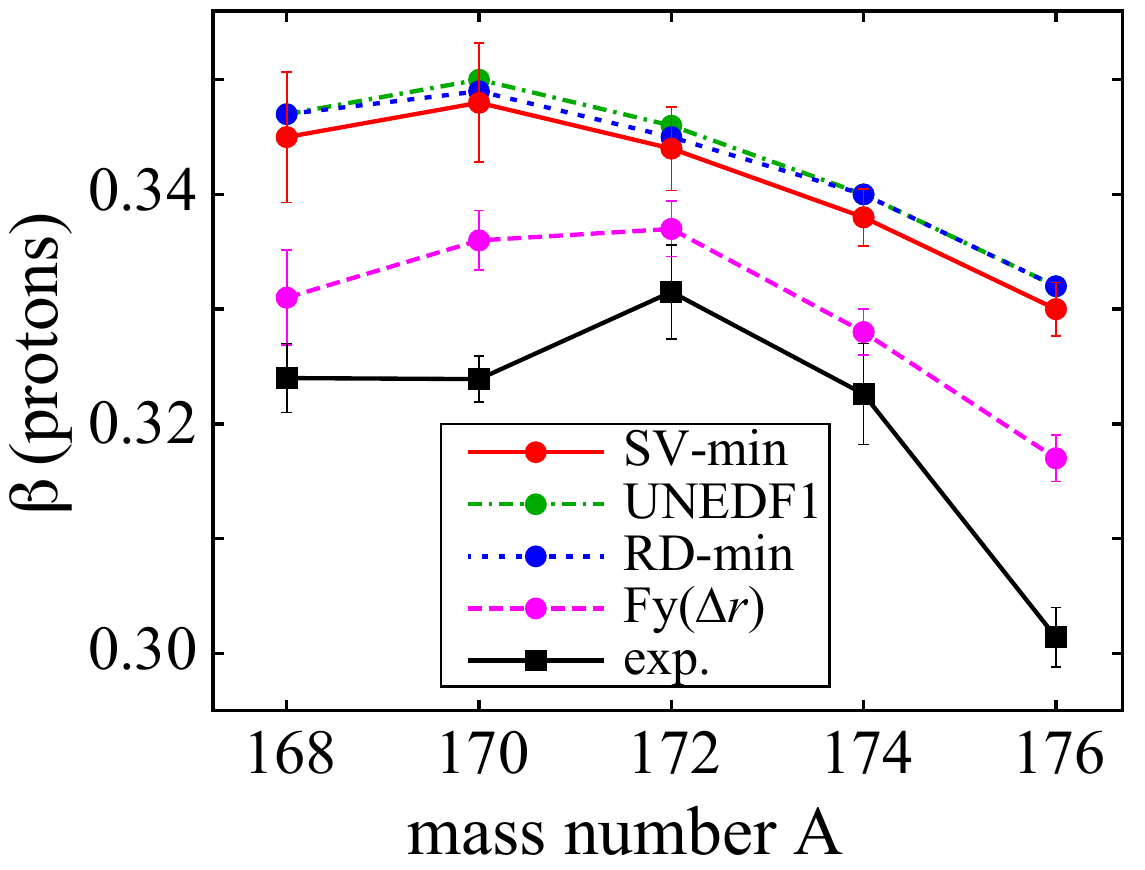}
			\caption{\label{fig:deformations} Quadrupole ground-state deformations $\beta$ for $^{168,170,172,174,176}$Yb obtained in nuclear DFT with different EDFs compared to empirical values \cite{Pritychenko2016}.
			}
		\end{figure}
		
		\begin{table*}
			\caption{\label{tab:NuclearMoments} Upper table: Theoretical and experimental values of difference in nuclear charge moments $\drt$ and $\drf$ between isotopes. The values for nuclear DFT calculations using SV-min, RD-min, UNEDF1, and \Fy{} EDFs are listed in columns 2 -- 5 and 8 -- 11. Columns 6 and 7 tabulate the values of $\drt$ from measured ISs in $\alpha$ transition and calculated $F_\alpha$ and $K_\alpha$ for GRASP2018 \cite{FroeseFischer2018} and \ambit{} \cite{Kahl2019}, respectively (see Table~\ref{tab:el_factors_1}). 
				Lower table: Theoretical values of $\mr{2}^{A}$, $\mr{4}^{A}$, and  quadrupole deformation $\beta^{A}$.  For  $\beta^{A}$, we show also the experimental values derived from the measured B(E2) values \cite{Pritychenko2016}; see also Fig.~\ref{fig:deformations}.
			}
			\begin{ruledtabular}
				\begin{tabular}{c|D{.}{.}{2.6}D{.}{.}{2.7}D{.}{.}{2.3}D{.}{.}{2.3}|D{.}{.}{2.3}D{.}{.}{2.3}|D{.}{.}{3.4}D{.}{.}{2.4}D{.}{.}{3.1}D{.}{.}{3.1}}
					\multirow{3}{*}{\makecell{Isotope pair\\$(A, A')$}} & \multicolumn{6}{c}{$\drt^{AA'}$ [fm$^2$]} & \multicolumn{4}{c}{$\drf^{AA'}$ [fm$^4$]} \\ 
					& \multicolumn{4}{c}{Nuclear DFT} & \multicolumn{2}{c|}{Measured $\nu_\alpha$} & \multicolumn{4}{c}{Nuclear DFT}\\
					& \multicolumn{1}{c}{SV-min} & \multicolumn{1}{c}{RD-min} & \multicolumn{1}{c}{UNEDF1} & \multicolumn{1}{c|}{\Fy{}} & \multicolumn{1}{c}{GRASP} & \multicolumn{1}{c|}{\ambit{}} & \multicolumn{1}{c}{SV-min} & \multicolumn{1}{c}{RD-min} & \multicolumn{1}{c}{UNEDF1} & \multicolumn{1}{c}{\Fy{}} \\
					\hline
					(168, 170) & -0.159(23) & -0.159(43)  & -0.175 & -0.203 & -0.145 & -0.154 & -10.6(2.1) & -10.6(3.8) & -11.8 & -14.3 \\
					(170, 172) & -0.125(29) & -0.128(65)  & -0.139 & -0.169 & -0.136 & -0.145 &  -7.4(3.1) &  -7.6(6.7) &  -8.3 & -11.4 \\
					(172, 174) & -0.119(48) & -0.127(100) & -0.135 & -0.120 & -0.107 & -0.113 &  -6.8(4.8) &  -7.4(10.) &  -8.1 &  -6.8 \\
					(174, 176) & -0.126(32) & -0.134(50)  & -0.134 & -0.134 & -0.102 & -0.108 &  -7.2(3.7) &  -7.8(5.2) &  -8.0 &  -7.2
				\end{tabular}
				\begin{tabular}{c|cccc|cccc|ccccc}
					Isotope & \multicolumn{4}{c}{$\mr{2}^{A}$ [fm$^2$]} & \multicolumn{4}{c}{$\mr{4}^{A}$ [fm$^4$]} & \multicolumn{5}{c}{$\beta^{A}$} \\ 
					$A$ & SV-min & RD-min & UNEDF1 & \Fy{} & SV-min & RD-min & UNEDF1 & \Fy{} & SV-min & RD-min & UNEDF1 & \Fy{} & Exp. \\
					\hline
					168 & 27.769 & 27.776 & 27.939 & 27.494 & 1012.9 &  1012.5 &  1021.4 &  991.03 &   0.345 & 0.347 &  0.347 &   0.331  &  0.324\\
					170 & 27.927 & 27.935 & 28.113 & 27.697 & 1023.5 &  1023.1 &  1033.2 &  1005.4&   0.348 & 0.349 &  0.350 &   0.336  &  0.324\\
					172 & 28.052 & 28.064 & 28.252 & 27.866 & 1030.9 &  1030.7 &  1041.5 &  1016.8&   0.344 & 0.345 &  0.346 &   0.337  &  0.332\\
					174 & 28.171 & 28.190 & 28.388 & 27.986 & 1037.6 &  1038.1 &  1049.6 &  1023.6&   0.338 & 0.340 &  0.340 &   0.328  &  0.323\\
					176 & 28.297 & 28.325 & 28.522 & 28.120 & 1044.8 &  1045.9 &  1057.6 &  1030.8&   0.330 & 0.332 &  0.332 &   0.317 &  0.301
				\end{tabular}
			\end{ruledtabular}
		\end{table*}
		
		\subsection{Radii from nuclear mean-field models}
		
		For the theoretical description of nuclear charge densities, we use here self-consistent mean-field models at the level of nuclear DFT \cite{Bender2003}. In particular, we employ the energy density functionals (EDFs) SV-min \cite{Klupfel2009}, RD-min \cite{Erler2010}, UNDEF1 \cite{Kortelainen2012}, and Fy($\Delta r$) \cite{Reinhard2017a}. SV-min and UNEDF1 are based on the standard Skyrme functional \cite{Bender2003}. RD-min replaces the power-law density dependence of the Skymre functional by a rational approximant. Fy($\Delta r$) uses the Fayans functional which has additionally gradient terms in pairing and surface energy. The model parameters of all four EDFs are calibrated to a large set of nuclear ground state data. SV-min, RD-min, and Fy($\Delta r)$ use a large set of data from spherical nuclei and information from the electromagnetic form factor  \cite{Klupfel2009}. In addition {\Fy} has  also been optimized to differential charge radii of Ca isotopes \cite{Reinhard2017a}.  The large dataset of UNEDF1 employs  energies and charge radii of spherical and deformed nuclei. 
		In all variants, we use the density-dependent pairing force treated in the BCS  approximation.  With these four EDFs we explore different functional forms as well as different optimization strategies. This should give an impression on these various influences, see the discussion of theoretical results in the main text.
		
		Another crucial aspect is the post-processing of the emerging proton and neutron density distributions to obtain a reliable charge density. This requires proper inclusion of the nucleons charge distribution, relativistic corrections, especially magnetic spin-orbit correction, which must be included in precision calculations of radial moments.
		Our DFT calculations take all these effects into account (see, e.g., Figs.~4, 6, and 7 of Ref.~\cite{Reinhard2021}). As there is no choice in that respect, all four EDFs are processed with that strategy.

		The considered Yb isotopes are all significantly deformed. Thus we use a DFT solver employing an axially symmetric grid in coordinate space which allows for reflection-symmetric deformations \cite{Reinhard2021c}.
		The  radial charge moments $\mr{n}$ are directly obtained from the calculated nuclear charge distribution $\rho_n({\mbox{\boldmath $r$}})$ (See Table~\ref{tab:NuclearMoments}).  Figure~\ref{fig:deformations}
		shows dimensionless ground-state quadrupole proton deformations   $\beta$ obtained in our DFT calculations  and compares them to the empirical values \cite{Pritychenko2016}. The deformations are defined in the usual way:  $\beta=4\pi Q_{20}/(3ZR_0^2)$, where $Q_{20}$ is the proton quadrupole moment and $R_0=1.2\,\mathrm{fm}\,A^{1/3}$. It is satisfactory to see that the calculations are consistent with experiment, considering the scale of $\beta$. In particular the maximum of $\beta$ is predicted by {\Fy} at $A=172$ in agreement with the experiment.

		\subsection{\label{sec:drf_nonlin_cal}Nonlinearity pattern from calculated $\drf$}
		
		Caution is necessary when calculating nonlinearity patterns from higher-order charge moments $\dmr{n}$ ($n > 2$) from nuclear calculations. $\dmr{n}$ and $\dmr{2}$ are obtained from difference in nuclear charge distributions $\delta\rho_{n}^{AA'}(r)$ between isotopes $A$ and $A'$ given by a nuclear calculation, and thus highly correlated to each other. 
		Since the FS, which is proportional to $\dmr{2}$, is the dominant source of total IS, \textit{calculated} ISs using $\dmr{2}$ from nuclear calculation should be used to ensure self-consistency as follows.
		It is especially important when the calculated $\dmr{2}$ do not reflect actual experimentally determined pattern (see Fig.~3(a) in the main text), as then the position of the points in King plot will be different, which changes the nonlinearity pattern significantly.
		
		$G^{(4)}_{\gamma\alpha}\overline{\drf}_{\perp}^{AA'}$, the nonlinearity from $\drf^{AA'}$, is given as the component of the vector
		\begin{equation}
			\begin{aligned}
				&G^{(4)}_{\gamma\alpha}\overline{\drf}^{AA'} = G^{(4)}_{\gamma\alpha}\frac{\drf^{AA'}}{\nu_{\alpha}^{AA'}} \\
				&~~~~= \frac{G^{(4)}_{\gamma\alpha}}{F_\alpha}\frac{\drf^{AA'}}{\drt^{AA'} + \frac{K_\alpha}{F_\alpha}\mu^{AA'} +  \frac{G^{(4)}_\alpha}{F_\alpha}\drf^{AA'}}
			\end{aligned}
		\end{equation}
		which is orthogonal to $\vec{1}$ and
		\begin{equation}
			\overline{\mu}^{AA'} \propto \frac{\mu^{AA'}}{\drt^{AA'} + \frac{K_\alpha}{F_\alpha}\mu^{AA'} +  \frac{G^{(4)}_\alpha}{F_\alpha}\drf^{AA'}}
		\end{equation}
		(see Sec.~\ref{sec:nonlin_analysis} for the vector notation). One can see that the nonlinearity arises mainly from the difference in $\drt^{AA'}$ and $\drf^{AA'}$'s patterns up to an overall scale, and it is thus important to use not only $\drf^{AA'}$ from nuclear calculations, but also the IS calculated using $\drt^{AA'}$ from the same nuclear calculation for self-consistency. 
		We have numerically verified that using measured values of $\nu_{\alpha}^{AA'}$ to normalize $\mu^{AA'}$ and $\drt^{AA'}$ results in a significantly different $\lambda_-/\lambda+$ ratio from the observed nonlinearity. 
		The change in the ratio $K_\alpha/F_\alpha$ and $G^{(4)}_\alpha/F_\alpha$ can tune the values of $\overline{\mu}^{AA'}$, $\overline{\drf}^{AA'}$, and thus the nonlinearity $\lambda_\pm$.
		
		The largest inset in Fig.~2 in the main text shows the nonlinearity $\lambda_\pm$ predicted by the nuclear DFT calculations.
		The solid lines across the symbols show the change in $\lambda_\pm$ when $G^{(4)}_\alpha/F_\alpha$ ratio changes by $\pm 50$\% of the calculated value. Changing $K_\alpha$ in between -2604.4~GHz$\cdot$u and +174.2~GHz$\cdot$u, which covers three times the difference in $K_\alpha$ values for the GRASP2018 and \ambit{} calculations, moves $\lambda_\pm$ points along the solid lines by smaller amounts.
		
		Calculations for all of the four nuclear DFTs predict a $\lambda_-/\lambda_+$ ratio fairly close to the measured ISs, despite the significant difference in the measured and calculated $\drt$.
		In particular, the \Fy{} functional predicts the $\lambda_-/\lambda_+$ ratio consistent with the measured ISs to within 2$\sigma$. It also predicts a reasonable magnitude of $\lambda_\pm$ when the results are combined with the calculated $G^{(4)}_{\gamma\alpha}/F_\alpha$ (see Tables~\ref{tab:el_factors_1} and \ref{tab:el_factors_2}). Note that the \Fy{} is also the only functional that predicts qualitatively correctly $\drt$ ratios out of the four functionals used in this work (see Fig.~3(a) in the main text), as well as the deformation parameter $\beta$ (Fig.~\ref{fig:deformations}).
		The effect of uncertainly in the calculated $G^{(4)}_{\gamma\alpha}/F_\alpha$ ratio is a mere scaling of the distance in $\lambda_\pm$ plane from the origin along the $\lambda_-/\lambda_+$ ratio line. Interestingly, it is numerically observed that the effects of change in the $K_\alpha/F_\alpha$ or $G^{(4)}_\alpha/F_\alpha$ ratios are similar to the change in $G^{(4)}_{\gamma\alpha}/F_\alpha$ ratio [i.e., the change of nonlinearity in $(\lambda_+,\lambda_-)$ plane is almost purely radial from the origin]. This suggests that the calculated $\lambda_-$/$\lambda_+$ ratios are robust with respect to the uncertainty in the calculated electronic factors.

		\clearpage

		\begingroup
		\begin{table}[b]
			\caption{\label{tab:el_factors_1}
				Calculated and experimental values of single-transition electronic factors $Z_\chi$ ($Z \in \{F, K, G^{(4)}, G^{(2)}, D\}$) for $\chi = \alpha$: ${}^2S_{1/2} \rightarrow {}^2D_{5/2}$ (411\,nm), $\beta$: ${}^2S_{1/2} \rightarrow {}^2D_{3/2}$ (436\,nm), and $\gamma$: ${}^2S_{1/2} \rightarrow {}^2F_{7/2}$ (467\,nm) transitions in Yb$^+$ ions; and $\delta$: ${}^1S_0 \rightarrow {}^3P_0$ (578\,nm), and $\epsilon$: ${}^1S_0 \rightarrow {}^1D_2$ (361\,nm) transitions in neutral Yb atoms. 
				$\omega_{\chi}/(2\pi)$ are transition frequencies. Other quantities are defined in the main text.
				Calculated values for each transition are obtained from CI method using GRASP2018 or \ambit{} (see Sec.~\ref{sec:ASC}). The units of $\omega_\chi/(2\pi)$, $F_\chi$, $K_\chi$, $G^{(4)}_\chi$, $G^{(2)}_\chi$, and $D_\chi$ are THz, GHz/fm$^2$, GHz$\cdot$u, MHz/fm$^4$, MHz/fm$^4$, and $10^3$ THz, respectively.}
			\begin{ruledtabular}
				\begin{tabular}{l|cccc}
					& GRASP & \ambit{} & Ref.~\cite{Figueroa2021} & Exp. \\
					\hline
					$\omega_\alpha/(2\pi)$ & 808.11 & 707.00 & & 729.47\footnotemark[1] \footnotemark[2]\\
					$\omega_\beta/(2\pi)$  & 770.13 & 679.86 & & 688.36\footnotemark[1] \footnotemark[3] \\
					$\omega_\gamma/(2\pi)$  & 580.12 & 1051.44 & & 642.12\footnotemark[1] \footnotemark[4] \\
					$\omega_\delta/(2\pi)$  & 458.36 & 522.78 & & 518.30\footnotemark[1] \footnotemark[5] \\
					$\omega_\epsilon/(2\pi)$  & & 819.47 & & 829.76\footnotemark[1] \footnotemark[6] \\
					\hline
					$F_\alpha$ & -15.852 & -14.715 & -17.604 &  \\
					$F_\beta$ & -16.094 & -14.968 & -18.003 &  \\
					$F_\gamma$ & 41.892 & 36.218 &  &  \\
					$F_\delta$ & -9.1508 & -9.719 &  &  \\
					$F_\epsilon$ &  & -13.528 & -14.437 &  \\
					\hline
					$K_\alpha$ & -1678.2 & -752 &  &  \\
					$K_\beta$ & -1638.5 & -661 &  &  \\
					$K_\gamma$ & 3127.6 & 12001 &  &  \\
					$K_\delta$ &  &  &  &  \\
					$K_\epsilon$ &  &  &  &  \\
					\hline
					$G^{(4)}_\alpha$ & 14.934 &  & 13.08 &  \\
					$G^{(4)}_\beta$ & 15.159 &  & 13.37 &  \\
					$G^{(4)}_\gamma$ & -39.422 &  &  &  \\
					$G^{(4)}_\delta$ & 8.951 &  &  &  \\
					$G^{(4)}_\epsilon$ &  &  & 10.42 &  \\
					\hline
					$G^{(2)}_\alpha$ & 42.565 & 81.908 & 28.53 &  \\
					$G^{(2)}_\beta$ & 43.204 & 83.247 & 28.53 &  \\
					$G^{(2)}_\gamma$ & -112.33 & -201.12 &  &  \\
					$G^{(2)}_\delta$ &  & 54.277 &  &  \\
					$G^{(2)}_\epsilon$ &  & 75.322 & 23.34 &  \\
					\hline
					$D_\alpha$\footnotemark[7] & 44.145 & 43.158 & 41.235 &  \\
					$D_\beta$\footnotemark[7] & 48.419 & 48.634 & 48.795 &  \\
					$D_\gamma$\footnotemark[7] & -730.4 & -352.38 &  &  \\
					$D_\delta$\footnotemark[7] & -55.729 & -42.855 &  &  \\
					$D_\epsilon$\footnotemark[7] &  & 5.6683 & 4.6238 &  
				\end{tabular}
			\end{ruledtabular}
			\footnotetext[1]{The exact value varies by the few-GHz isotope shifts.}
			\footnotetext[2]{Ref.~\cite{Taylor1997,Roberts1999}}
			\footnotetext[3]{Ref.~\cite{Tamm2009,Webster2010}}
			\footnotetext[4]{Ref.~\cite{Furst2020,Huntemann2012,King2012}}
			\footnotetext[5]{Ref.~\cite{Pizzocaro2020}}
			\footnotetext[6]{Ref.~\cite{NIST_ASD}}
			\footnotetext[7]{At $m_\phi = 1$~eV. Values over different $m_\phi$'s are shown in Fig.~\ref{fig:D1_vs_m}}
		\end{table}

		\begin{figure}
			\centering
			\includegraphics[width=\columnwidth]{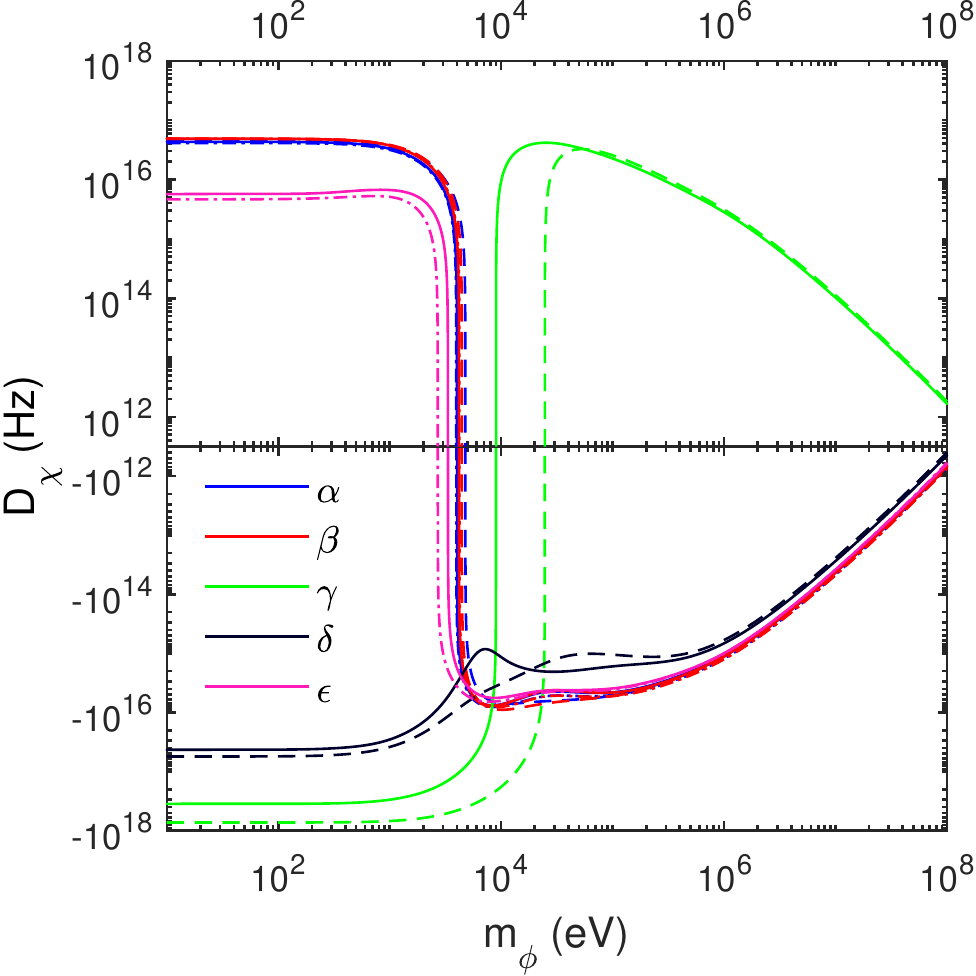}
			\caption{Single-transition factors $D_{\chi}$ vs new-boson mass $m_\phi$ for five transitions $\chi$ coded with different colors (see legend) derived from atomic structure calculations using CI method. Solid, dashed, and dash-dotted lines are for \ambit{}, GRASP2018, and Ref.~\cite{Figueroa2021}, respectively.
			}
			\label{fig:D1_vs_m}
		\end{figure}
		\endgroup

		\FloatBarrier
		
		\setlength{\LTcapwidth}{\textwidth}
		\begin{longtable*}{@{\extracolsep{\fill}}l|ccccc@{}}
			\caption{\label{tab:el_factors_2}
				Calculated and experimental values of two-transition electronic factors $f_{\kappa\chi}$ and $Z_{\kappa\chi}$ ($Z \in \{K, G^{(4)}, G^{(2)}, D\}$) for $\chi, \kappa \in \{\alpha, \beta, \gamma, \delta, \epsilon\}$. The values are calculated from the single-transition values in Table~\ref{tab:el_factors_1}.
				$f_{\chi\kappa}$ is dimensionless. The units of $K_{\kappa\chi}$, $G^{(4)}_{\kappa\chi}$, $G^{(2)}_{\kappa\chi}$, and $D_{\kappa\chi}$ are GHz$\cdot$u, kHz/fm$^4$, kHz/fm$^4$, and $10^3$ THz, respectively. 
				The last two columns (``Fit'') are for data from linear fit of corresponding 2D King plots $\overline{\nu}^{AA'}_{\kappa} = f_{\kappa\chi} + K_{\kappa\chi} \overline{\mu}^{AA'}$ with (``X corr.'') and without (``No X corr.'') uncertainties in and correlations between independent variables (see Sec.~\ref{sec:fitting}).
				$\chi^2_{\kappa\chi}$ and $s_{\kappa\chi}$ are $\chi^2$ and the significance of linear fit, respectively.
			} \\
			\noalign{\vspace{3pt}} %
			\toprule\rule{0pt}{12pt} %
			& \multirow{2}{*}{GRASP} & \multirow{2}{*}{\ambit{}} & \multirow{2}{*}{Ref.~\cite{Figueroa2021}} & \multicolumn{2}{c}{Fit} \\ 
			& & & & X corr. & No X corr. \\
			\hline
			\noalign{\nobreak\vspace{3pt}}%
			\endfirsthead %
			\multicolumn{6}{c}{\tablename~\thetable{} (continued)} \\ 
			\noalign{\nobreak\vspace{3pt}}%
			\hline
			\noalign{\nobreak\vspace{3pt}}%
			& \multirow{2}{*}{GRASP} & \multirow{2}{*}{\ambit{}} & \multirow{2}{*}{Ref.~\cite{Figueroa2021}} & \multicolumn{2}{c}{Fit} \\ 
			& & & & X Corr. & No X corr. \\ \hline
			\noalign{\nobreak\vspace{3pt}}%
			\endhead %
			\noalign{\nobreak\vspace{3pt}}%
			\hline
			\noalign{\nobreak\vspace{3pt}}%
			\multicolumn{6}{r}{Continued on the next page} \\
			\noalign{\nobreak\vspace{3pt}}%
			\hline
			\endfoot %
			\botrule
			\noalign{\nobreak\vspace{3pt}}%
			\multicolumn{6}{l}{\footnotesize \textsuperscript{a} At $m_\phi = 1$~eV. Values over different $m_\phi$'s are shown in Fig.~\ref{fig:D2_vs_m}}
			\endlastfoot
			$f_{\beta\alpha}$ & 1.0152 & 1.0172 & 1.0227 & 1.01141025(86) & 1.01141025(86) \\
			$f_{\gamma\alpha}$ & -2.6427 & -2.4613 &  & -2.2213082(14) & -2.2213084(13) \\
			$f_{\delta\alpha}$ & 0.57727 & 0.66048 &  & 0.61172988(34) & 0.61172995(35) \\
			$f_{\epsilon\alpha}$ &  & 0.91933 & 0.8201 & 0.81761175(80) & 0.81761175(80) \\
			$f_{\gamma\beta}$ & -2.603 & -2.4197 &  & -2.1962536(14) & -2.1962537(13) \\
			$f_{\delta\beta}$ & 0.5686 & 0.64932 &  & 0.60482313(37) & 0.60482322(37) \\
			$f_{\epsilon\beta}$ &  & 0.90379 & 0.80192 & 0.80838924(76) & 0.80838924(76) \\
			$f_{\delta\gamma}$ & -0.21844 & -0.26835 &  & -0.275391225(69) & -0.275391430(78) \\
			$f_{\epsilon\gamma}$ &  & -0.37352 &  & -0.36807660(27) & -0.36807657(28) \\
			$f_{\epsilon\delta}$ &  & 1.3919 &  & 1.33656619(92) & 1.33656619(92) \\
			\hline
			$K_{\beta\alpha}$ & 65.306 & 103.92 &  & 120.208(23) & 120.208(23) \\
			$K_{\gamma\alpha}$ & -1307.6 & 10150 &  & 5737.593(39) & 5737.595(35) \\
			$K_{\delta\alpha}$ &  &  &  & 363.1350(94) & 363.1332(97) \\
			$K_{\epsilon\alpha}$ &  &  &  & 1.811(21) & 1.811(21) \\
			$K_{\gamma\beta}$ & -1137.6 & 10402 &  & 6001.679(38) & 6001.683(35) \\
			$K_{\delta\beta}$ &  &  &  & 290.5263(97) & 290.5242(99) \\
			$K_{\epsilon\beta}$ &  &  &  & -95.402(20) & -95.402(20) \\
			$K_{\delta\gamma}$ &  &  &  & 1943.2126(37) & 1943.2019(43) \\
			$K_{\epsilon\gamma}$ &  &  &  & 2113.679(14) & 2113.681(14) \\
			$K_{\epsilon\delta}$ &  &  &  & -483.666(15) & -483.666(15) \\
			\hline
			$G^{(4)}_{\beta\alpha}$ & -3.5056 &  & -6.4622 &  \\
			$G^{(4)}_{\gamma\alpha}$ & 45.789 &  &  &  \\
			$G^{(4)}_{\delta\alpha}$ & 329.81 &  &  &  \\
			$G^{(4)}_{\epsilon\alpha}$ &  &  & -306.88 &  \\
			$G^{(4)}_{\gamma\beta}$ & 36.664 &  &  &  \\
			$G^{(4)}_{\delta\beta}$ & 331.8 &  &  &  \\
			$G^{(4)}_{\epsilon\beta}$ &  &  & -301.7 &  \\
			$G^{(4)}_{\delta\gamma}$ & 339.81 &  &  &  \\
			$G^{(4)}_{\epsilon\gamma}$ &  &  &  &  \\
			$G^{(4)}_{\epsilon\delta}$ &  &  &  &  \\
			\hline
			\pagebreak
			$G^{(2)}_{\beta\alpha}$ & -10.442 & -68.645 & -646.64 &  &  \\
			$G^{(2)}_{\gamma\alpha}$ & 162.69 & 471.33 &  &  &  \\
			$G^{(2)}_{\delta\alpha}$ &  & 181.24 &  &  &  \\
			$G^{(2)}_{\epsilon\alpha}$ &  & 22.9 & -57.388 &  &  \\
			$G^{(2)}_{\gamma\beta}$ & 135.51 & 305.24 &  &  &  \\
			$G^{(2)}_{\delta\beta}$ &  & 225.81 &  &  &  \\
			$G^{(2)}_{\epsilon\beta}$ &  & 84.94 & 461.17 &  &  \\
			$G^{(2)}_{\delta\gamma}$ &  & 307.72 &  &  &  \\
			$G^{(2)}_{\epsilon\gamma}$ &  & 198.95 &  &  &  \\
			$G^{(2)}_{\epsilon\delta}$ &  & -229.38 &  &  &  \\
			\hline
			$D_{\beta\alpha}$\textsuperscript{a} & 3.6016 & 4.7337 & 6.6257 &  &  \\
			$D_{\gamma\alpha}$\textsuperscript{a} & -613.74 & -246.15 &  &  &  \\
			$D_{\delta\alpha}$\textsuperscript{a} & -81.212 & -71.359 &  &  &  \\
			$D_{\epsilon\alpha}$\textsuperscript{a} &  & -34.008 & -29.464 &  &  \\
			$D_{\gamma\beta}$\textsuperscript{a} & -604.37 & -234.7 &  &  &  \\
			$D_{\delta\beta}$\textsuperscript{a} & -83.26 & -74.433 &  &  &  \\
			$D_{\epsilon\beta}$\textsuperscript{a} &  & -38.286 & -34.82 &  &  \\
			$D_{\delta\gamma}$\textsuperscript{a} & -215.28 & -137.41 &  &  &  \\
			$D_{\epsilon\gamma}$\textsuperscript{a} &  & -125.95 &  &  &  \\
			$D_{\epsilon\delta}$\textsuperscript{a} &  & 65.321 &  &  &  \\
			\hline
			$\chi^2_{\beta\alpha}$ &  &  &  & 11.792 & 11.738 \\
			$\chi^2_{\gamma\alpha}$ &  &  &  & 1755.2 & 2057 \\
			$\chi^2_{\delta\alpha}$ &  &  &  & 10504 & 10010 \\
			$\chi^2_{\epsilon\alpha}$ &  &  &  & 74.581 & 74.575 \\
			$\chi^2_{\gamma\beta}$ &  &  &  & 2220.6 & 2546 \\
			$\chi^2_{\delta\beta}$ &  &  &  & 16555 & 15916 \\
			$\chi^2_{\epsilon\beta}$ &  &  &  & 137.48 & 137.91 \\
			$\chi^2_{\delta\gamma}$ &  &  &  & 57854 & 43986 \\
			$\chi^2_{\epsilon\gamma}$ &  &  &  & 2040.2 & 1920.7 \\
			$\chi^2_{\epsilon\delta}$ &  &  &  & 4511.9 & 4512 \\
			\hline
			$s_{\beta\alpha}$ &  &  &  & 2.99$\sigma$ & 2.99$\sigma$ \\
			$s_{\gamma\alpha}$ &  &  &  & 41.8$\sigma$ & 45.3$\sigma$ \\
			$s_{\delta\alpha}$ &  &  &  & 102$\sigma$ & 100$\sigma$ \\
			$s_{\epsilon\alpha}$ &  &  &  & 8.36$\sigma$ & 8.36$\sigma$ \\
			$s_{\gamma\beta}$ &  &  &  & 47$\sigma$ & 50.4$\sigma$ \\
			$s_{\delta\beta}$ &  &  &  & 129$\sigma$ & 126$\sigma$ \\
			$s_{\epsilon\beta}$ &  &  &  & 11.5$\sigma$ & 11.5$\sigma$ \\
			$s_{\delta\gamma}$ &  &  &  & 241$\sigma$ & 210$\sigma$ \\
			$s_{\epsilon\gamma}$ &  &  &  & 45.1$\sigma$ & 43.7$\sigma$ \\
			$s_{\epsilon\delta}$ &  &  &  & 67.1$\sigma$ & 67.1$\sigma$ 
		\end{longtable*}
		
		\begin{figure*}
			\centering
			\includegraphics[width=\textwidth]{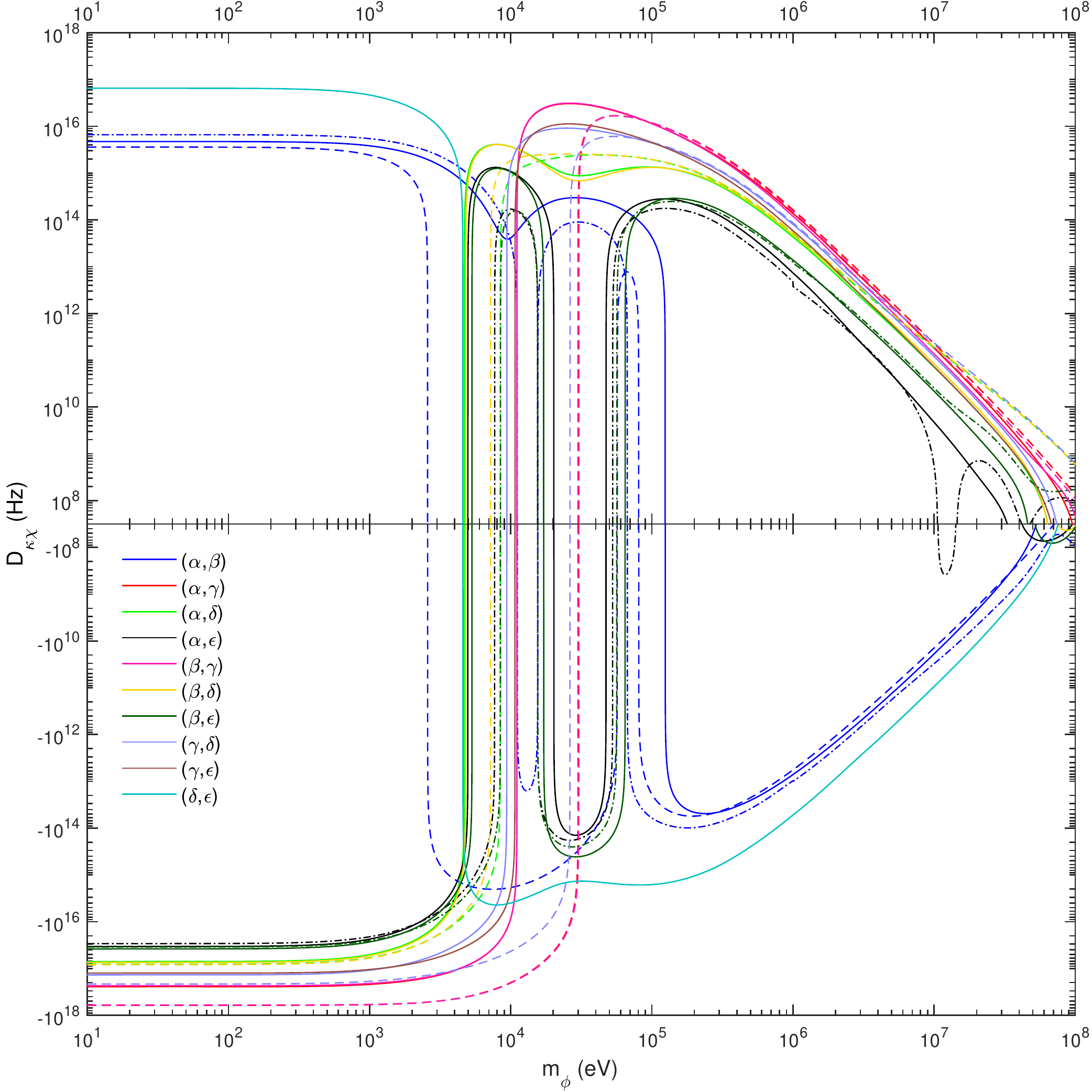}
			\caption{Two-transition factors $D_{\kappa\chi}$ vs new-boson mass $m_\phi$ for variable transition pairs ($\chi$,$\kappa$) coded with different colors (see legend) calculated using $D_{\chi}$ and $D_{\kappa}$ in Fig.~\ref{fig:D1_vs_m}. Solid, dashed, and dash-dotted lines are for \ambit{}, GRASP2018, and Ref.~\cite{Figueroa2021}, respectively (some of dashed and dash-dotted lines are missing as the corresponding $D_{\kappa\chi}$ are not available; see Table~\ref{tab:el_factors_2}).}
			\label{fig:D2_vs_m}
		\end{figure*}

		\setlength{\LTcapwidth}{\textwidth}
		\begin{longtable*}{@{\extracolsep{\fill}}l|cc|cc|cc|ccc@{}}
			\caption{\label{tab:el_factors_3}
				Calculated and experimental values of three-transition electronic factors $f_{\eta\kappa\chi}$ and $Z_{\eta\kappa\chi}$ ($Z \in \{K, G^{(2)}, D\}$) for $\chi, \kappa, \eta \in \{\alpha, \beta, \gamma, \delta, \epsilon\}$. The quantities are defined in Sec.~\ref{sec:3Dking}. 
				The values are calculated from the two-transition values in Table~\ref{tab:el_factors_2}.
				$f_{\eta\kappa\chi}$ is dimensionless. The units of $K_{\eta\kappa\chi}$, $G^{(2)}_{\eta\kappa\chi}$, $D_{\eta\kappa\chi}$, and $\upsilon_{ne}D_{\chi\kappa}$ are GHz$\cdot$u, kHz/fm$^4$, kHz/fm$^4$, $10^3$ THz, and kHz, respectively. 
				The last three columns (``Fit'') are for data from fit of corresponding 3D King plots $ \bbar{\nu}^{AA'}_{\eta} = K_{\eta\kappa\chi} + f_{\eta\kappa\chi} \bbar{\nu}^{AA'}_{\chi} + f_{\eta\chi\kappa} \bbar{\nu}^{AA'}_{\kappa}$ (``Linear``), and $G^{(2)}_{\eta\kappa\chi} \bbar{\drtsq}^{AA'}$ (``QFS'') or $\upsilon_{ne}D_{\eta\kappa\chi} \bbar{a}^{AA'}$ (``New boson'') terms in addition to the relation.
				$\chi^2_{\eta\kappa\chi}$ and $s_{\eta\kappa\chi}$ are $\chi^2$ and the significance of fit, respectively.
			} \\
			\noalign{\vspace{3pt}} %
			\toprule\rule{0pt}{12pt} %
			& \multicolumn{2}{c}{GRASP} & \multicolumn{2}{c}{\ambit{}} & \multicolumn{2}{c}{Ref.~\cite{Figueroa2021}} & \multicolumn{3}{c}{Fit} \\ 
			& Cal. & Exp. & Cal. & Exp. & Cal. & Exp. & Linear & QFS & New boson \\
			\hline
			\noalign{\nobreak\vspace{3pt}}%
			\endfirsthead %
			\multicolumn{10}{c}{\tablename~\thetable{} (continued)} \\ 
			\noalign{\nobreak\vspace{3pt}}%
			\hline
			\noalign{\nobreak\vspace{3pt}}%
			& \multicolumn{2}{c}{GRASP} & \multicolumn{2}{c}{\ambit{}} & \multicolumn{2}{c}{Ref.~\cite{Figueroa2021}} & \multicolumn{3}{c}{Fit} \\ 
			& Cal. & Exp. & Cal. & Exp. & Cal. & Exp. & Linear & QFS & New boson \\ \hline
			\noalign{\nobreak\vspace{3pt}}%
			\endhead %
			\noalign{\nobreak\vspace{3pt}}%
			\hline
			\noalign{\nobreak\vspace{3pt}}%
			\multicolumn{10}{r}{Continued on the next page} \\
			\noalign{\nobreak\vspace{3pt}}%
			\hline
			\endfoot %
			\botrule
			\multicolumn{6}{l}{\footnotesize \textsuperscript{a} At $m_\phi = 1$~eV. Values over different $m_\phi$'s are shown in Fig.~\ref{fig:D3_vs_m}}
			\endlastfoot
			$f_{\beta\gamma\alpha}$ & 0.81292 &  &  &  &  &  & 0.978(26) & 0.998(27) & 1.052(36) \\
			$f_{\beta\delta\alpha}$ & 1.0214 &  &  &  &  &  & 1.023(13) & 1.018(13) & 0.993(16) \\
			$f_{\beta\epsilon\alpha}$ &  &  &  &  & 1.0054 &  & 1.14(10) & 1.058(99) & 0.86(12) \\
			$f_{\delta\gamma\alpha}$ & 19.612 &  &  &  &  &  & 1.867(41) & 1.877(41) & 1.965(51) \\
			$f_{\epsilon\gamma\alpha}$ &  &  &  &  &  &  & 1.049(30) & 1.046(33) & 1.040(45) \\
			$f_{\epsilon\delta\alpha}$ &  &  &  &  &  &  & 0.701(13) & 0.707(14) & 0.717(19) \\
			$f_{\delta\gamma\beta}$ & 24.126 &  &  &  &  &  & 1.885(35) & 1.880(37) & 1.868(49) \\
			$f_{\epsilon\gamma\beta}$ &  &  &  &  &  &  & 1.090(28) & 1.047(31) & 0.989(43) \\
			$f_{\epsilon\delta\beta}$ &  &  &  &  &  &  & 0.673(11) & 0.695(13) & 0.722(19) \\
			$f_{\epsilon\delta\gamma}$ &  &  &  &  &  &  & -0.2082(32) & -0.2146(38) & -0.2223(54) \\
			\hline
			$f_{\beta\alpha\gamma}$ & -0.076559 &  &  &  &  &  & -0.015(12) & -0.006(12) & 0.018(16) \\
			$f_{\beta\alpha\delta}$ & -0.010629 &  &  &  &  &  & -0.019(21) & -0.010(21) & 0.030(26) \\
			$f_{\beta\alpha\epsilon}$ &  &  &  &  & 0.021058 &  & -0.15(12) & -0.06(12) & 0.18(15) \\
			$f_{\delta\alpha\gamma}$ & 7.2027 &  &  &  &  &  & 0.565(18) & 0.570(19) & 0.609(23) \\
			$f_{\epsilon\alpha\gamma}$ &  &  &  &  &  &  & 0.104(14) & 0.103(15) & 0.100(20) \\
			$f_{\epsilon\alpha\delta}$ &  &  &  &  &  &  & 0.191(21) & 0.180(23) & 0.165(31) \\
			$f_{\delta\beta\gamma}$ & 9.0498 &  &  &  &  &  & 0.583(16) & 0.580(17) & 0.575(22) \\
			$f_{\epsilon\beta\gamma}$ &  &  &  &  &  &  & 0.128(13) & 0.109(14) & 0.082(20) \\
			$f_{\epsilon\beta\delta}$ &  &  &  &  &  &  & 0.223(19) & 0.187(22) & 0.143(32) \\
			$f_{\epsilon\gamma\delta}$ &  &  &  &  &  &  & 0.580(12) & 0.557(14) & 0.529(20) \\
			\hline
			$K_{\beta\gamma\alpha}$ & -34.804 & 89(21) &  & -80(160) &  &  & 206(66) & 154(69) & 17(91) \\
			$K_{\beta\delta\alpha}$ &  &  &  &  &  &  & 126.9(7.7) & 124.2(7.8) & 111.1(9.0) \\
			$K_{\beta\epsilon\alpha}$ &  &  &  &  &  &  & 120.48(23) & 120.53(21) & 121.79(43) \\
			$K_{\delta\gamma\alpha}$ &  &  &  &  &  &  & -2880(110) & -2900(110) & -3130(130) \\
			$K_{\epsilon\gamma\alpha}$ &  &  &  &  &  &  & -596(79) & -587(85) & -570(120) \\
			$K_{\epsilon\delta\alpha}$ &  &  &  &  &  &  & -67.6(7.6) & -63.8(8.3) & -59(11) \\
			$K_{\delta\gamma\beta}$ &  &  &  &  &  &  & -3207(97) & -3190(100) & -3160(130) \\
			$K_{\epsilon\gamma\beta}$ &  &  &  &  &  &  & -865(77) & -749(86) & -590(120) \\
			$K_{\epsilon\delta\beta}$ &  &  &  &  &  &  & -160.3(5.5) & -150.1(6.4) & -138.8(8.8) \\
			$K_{\epsilon\delta\gamma}$ &  &  &  &  &  &  & 986(23) & 1031(27) & 1083(38) \\
			\hline
			\pagebreak
			$G^{(2)}_{\beta\gamma\alpha}$ & 2.0139 & -13.4(2.6) &  & -77.3(7.5) &  &  &  & 57(18) &  \\
			$G^{(2)}_{\beta\delta\alpha}$ &  &  &  & -74.1(4.6) &  &  &  & 58(18) &  \\
			$G^{(2)}_{\beta\epsilon\alpha}$ &  &  &  & -72.8(3.4) & -645.43 & -636.1(8.6) &  & 57(18) &  \\
			$G^{(2)}_{\delta\gamma\alpha}$ &  &  &  & -106(11) &  &  &  & 94(24) &  \\
			$G^{(2)}_{\epsilon\gamma\alpha}$ &  &  &  & -24.4(9.6) &  &  &  & -6(22) &  \\
			$G^{(2)}_{\epsilon\delta\alpha}$ &  &  &  & -6.9(5.6) &  &  &  & -23(19) &  \\
			$G^{(2)}_{\delta\gamma\beta}$ &  &  &  & 50.2(6.8) &  &  &  & -13(27) &  \\
			$G^{(2)}_{\epsilon\gamma\beta}$ &  &  &  & 59.9(6.0) &  &  &  & -65(22) &  \\
			$G^{(2)}_{\epsilon\delta\beta}$ &  &  &  & 52.7(7.2) &  &  &  & -63(20) &  \\
			$G^{(2)}_{\epsilon\delta\gamma}$ &  &  &  & 36.1(6.1) &  &  &  & -58(18) &  \\
			\hline
			$D_{\beta\gamma\alpha}$\textsuperscript{a} & -43.386 & 14.8(9.8) &  & 9.2(3.9) &  &  &  &  &  \\
			$D_{\beta\delta\alpha}$\textsuperscript{a} & 2.7384 & 6.0(2.1) &  & 6.9(1.8) &  &  &  &  &  \\
			$D_{\beta\epsilon\alpha}$\textsuperscript{a} &  &  &  & 11.0(5.1) & 7.2462 & 12.0(4.4) &  &  &  \\
			$D_{\delta\gamma\alpha}$\textsuperscript{a} & 4339.4 & 293(14) &  & 78.7(5.6) &  &  &  &  &  \\
			$D_{\epsilon\gamma\alpha}$\textsuperscript{a} &  &  &  & -9.3(5.0) &  &  &  &  &  \\
			$D_{\epsilon\delta\alpha}$\textsuperscript{a} &  &  &  & -22.3(2.2) &  &  &  &  &  \\
			$D_{\delta\gamma\beta}$\textsuperscript{a} & 5386.1 & 264(13) &  & 60.6(5.2) &  &  &  &  &  \\
			$D_{\epsilon\gamma\beta}$\textsuperscript{a} &  &  &  & -19.0(4.6) &  &  &  &  &  \\
			$D_{\epsilon\delta\beta}$\textsuperscript{a} &  &  &  & -27.7(2.4) &  &  &  &  &  \\
			$D_{\epsilon\delta\gamma}$\textsuperscript{a} &  &  &  & -53.2(2.7) &  &  &  &  &  \\
			\hline
			$\upsilon_{ne}D_{\beta\gamma\alpha}$ &  &  &  &  &  &  &  &  & 54(17) \\
			$\upsilon_{ne}D_{\beta\delta\alpha}$ &  &  &  &  &  &  &  &  & 51(15) \\
			$\upsilon_{ne}D_{\beta\epsilon\alpha}$ &  &  &  &  &  &  &  &  & 55(17) \\
			$\upsilon_{ne}D_{\delta\gamma\alpha}$ &  &  &  &  &  &  &  &  & 88(24) \\
			$\upsilon_{ne}D_{\epsilon\gamma\alpha}$ &  &  &  &  &  &  &  &  & -5(21) \\
			$\upsilon_{ne}D_{\epsilon\delta\alpha}$ &  &  &  &  &  &  &  &  & -20(17) \\
			$\upsilon_{ne}D_{\delta\gamma\beta}$ &  &  &  &  &  &  &  &  & -12(23) \\
			$\upsilon_{ne}D_{\epsilon\gamma\beta}$ &  &  &  &  &  &  &  &  & -58(19) \\
			$\upsilon_{ne}D_{\epsilon\delta\beta}$ &  &  &  &  &  &  &  &  & -57(18) \\
			$\upsilon_{ne}D_{\epsilon\delta\gamma}$ &  &  &  &  &  &  &  &  & -52(16) \\
			\hline
			$\chi^2_{\beta\gamma\alpha}$ &  &  &  &  &  &  & 10.532 &  &  \\
			$\chi^2_{\beta\delta\alpha}$ &  &  &  &  &  &  & 10.9 &  &  \\
			$\chi^2_{\beta\epsilon\alpha}$ &  &  &  &  &  &  & 8.724 &  &  \\
			$\chi^2_{\delta\gamma\alpha}$ &  &  &  &  &  &  & 15.221 &  &  \\
			$\chi^2_{\epsilon\gamma\alpha}$ &  &  &  &  &  &  & 0.065554 &  &  \\
			$\chi^2_{\epsilon\delta\alpha}$ &  &  &  &  &  &  & 1.4067 &  &  \\
			$\chi^2_{\delta\gamma\beta}$ &  &  &  &  &  &  & 0.23876 &  &  \\
			$\chi^2_{\epsilon\gamma\beta}$ &  &  &  &  &  &  & 8.3928 &  &  \\
			$\chi^2_{\epsilon\delta\beta}$ &  &  &  &  &  &  & 10.248 &  &  \\
			$\chi^2_{\epsilon\delta\gamma}$ &  &  &  &  &  &  & 10.481 &  &  \\
			\hline
			\pagebreak
			$s_{\beta\gamma\alpha}$ &  &  &  &  &  &  & 3.2454$\sigma$ &  &  \\
			$s_{\beta\delta\alpha}$ &  &  &  &  &  &  & 3.3015$\sigma$ &  &  \\
			$s_{\beta\epsilon\alpha}$ &  &  &  &  &  &  & 2.9536$\sigma$ &  &  \\
			$s_{\delta\gamma\alpha}$ &  &  &  &  &  &  & 3.9014$\sigma$ &  &  \\
			$s_{\epsilon\gamma\alpha}$ &  &  &  &  &  &  & 0.25604$\sigma$ &  &  \\
			$s_{\epsilon\delta\alpha}$ &  &  &  &  &  &  & 1.1861$\sigma$ &  &  \\
			$s_{\delta\gamma\beta}$ &  &  &  &  &  &  & 0.48863$\sigma$ &  &  \\
			$s_{\epsilon\gamma\beta}$ &  &  &  &  &  &  & 2.897$\sigma$ &  &  \\
			$s_{\epsilon\delta\beta}$ &  &  &  &  &  &  & 3.2012$\sigma$ &  &  \\
			$s_{\epsilon\delta\gamma}$ &  &  &  &  &  &  & 3.2375$\sigma$ &  &  
		\end{longtable*}

		\begin{figure*}
			\centering
			\subfloat[$(\alpha,\gamma,\beta)$ transitions]{
				\includegraphics[width=0.95\columnwidth]{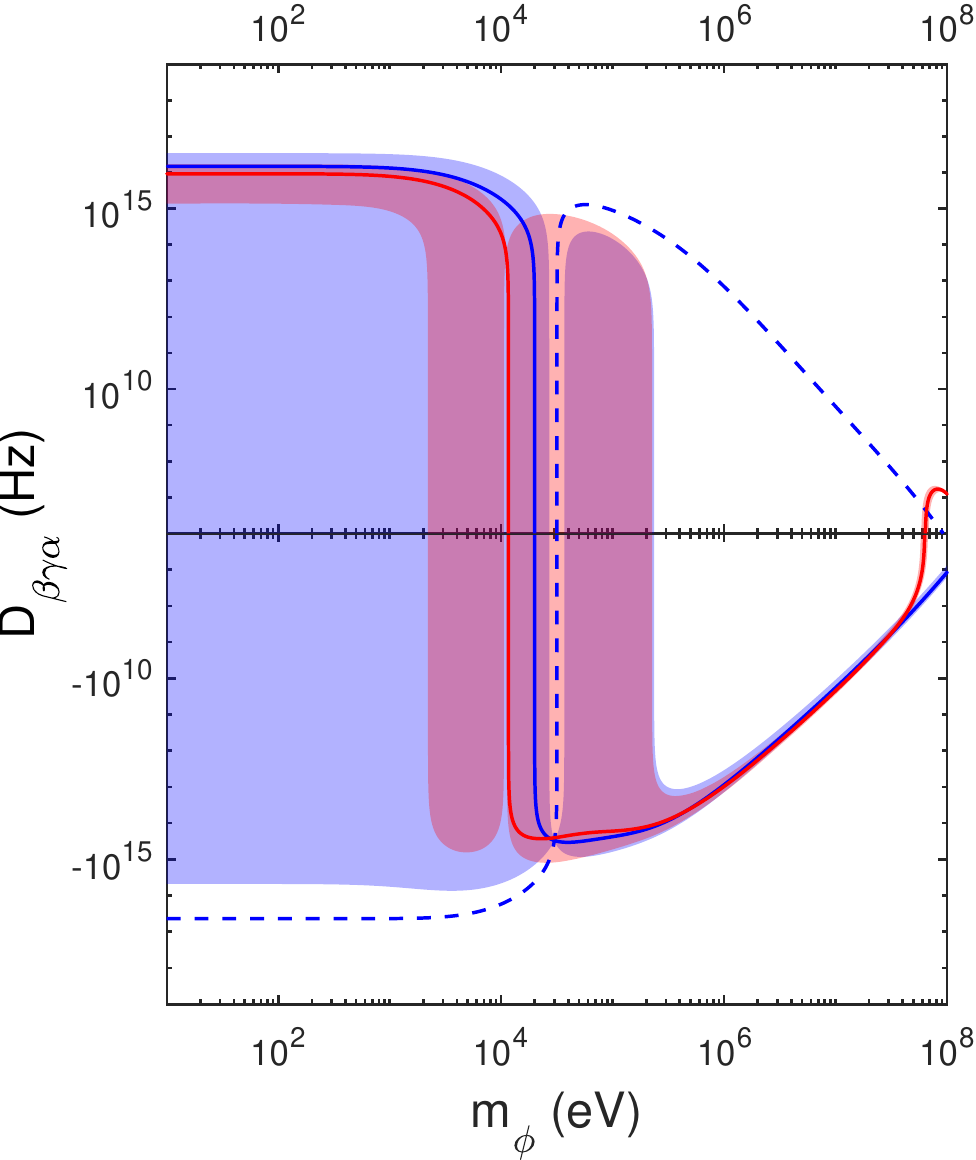} 
			}\hfill
			\subfloat[$(\alpha,\delta,\beta)$ transitions]{
				\includegraphics[width=0.95\columnwidth]{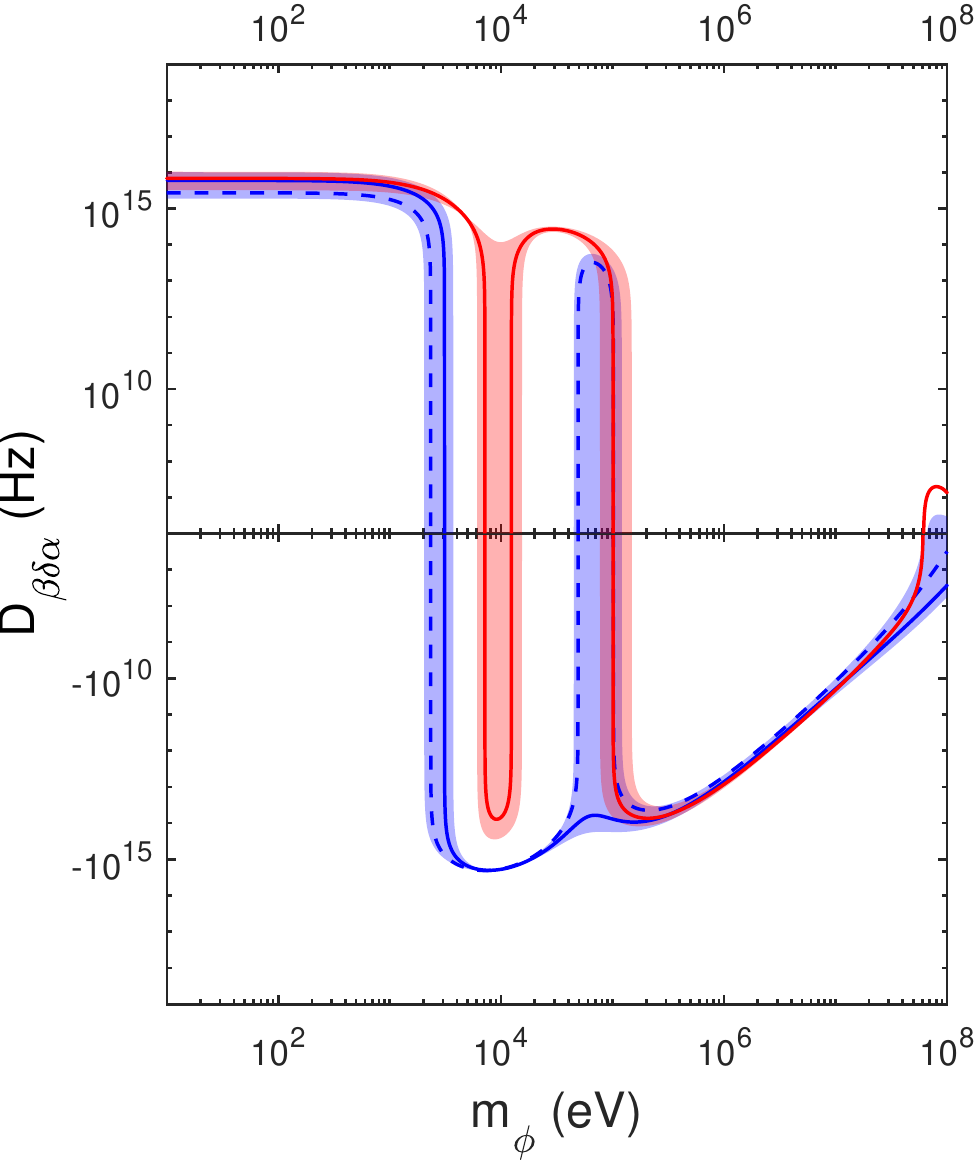} 
			}\hfill
			\subfloat[$(\alpha,\epsilon,\beta)$ transitions]{
				\includegraphics[width=0.95\columnwidth]{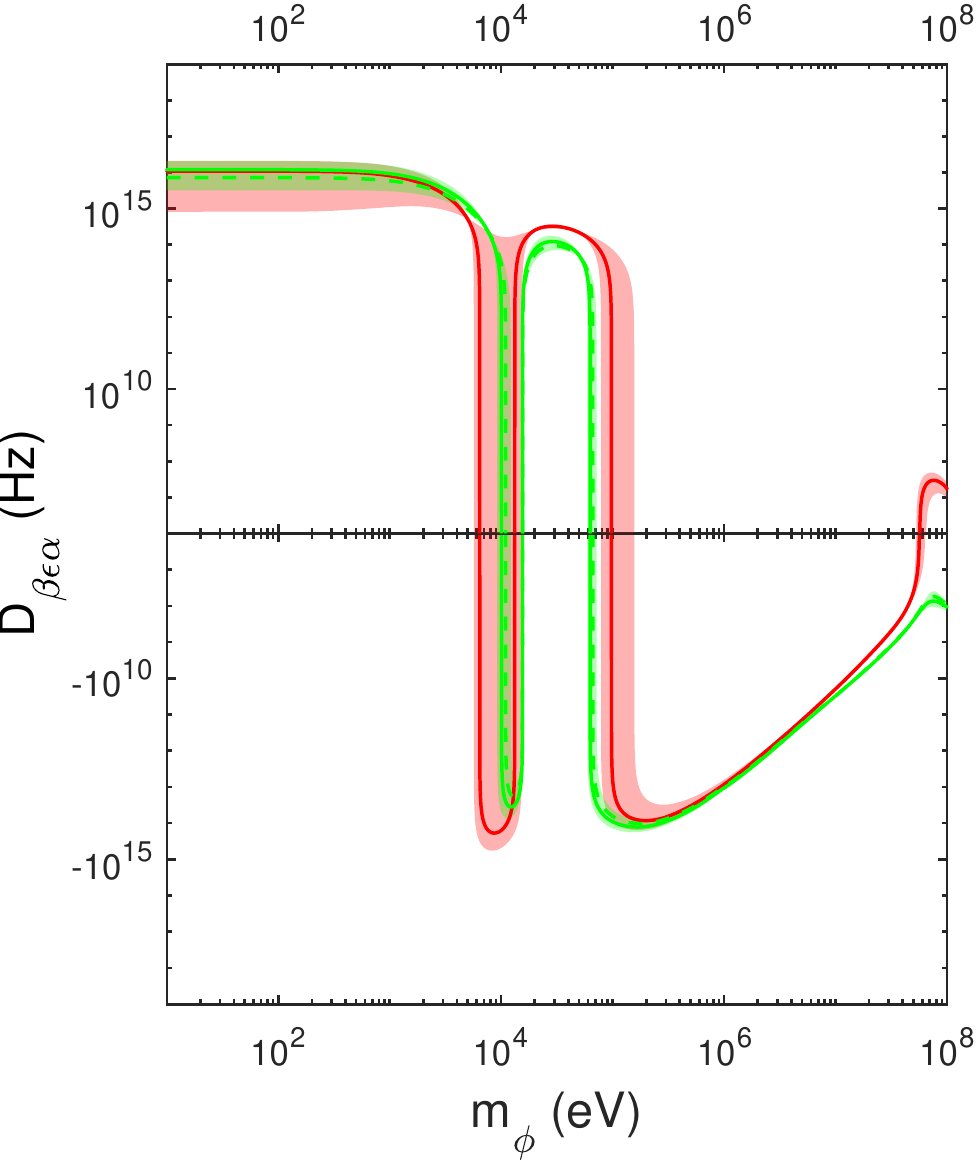}
			}\hfill
			\subfloat[$(\alpha,\gamma,\delta)$ transitions]{
				\includegraphics[width=0.95\columnwidth]{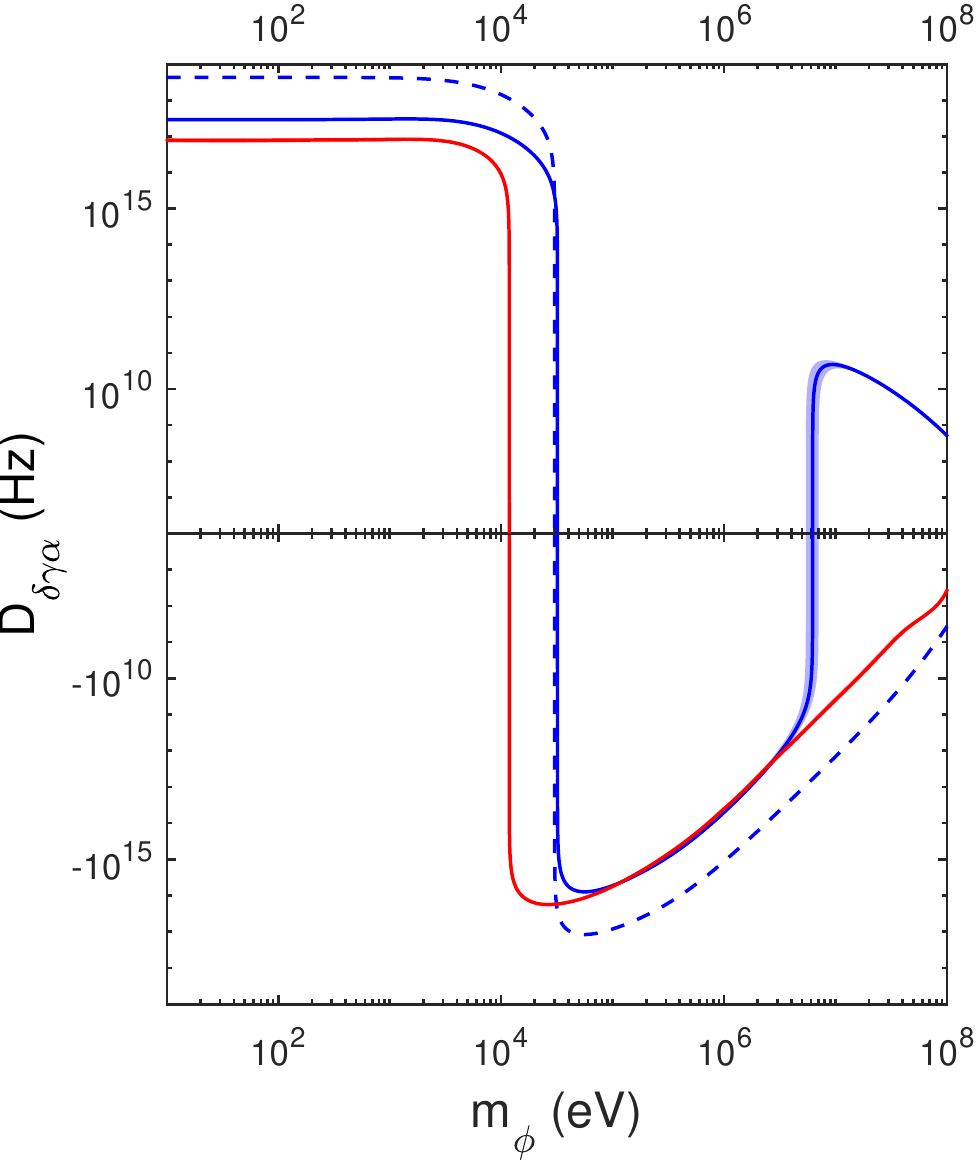}
			}\hfill
			\caption{Calculated $D_{\eta\kappa\chi}$ vs new-boson mass $m_\phi$ for all different choices of three transitions $(\chi,\kappa,\eta)$ out of five available transitions $\alpha$, $\beta$, $\gamma$, $\delta$, and $\epsilon$, each corresponding to one of the subfigures (a -- j). Solid lines correspond to the $D_{\eta\kappa\chi}$ obtained from $D_{\kappa\chi}$ and $D_{\eta\chi}$ in Fig.~\ref{fig:D2_vs_m}, and $f_{\eta\chi\kappa} = G^{(4)}_{\eta\chi}/G^{(4)}_{\kappa\chi}$ ratio from the linear fit in 3D King plot (see Table~\ref{tab:el_factors_3}). Shaded regions for $D_{\eta\kappa\chi}$ indicate 95\% confidence intervals that arise from fitted $f_{\eta\chi\kappa}$'s uncertainty. (Figures and caption continue on the next page.)}
			\label{fig:D3_vs_m}
		\end{figure*}
		
		\addtocounter{figure}{-1}
		\begin{figure*}
			\centering
			\addtocounter{subfigure}{4}
			\subfloat[$(\alpha,\gamma,\epsilon)$ transitions]{
				\includegraphics[width=0.95\columnwidth]{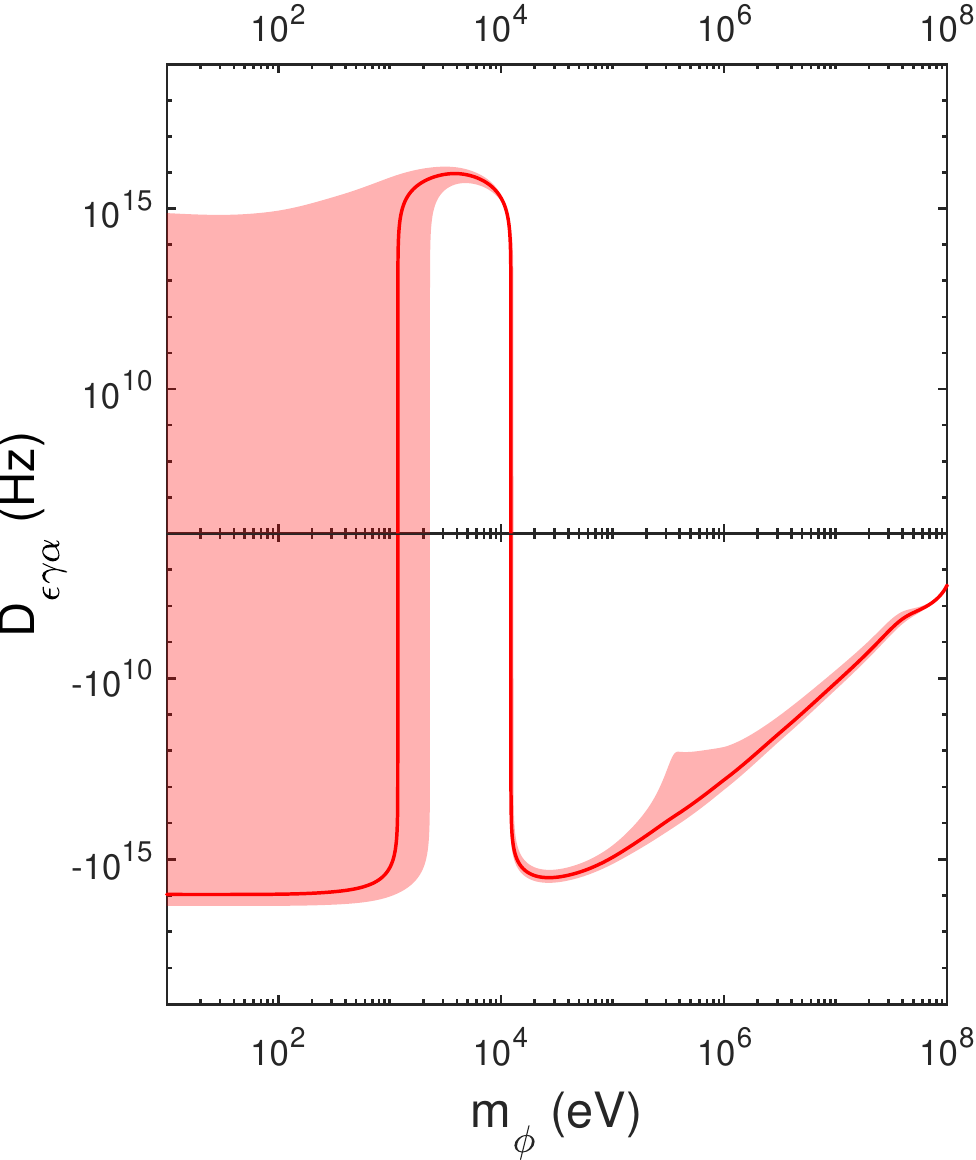} 
			}\hfill
			\subfloat[$(\alpha,\delta,\epsilon)$ transitions]{
				\includegraphics[width=0.95\columnwidth]{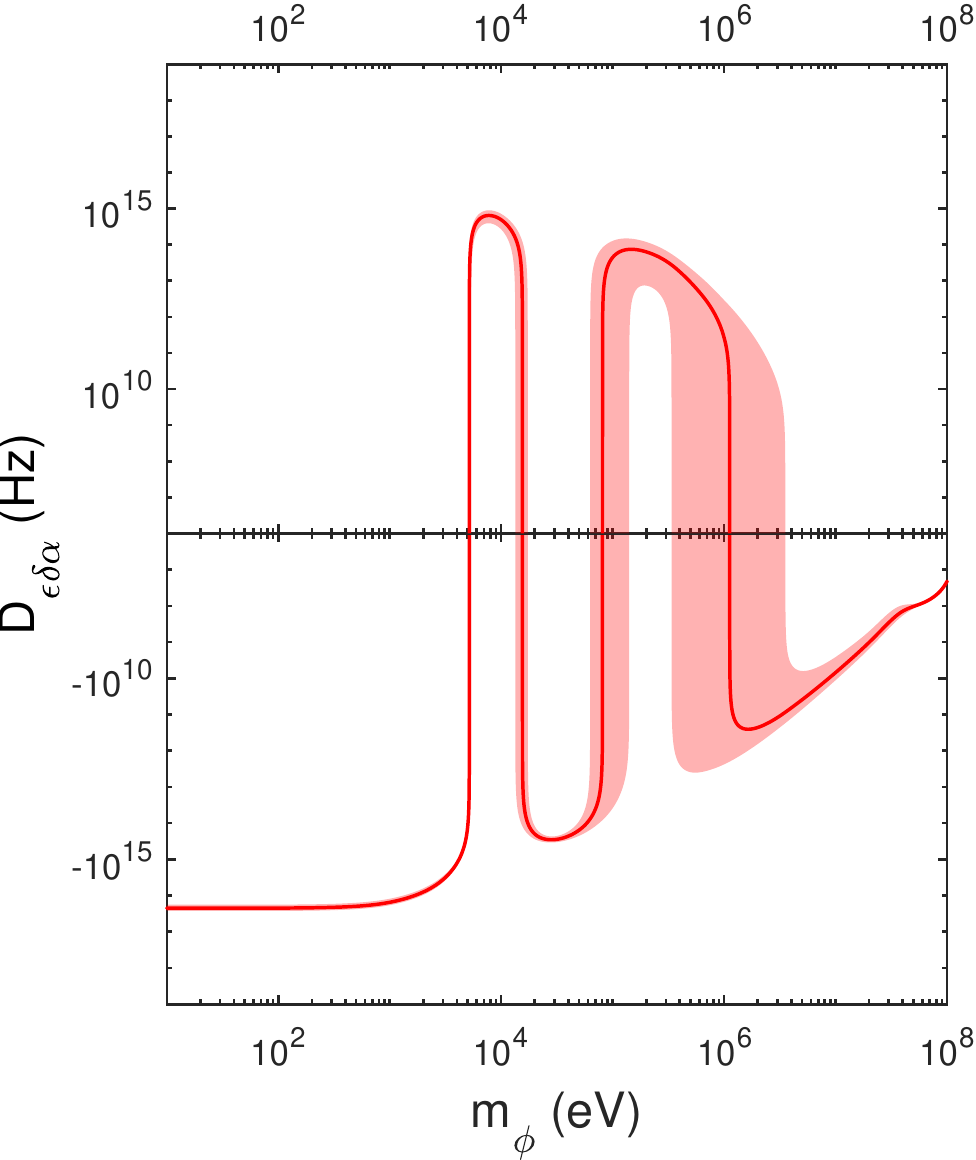} 
			}\hfill
			\subfloat[$(\beta,\gamma,\delta)$ transitions]{
				\includegraphics[width=0.95\columnwidth]{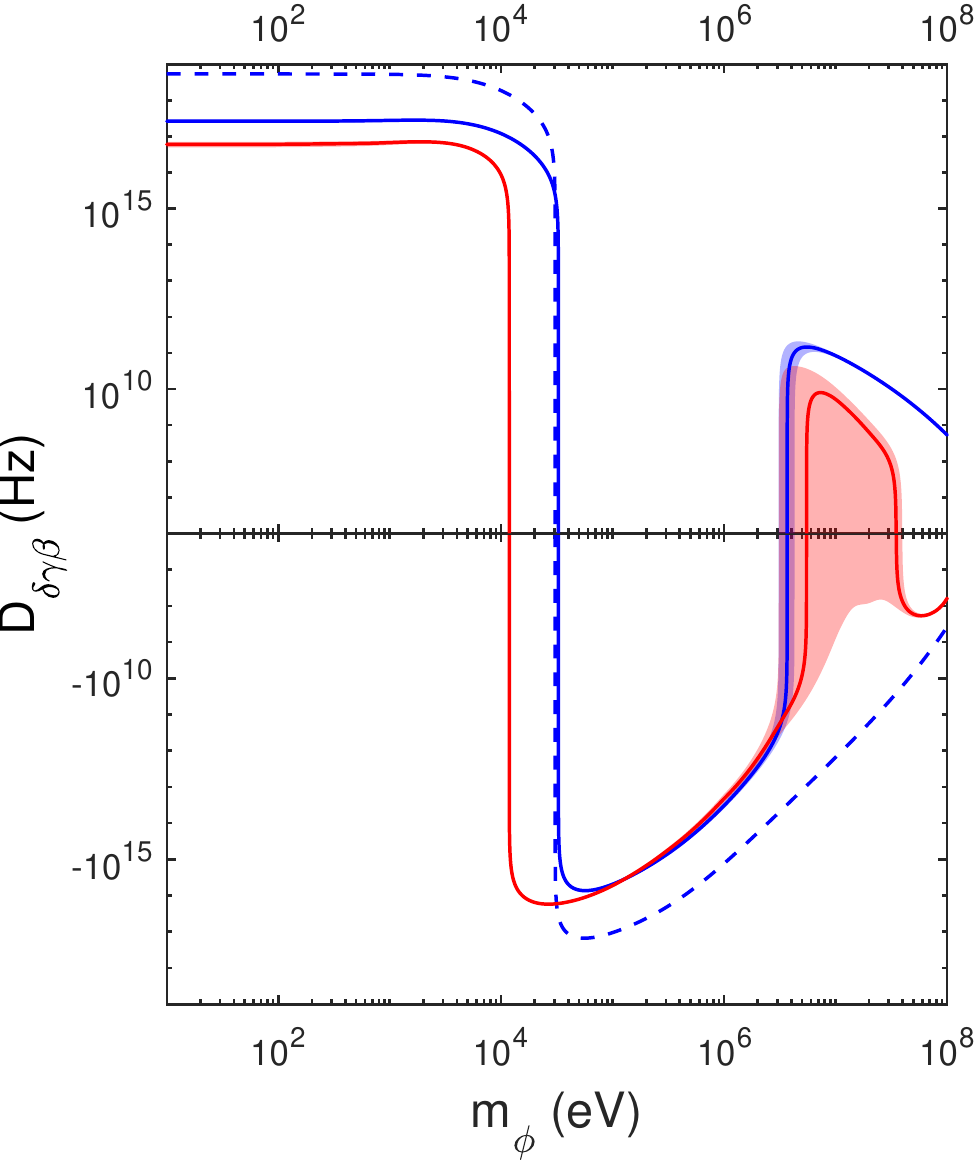}
			}\hfill
			\subfloat[$(\beta,\gamma,\epsilon)$ transitions]{
				\includegraphics[width=0.95\columnwidth]{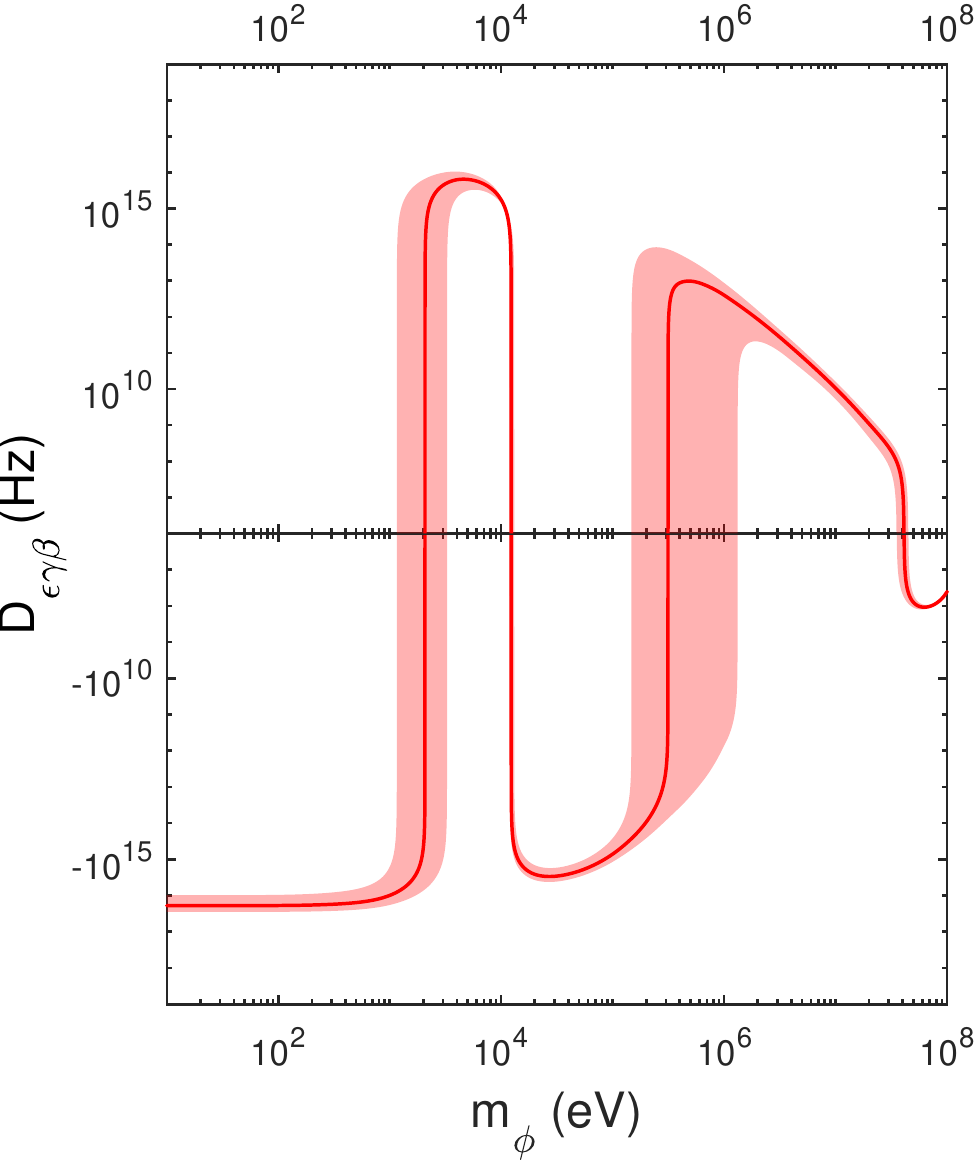}
			}\hfill
			\caption{(Continued) Dashed lines show $D_{\eta\kappa\chi}$ calculated purely from ASCs (i.e., using calculated $f_{\eta\chi\kappa}$). Blue, red, and green colors correspond to ASCs performed using GRASP2018, \ambit{}, and in Ref.~\cite{Figueroa2021}, respectively. (Figures continued on the next page.)}
		\end{figure*}
		
		\addtocounter{figure}{-1}
		\begin{figure*}
			\centering
			\addtocounter{subfigure}{8}
			\subfloat[$(\beta,\delta,\epsilon)$ transitions]{
				\includegraphics[width=0.95\columnwidth]{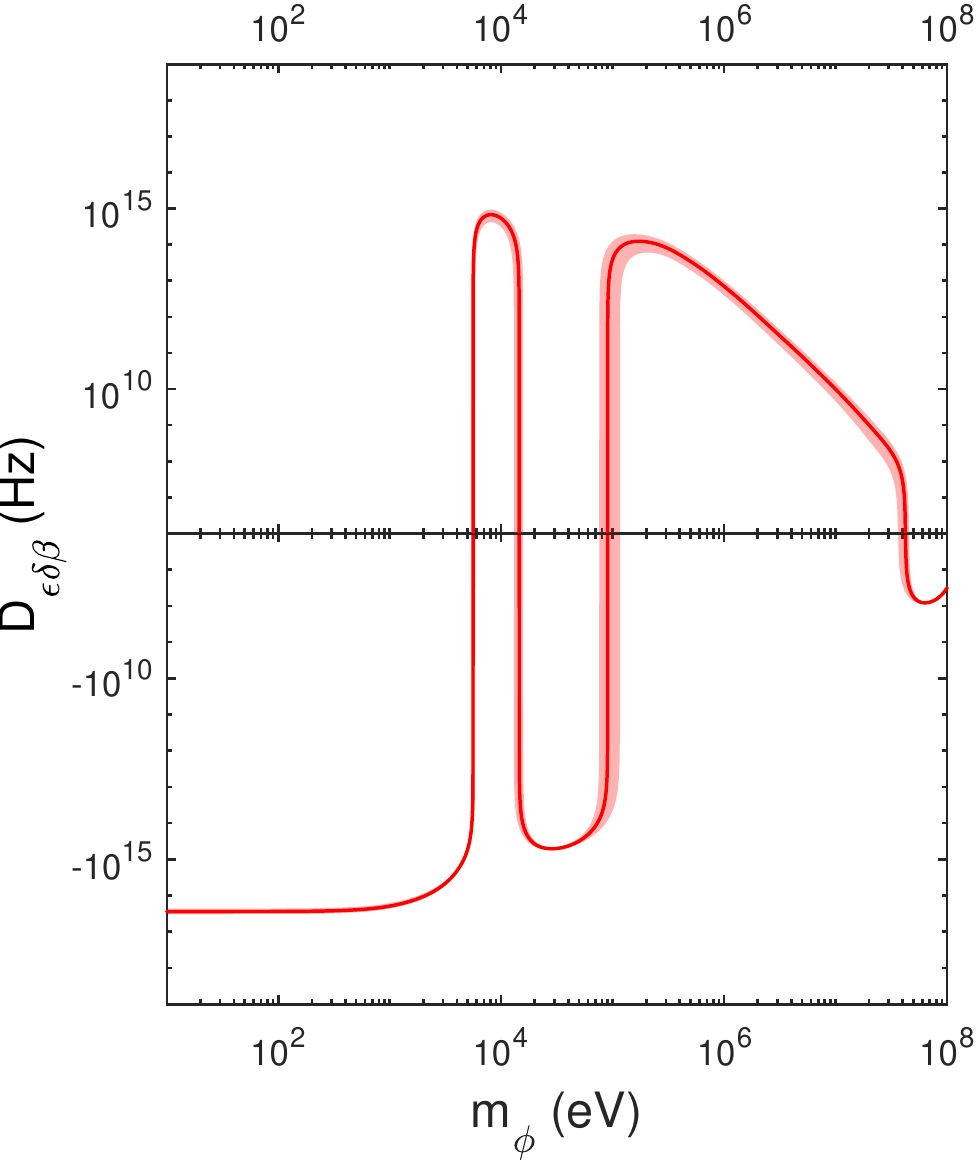}
			}\hfill
			\subfloat[$(\gamma,\delta,\epsilon)$ transitions]{
				\includegraphics[width=0.95\columnwidth]{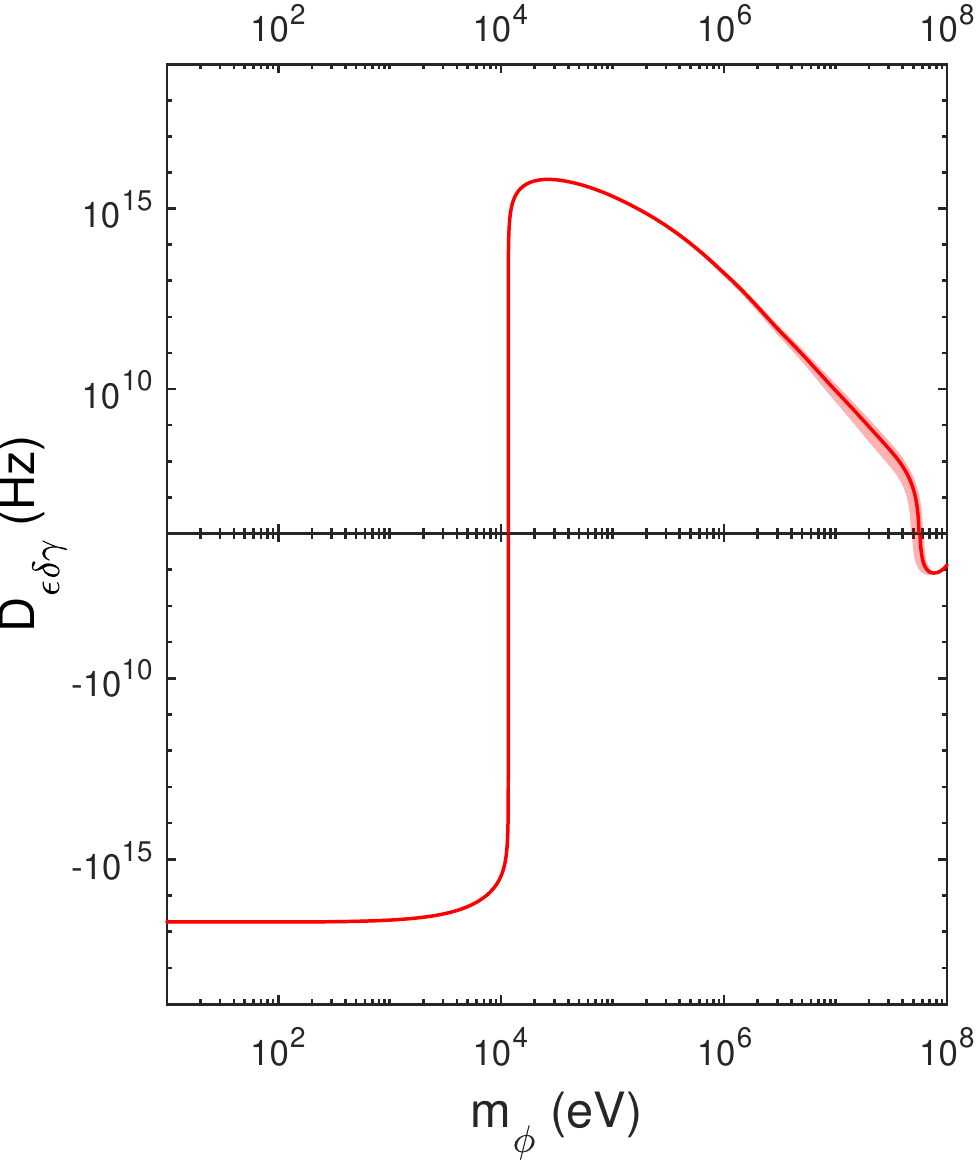}
			}\hfill
			\caption{(Continued)}
		\end{figure*}


\begin{thebibliography}{86}%
		\makeatletter
		\providecommand \@ifxundefined [1]{%
			\@ifx{#1\undefined}
		}%
		\providecommand \@ifnum [1]{%
			\ifnum #1\expandafter \@firstoftwo
			\else \expandafter \@secondoftwo
			\fi
		}%
		\providecommand \@ifx [1]{%
			\ifx #1\expandafter \@firstoftwo
			\else \expandafter \@secondoftwo
			\fi
		}%
		\providecommand \natexlab [1]{#1}%
		\providecommand \enquote  [1]{``#1''}%
		\providecommand \bibnamefont  [1]{#1}%
		\providecommand \bibfnamefont [1]{#1}%
		\providecommand \citenamefont [1]{#1}%
		\providecommand \href@noop [0]{\@secondoftwo}%
		\providecommand \href [0]{\begingroup \@sanitize@url \@href}%
		\providecommand \@href[1]{\@@startlink{#1}\@@href}%
		\providecommand \@@href[1]{\endgroup#1\@@endlink}%
		\providecommand \@sanitize@url [0]{\catcode `\\12\catcode `\$12\catcode
			`\&12\catcode `\#12\catcode `\^12\catcode `\_12\catcode `\%12\relax}%
		\providecommand \@@startlink[1]{}%
		\providecommand \@@endlink[0]{}%
		\providecommand \url  [0]{\begingroup\@sanitize@url \@url }%
		\providecommand \@url [1]{\endgroup\@href {#1}{\urlprefix }}%
		\providecommand \urlprefix  [0]{URL }%
		\providecommand \Eprint [0]{\href }%
		\providecommand \doibase [0]{https://doi.org/}%
		\providecommand \selectlanguage [0]{\@gobble}%
		\providecommand \bibinfo  [0]{\@secondoftwo}%
		\providecommand \bibfield  [0]{\@secondoftwo}%
		\providecommand \translation [1]{[#1]}%
		\providecommand \BibitemOpen [0]{}%
		\providecommand \bibitemStop [0]{}%
		\providecommand \bibitemNoStop [0]{.\EOS\space}%
		\providecommand \EOS [0]{\spacefactor3000\relax}%
		\providecommand \BibitemShut  [1]{\csname bibitem#1\endcsname}%
		\let\auto@bib@innerbib\@empty
		\bibitem [{\citenamefont {Rubin}\ \emph {et~al.}(1980)\citenamefont {Rubin},
			\citenamefont {Ford},\ and\ \citenamefont {Thonnard}}]{GallaxyRotation}%
		\BibitemOpen
		\bibfield  {author} {\bibinfo {author} {\bibfnamefont {V.~C.}\ \bibnamefont
				{Rubin}}, \bibinfo {author} {\bibfnamefont {W.~K.}\ \bibnamefont {Ford}},\
			and\ \bibinfo {author} {\bibfnamefont {N.}~\bibnamefont {Thonnard}},\
		}\bibfield  {title} {\bibinfo {title} {Rotational properties of 21 {Sc}
				galaxies with a large range of luminosities and radii, from {NGC} 4605 ({$R =
					4$ kpc}) to {UGC} 2885 ({$R = 122$ kpc})},\ }\href
		{https://doi.org/10.1086/158003} {\bibfield  {journal} {\bibinfo  {journal}
				{Astrophys. J.}\ }\textbf {\bibinfo {volume} {238}},\ \bibinfo {pages} {471}
			(\bibinfo {year} {1980})}\BibitemShut {NoStop}%
		\bibitem [{\citenamefont {Clowe}\ \emph {et~al.}(2006)\citenamefont {Clowe},
			\citenamefont {Brada{\v{c}}}, \citenamefont {Gonzalez}, \citenamefont
			{Markevitch}, \citenamefont {Randall}, \citenamefont {Jones},\ and\
			\citenamefont {Zaritsky}}]{GallaxyCollision}%
		\BibitemOpen
		\bibfield  {author} {\bibinfo {author} {\bibfnamefont {D.}~\bibnamefont
				{Clowe}}, \bibinfo {author} {\bibfnamefont {M.}~\bibnamefont {Brada{\v{c}}}},
			\bibinfo {author} {\bibfnamefont {A.~H.}\ \bibnamefont {Gonzalez}}, \bibinfo
			{author} {\bibfnamefont {M.}~\bibnamefont {Markevitch}}, \bibinfo {author}
			{\bibfnamefont {S.~W.}\ \bibnamefont {Randall}}, \bibinfo {author}
			{\bibfnamefont {C.}~\bibnamefont {Jones}},\ and\ \bibinfo {author}
			{\bibfnamefont {D.}~\bibnamefont {Zaritsky}},\ }\bibfield  {title} {\bibinfo
			{title} {A direct empirical proof of the existence of dark matter},\ }\href
		{https://doi.org/10.1086/508162} {\bibfield  {journal} {\bibinfo  {journal}
				{Astrophys. J.}\ }\textbf {\bibinfo {volume} {648}},\ \bibinfo {pages} {L109}
			(\bibinfo {year} {2006})}\BibitemShut {NoStop}%
		\bibitem [{\citenamefont {Massey}\ \emph {et~al.}(2010)\citenamefont {Massey},
			\citenamefont {Kitching},\ and\ \citenamefont {Richard}}]{GravLensing}%
		\BibitemOpen
		\bibfield  {author} {\bibinfo {author} {\bibfnamefont {R.}~\bibnamefont
				{Massey}}, \bibinfo {author} {\bibfnamefont {T.}~\bibnamefont {Kitching}},\
			and\ \bibinfo {author} {\bibfnamefont {J.}~\bibnamefont {Richard}},\
		}\bibfield  {title} {\bibinfo {title} {The dark matter of gravitational
				lensing},\ }\href {https://doi.org/10.1088/0034-4885/73/8/086901} {\bibfield
			{journal} {\bibinfo  {journal} {Rep. Prog. Phys.}\ }\textbf {\bibinfo
				{volume} {73}},\ \bibinfo {pages} {086901} (\bibinfo {year}
			{2010})}\BibitemShut {NoStop}%
		\bibitem [{\citenamefont {Aghanim}\ \emph {et~al.}(2020)\citenamefont
			{Aghanim}, \citenamefont {Akrami}, \citenamefont {Arroja}, \citenamefont
			{Ashdown}, \citenamefont {Aumont}, \citenamefont {Baccigalupi}, \citenamefont
			{Ballardini}, \citenamefont {Banday}, \citenamefont {Barreiro},\ and\
			\citenamefont {et~al.}}]{CMB}%
		\BibitemOpen
		\bibfield  {author} {\bibinfo {author} {\bibfnamefont {N.}~\bibnamefont
				{Aghanim}}, \bibinfo {author} {\bibfnamefont {Y.}~\bibnamefont {Akrami}},
			\bibinfo {author} {\bibfnamefont {F.}~\bibnamefont {Arroja}}, \bibinfo
			{author} {\bibfnamefont {M.}~\bibnamefont {Ashdown}}, \bibinfo {author}
			{\bibfnamefont {J.}~\bibnamefont {Aumont}}, \bibinfo {author} {\bibfnamefont
				{C.}~\bibnamefont {Baccigalupi}}, \bibinfo {author} {\bibfnamefont
				{M.}~\bibnamefont {Ballardini}}, \bibinfo {author} {\bibfnamefont {A.~J.}\
				\bibnamefont {Banday}}, \bibinfo {author} {\bibfnamefont {R.~B.}\
				\bibnamefont {Barreiro}},\ and\ \bibinfo {author} {\bibnamefont {et~al.}},\
		}\bibfield  {title} {\bibinfo {title} {Planck2018 results},\ }\href
		{https://doi.org/10.1051/0004-6361/201833880} {\bibfield  {journal} {\bibinfo
				{journal} {Astron. Astrophys.}\ }\textbf {\bibinfo {volume} {641}},\
			\bibinfo {pages} {A1} (\bibinfo {year} {2020})}\BibitemShut {NoStop}%
		\bibitem [{\citenamefont {Zyla}\ \emph {et~al.}(2020)\citenamefont {Zyla} \emph
			{et~al.}}]{PDG}%
		\BibitemOpen
		\bibfield  {author} {\bibinfo {author} {\bibfnamefont {P.}~\bibnamefont
				{Zyla}} \emph {et~al.} (\bibinfo {collaboration} {Particle Data Group}),\
		}\bibfield  {title} {\bibinfo {title} {{Review of Particle Physics}},\ }\href
		{https://doi.org/10.1093/ptep/ptaa104} {\bibfield  {journal} {\bibinfo
				{journal} {Prog. Theor. Exp. Phys.}\ }\textbf {\bibinfo {volume} {2020}},\
			\bibinfo {pages} {083C01} (\bibinfo {year} {2020})}\BibitemShut {NoStop}%
		\bibitem [{\citenamefont {Choi}\ \emph {et~al.}(2021)\citenamefont {Choi},
			\citenamefont {Im},\ and\ \citenamefont {Shin}}]{ALPs}%
		\BibitemOpen
		\bibfield  {author} {\bibinfo {author} {\bibfnamefont {K.}~\bibnamefont
				{Choi}}, \bibinfo {author} {\bibfnamefont {S.~H.}\ \bibnamefont {Im}},\ and\
			\bibinfo {author} {\bibfnamefont {C.~S.}\ \bibnamefont {Shin}},\ }\bibfield
		{title} {\bibinfo {title} {Recent progress in the physics of axions and
				axion-like particles},\ }\href
		{https://doi.org/10.1146/annurev-nucl-120720-031147} {\bibfield  {journal}
			{\bibinfo  {journal} {Annu. Rev. Nucl. Part. Sci.}\ }\textbf {\bibinfo
				{volume} {71}},\ \bibinfo {pages} {225} (\bibinfo {year} {2021})}\BibitemShut
		{NoStop}%
		\bibitem [{\citenamefont {Chupp}\ \emph {et~al.}(2019)\citenamefont {Chupp},
			\citenamefont {Fierlinger}, \citenamefont {Ramsey-Musolf},\ and\
			\citenamefont {Singh}}]{EDM}%
		\BibitemOpen
		\bibfield  {author} {\bibinfo {author} {\bibfnamefont {T.~E.}\ \bibnamefont
				{Chupp}}, \bibinfo {author} {\bibfnamefont {P.}~\bibnamefont {Fierlinger}},
			\bibinfo {author} {\bibfnamefont {M.~J.}\ \bibnamefont {Ramsey-Musolf}},\
			and\ \bibinfo {author} {\bibfnamefont {J.~T.}\ \bibnamefont {Singh}},\
		}\bibfield  {title} {\bibinfo {title} {Electric dipole moments of atoms,
				molecules, nuclei, and particles},\ }\href
		{https://doi.org/10.1103/RevModPhys.91.015001} {\bibfield  {journal}
			{\bibinfo  {journal} {Rev. Mod. Phys.}\ }\textbf {\bibinfo {volume} {91}},\
			\bibinfo {pages} {015001} (\bibinfo {year} {2019})}\BibitemShut {NoStop}%
		\bibitem [{\citenamefont {Safronova}\ \emph {et~al.}(2018)\citenamefont
			{Safronova}, \citenamefont {Budker}, \citenamefont {DeMille}, \citenamefont
			{Kimball}, \citenamefont {Derevianko},\ and\ \citenamefont
			{Clark}}]{BSMwAMO}%
		\BibitemOpen
		\bibfield  {author} {\bibinfo {author} {\bibfnamefont {M.~S.}\ \bibnamefont
				{Safronova}}, \bibinfo {author} {\bibfnamefont {D.}~\bibnamefont {Budker}},
			\bibinfo {author} {\bibfnamefont {D.}~\bibnamefont {DeMille}}, \bibinfo
			{author} {\bibfnamefont {D.~F.~J.}\ \bibnamefont {Kimball}}, \bibinfo
			{author} {\bibfnamefont {A.}~\bibnamefont {Derevianko}},\ and\ \bibinfo
			{author} {\bibfnamefont {C.~W.}\ \bibnamefont {Clark}},\ }\bibfield  {title}
		{\bibinfo {title} {Search for new physics with atoms and molecules},\ }\href
		{https://doi.org/10.1103/RevModPhys.90.025008} {\bibfield  {journal}
			{\bibinfo  {journal} {Rev. Mod. Phys.}\ }\textbf {\bibinfo {volume} {90}},\
			\bibinfo {pages} {025008} (\bibinfo {year} {2018})}\BibitemShut {NoStop}%
		\bibitem [{\citenamefont {Delaunay}\ \emph {et~al.}(2017)\citenamefont
			{Delaunay}, \citenamefont {Ozeri}, \citenamefont {Perez},\ and\ \citenamefont
			{Soreq}}]{Delaunay2017}%
		\BibitemOpen
		\bibfield  {author} {\bibinfo {author} {\bibfnamefont {C.}~\bibnamefont
				{Delaunay}}, \bibinfo {author} {\bibfnamefont {R.}~\bibnamefont {Ozeri}},
			\bibinfo {author} {\bibfnamefont {G.}~\bibnamefont {Perez}},\ and\ \bibinfo
			{author} {\bibfnamefont {Y.}~\bibnamefont {Soreq}},\ }\bibfield  {title}
		{\bibinfo {title} {Probing atomic {Higgs}-like forces at the precision
				frontier},\ }\href {https://doi.org/10.1103/PhysRevD.96.093001} {\bibfield
			{journal} {\bibinfo  {journal} {Phys. Rev. D}\ }\textbf {\bibinfo {volume}
				{96}},\ \bibinfo {pages} {093001} (\bibinfo {year} {2017})}\BibitemShut
		{NoStop}%
		\bibitem [{\citenamefont {Berengut}\ \emph {et~al.}(2018)\citenamefont
			{Berengut}, \citenamefont {Budker}, \citenamefont {Delaunay}, \citenamefont
			{Flambaum}, \citenamefont {Frugiuele}, \citenamefont {Fuchs}, \citenamefont
			{Grojean}, \citenamefont {Harnik}, \citenamefont {Ozeri}, \citenamefont
			{Perez},\ and\ \citenamefont {Soreq}}]{Berengut2018}%
		\BibitemOpen
		\bibfield  {author} {\bibinfo {author} {\bibfnamefont {J.~C.}\ \bibnamefont
				{Berengut}}, \bibinfo {author} {\bibfnamefont {D.}~\bibnamefont {Budker}},
			\bibinfo {author} {\bibfnamefont {C.}~\bibnamefont {Delaunay}}, \bibinfo
			{author} {\bibfnamefont {V.~V.}\ \bibnamefont {Flambaum}}, \bibinfo {author}
			{\bibfnamefont {C.}~\bibnamefont {Frugiuele}}, \bibinfo {author}
			{\bibfnamefont {E.}~\bibnamefont {Fuchs}}, \bibinfo {author} {\bibfnamefont
				{C.}~\bibnamefont {Grojean}}, \bibinfo {author} {\bibfnamefont
				{R.}~\bibnamefont {Harnik}}, \bibinfo {author} {\bibfnamefont
				{R.}~\bibnamefont {Ozeri}}, \bibinfo {author} {\bibfnamefont
				{G.}~\bibnamefont {Perez}},\ and\ \bibinfo {author} {\bibfnamefont
				{Y.}~\bibnamefont {Soreq}},\ }\bibfield  {title} {\bibinfo {title} {Probing
				new long-range interactions by isotope shift spectroscopy},\ }\href
		{https://doi.org/10.1103/PhysRevLett.120.091801} {\bibfield  {journal}
			{\bibinfo  {journal} {Phys. Rev. Lett.}\ }\textbf {\bibinfo {volume} {120}},\
			\bibinfo {pages} {091801} (\bibinfo {year} {2018})}\BibitemShut {NoStop}%
		\bibitem [{\citenamefont {King}(1984)}]{King1984}%
		\BibitemOpen
		\bibfield  {author} {\bibinfo {author} {\bibfnamefont {W.~H.}\ \bibnamefont
				{King}},\ }\href@noop {} {\emph {\bibinfo {title} {Isotope Shifts in Atomic
					Spectra}}}\ (\bibinfo  {publisher} {Plenum Press},\ \bibinfo {year}
		{1984})\BibitemShut {NoStop}%
		\bibitem [{\citenamefont {Flambaum}\ \emph {et~al.}(2018)\citenamefont
			{Flambaum}, \citenamefont {Geddes},\ and\ \citenamefont
			{Viatkina}}]{Flambaum2018}%
		\BibitemOpen
		\bibfield  {author} {\bibinfo {author} {\bibfnamefont {V.~V.}\ \bibnamefont
				{Flambaum}}, \bibinfo {author} {\bibfnamefont {A.~J.}\ \bibnamefont
				{Geddes}},\ and\ \bibinfo {author} {\bibfnamefont {A.~V.}\ \bibnamefont
				{Viatkina}},\ }\bibfield  {title} {\bibinfo {title} {Isotope shift,
				nonlinearity of {King} plots, and the search for new particles},\ }\href
		{https://doi.org/10.1103/PhysRevA.97.032510} {\bibfield  {journal} {\bibinfo
				{journal} {Phys. Rev. A}\ }\textbf {\bibinfo {volume} {97}},\ \bibinfo
			{pages} {032510} (\bibinfo {year} {2018})}\BibitemShut {NoStop}%
		\bibitem [{\citenamefont {Allehabi}\ \emph {et~al.}(2021)\citenamefont
			{Allehabi}, \citenamefont {Dzuba}, \citenamefont {Flambaum},\ and\
			\citenamefont {Afanasjev}}]{Allehabi2021}%
		\BibitemOpen
		\bibfield  {author} {\bibinfo {author} {\bibfnamefont {S.~O.}\ \bibnamefont
				{Allehabi}}, \bibinfo {author} {\bibfnamefont {V.~A.}\ \bibnamefont {Dzuba}},
			\bibinfo {author} {\bibfnamefont {V.~V.}\ \bibnamefont {Flambaum}},\ and\
			\bibinfo {author} {\bibfnamefont {A.~V.}\ \bibnamefont {Afanasjev}},\
		}\bibfield  {title} {\bibinfo {title} {Nuclear deformation as a source of the
				nonlinearity of the {King} plot in the {Yb$^{+}$} ion},\ }\href
		{https://doi.org/10.1103/PhysRevA.103.L030801} {\bibfield  {journal}
			{\bibinfo  {journal} {Phys. Rev. A}\ }\textbf {\bibinfo {volume} {103}},\
			\bibinfo {pages} {L030801} (\bibinfo {year} {2021})}\BibitemShut {NoStop}%
		\bibitem [{\citenamefont {Mikami}\ \emph {et~al.}(2017)\citenamefont {Mikami},
			\citenamefont {Tanaka},\ and\ \citenamefont {Yamamoto}}]{Mikami2017}%
		\BibitemOpen
		\bibfield  {author} {\bibinfo {author} {\bibfnamefont {K.}~\bibnamefont
				{Mikami}}, \bibinfo {author} {\bibfnamefont {M.}~\bibnamefont {Tanaka}},\
			and\ \bibinfo {author} {\bibfnamefont {Y.}~\bibnamefont {Yamamoto}},\
		}\bibfield  {title} {\bibinfo {title} {Probing new intra-atomic force with
				isotope shifts},\ }\href {https://doi.org/10.1140/epjc/s10052-017-5467-4}
		{\bibfield  {journal} {\bibinfo  {journal} {Eur. Phys. J. C}\ }\textbf
			{\bibinfo {volume} {77}},\ \bibinfo {pages} {896} (\bibinfo {year}
			{2017})}\BibitemShut {NoStop}%
		\bibitem [{\citenamefont {Tanaka}\ and\ \citenamefont
			{Yamamoto}(2020)}]{Tanaka2019}%
		\BibitemOpen
		\bibfield  {author} {\bibinfo {author} {\bibfnamefont {M.}~\bibnamefont
				{Tanaka}}\ and\ \bibinfo {author} {\bibfnamefont {Y.}~\bibnamefont
				{Yamamoto}},\ }\bibfield  {title} {\bibinfo {title} {Relativistic effects in
				the search for new intra-atomic force with isotope shifts},\ }\href
		{https://doi.org/10.1093/ptep/ptaa121} {\bibfield  {journal} {\bibinfo
				{journal} {Prog. Theor. Exp. Phys.}\ }\textbf {\bibinfo {volume} {2020}},\
			\bibinfo {pages} {103B02} (\bibinfo {year} {2020})}\BibitemShut {NoStop}%
		\bibitem [{\citenamefont {Reinhard}\ \emph {et~al.}(2020)\citenamefont
			{Reinhard}, \citenamefont {Nazarewicz},\ and\ \citenamefont
			{Garcia~Ruiz}}]{Reinhard2020}%
		\BibitemOpen
		\bibfield  {author} {\bibinfo {author} {\bibfnamefont {P.-G.}\ \bibnamefont
				{Reinhard}}, \bibinfo {author} {\bibfnamefont {W.}~\bibnamefont
				{Nazarewicz}},\ and\ \bibinfo {author} {\bibfnamefont {R.~F.}\ \bibnamefont
				{Garcia~Ruiz}},\ }\bibfield  {title} {\bibinfo {title} {Beyond the charge
				radius: The information content of the fourth radial moment},\ }\href
		{https://doi.org/10.1103/PhysRevC.101.021301} {\bibfield  {journal} {\bibinfo
				{journal} {Phys. Rev. C}\ }\textbf {\bibinfo {volume} {101}},\ \bibinfo
			{pages} {021301(R)} (\bibinfo {year} {2020})}\BibitemShut {NoStop}%
		\bibitem [{\citenamefont {Allehabi}\ \emph {et~al.}(2020)\citenamefont
			{Allehabi}, \citenamefont {Dzuba}, \citenamefont {Flambaum}, \citenamefont
			{Afanasjev},\ and\ \citenamefont {Agbemava}}]{Allehabi2020}%
		\BibitemOpen
		\bibfield  {author} {\bibinfo {author} {\bibfnamefont {S.~O.}\ \bibnamefont
				{Allehabi}}, \bibinfo {author} {\bibfnamefont {V.~A.}\ \bibnamefont {Dzuba}},
			\bibinfo {author} {\bibfnamefont {V.~V.}\ \bibnamefont {Flambaum}}, \bibinfo
			{author} {\bibfnamefont {A.~V.}\ \bibnamefont {Afanasjev}},\ and\ \bibinfo
			{author} {\bibfnamefont {S.~E.}\ \bibnamefont {Agbemava}},\ }\bibfield
		{title} {\bibinfo {title} {Using isotope shift for testing nuclear theory:
				The case of nobelium isotopes},\ }\href
		{https://doi.org/10.1103/PhysRevC.102.024326} {\bibfield  {journal} {\bibinfo
				{journal} {Phys. Rev. C}\ }\textbf {\bibinfo {volume} {102}},\ \bibinfo
			{pages} {024326} (\bibinfo {year} {2020})}\BibitemShut {NoStop}%
		\bibitem [{\citenamefont {M\"uller}\ \emph {et~al.}(2021)\citenamefont
			{M\"uller}, \citenamefont {Yerokhin}, \citenamefont {Artemyev},\ and\
			\citenamefont {Surzhykov}}]{Muller2021}%
		\BibitemOpen
		\bibfield  {author} {\bibinfo {author} {\bibfnamefont {R.~A.}\ \bibnamefont
				{M\"uller}}, \bibinfo {author} {\bibfnamefont {V.~A.}\ \bibnamefont
				{Yerokhin}}, \bibinfo {author} {\bibfnamefont {A.~N.}\ \bibnamefont
				{Artemyev}},\ and\ \bibinfo {author} {\bibfnamefont {A.}~\bibnamefont
				{Surzhykov}},\ }\bibfield  {title} {\bibinfo {title} {Nonlinearities of
				{King}'s plot and their dependence on nuclear radii},\ }\href
		{https://doi.org/10.1103/PhysRevA.104.L020802} {\bibfield  {journal}
			{\bibinfo  {journal} {Phys. Rev. A}\ }\textbf {\bibinfo {volume} {104}},\
			\bibinfo {pages} {L020802} (\bibinfo {year} {2021})}\BibitemShut {NoStop}%
		\bibitem [{\citenamefont {Counts}\ \emph {et~al.}(2020)\citenamefont {Counts},
			\citenamefont {Hur}, \citenamefont {Aude~Craik}, \citenamefont {Jeon},
			\citenamefont {Leung}, \citenamefont {Berengut}, \citenamefont {Geddes},
			\citenamefont {Kawasaki}, \citenamefont {Jhe},\ and\ \citenamefont
			{Vuleti\ifmmode~\acute{c}\else \'{c}\fi{}}}]{counts2020}%
		\BibitemOpen
		\bibfield  {author} {\bibinfo {author} {\bibfnamefont {I.}~\bibnamefont
				{Counts}}, \bibinfo {author} {\bibfnamefont {J.}~\bibnamefont {Hur}},
			\bibinfo {author} {\bibfnamefont {D.~P.~L.}\ \bibnamefont {Aude~Craik}},
			\bibinfo {author} {\bibfnamefont {H.}~\bibnamefont {Jeon}}, \bibinfo {author}
			{\bibfnamefont {C.}~\bibnamefont {Leung}}, \bibinfo {author} {\bibfnamefont
				{J.~C.}\ \bibnamefont {Berengut}}, \bibinfo {author} {\bibfnamefont
				{A.}~\bibnamefont {Geddes}}, \bibinfo {author} {\bibfnamefont
				{A.}~\bibnamefont {Kawasaki}}, \bibinfo {author} {\bibfnamefont
				{W.}~\bibnamefont {Jhe}},\ and\ \bibinfo {author} {\bibfnamefont
				{V.}~\bibnamefont {Vuleti\ifmmode~\acute{c}\else \'{c}\fi{}}},\ }\bibfield
		{title} {\bibinfo {title} {Evidence for nonlinear isotope shift in {Yb$^+$}
				search for new boson},\ }\href
		{https://doi.org/10.1103/PhysRevLett.125.123002} {\bibfield  {journal}
			{\bibinfo  {journal} {Phys. Rev. Lett.}\ }\textbf {\bibinfo {volume} {125}},\
			\bibinfo {pages} {123002} (\bibinfo {year} {2020})}\BibitemShut {NoStop}%
		\bibitem [{\citenamefont {Solaro}\ \emph {et~al.}(2020)\citenamefont {Solaro},
			\citenamefont {Meyer}, \citenamefont {Fisher}, \citenamefont {Berengut},
			\citenamefont {Fuchs},\ and\ \citenamefont {Drewsen}}]{Solaro2020}%
		\BibitemOpen
		\bibfield  {author} {\bibinfo {author} {\bibfnamefont {C.}~\bibnamefont
				{Solaro}}, \bibinfo {author} {\bibfnamefont {S.}~\bibnamefont {Meyer}},
			\bibinfo {author} {\bibfnamefont {K.}~\bibnamefont {Fisher}}, \bibinfo
			{author} {\bibfnamefont {J.~C.}\ \bibnamefont {Berengut}}, \bibinfo {author}
			{\bibfnamefont {E.}~\bibnamefont {Fuchs}},\ and\ \bibinfo {author}
			{\bibfnamefont {M.}~\bibnamefont {Drewsen}},\ }\bibfield  {title} {\bibinfo
			{title} {Improved isotope-shift-based bounds on bosons beyond the standard
				model through measurements of the {${}^{2}\mathrm{D}_{3/2}\ensuremath{-}
					{}^{2}\mathrm{D}_{5/2}$} interval in {$\mathrm{Ca}^{+}$}},\ }\href
		{https://doi.org/10.1103/PhysRevLett.125.123003} {\bibfield  {journal}
			{\bibinfo  {journal} {Phys. Rev. Lett.}\ }\textbf {\bibinfo {volume} {125}},\
			\bibinfo {pages} {123003} (\bibinfo {year} {2020})}\BibitemShut {NoStop}%
		\bibitem [{\citenamefont {Ono}\ \emph {et~al.}(2021)\citenamefont {Ono},
			\citenamefont {Saito}, \citenamefont {Ishiyama}, \citenamefont {Higomoto},
			\citenamefont {Takano}, \citenamefont {Takasu}, \citenamefont {Yamamoto},
			\citenamefont {Tanaka},\ and\ \citenamefont {Takahashi}}]{ono2021}%
		\BibitemOpen
		\bibfield  {author} {\bibinfo {author} {\bibfnamefont {K.}~\bibnamefont
				{Ono}}, \bibinfo {author} {\bibfnamefont {Y.}~\bibnamefont {Saito}}, \bibinfo
			{author} {\bibfnamefont {T.}~\bibnamefont {Ishiyama}}, \bibinfo {author}
			{\bibfnamefont {T.}~\bibnamefont {Higomoto}}, \bibinfo {author}
			{\bibfnamefont {T.}~\bibnamefont {Takano}}, \bibinfo {author} {\bibfnamefont
				{Y.}~\bibnamefont {Takasu}}, \bibinfo {author} {\bibfnamefont
				{Y.}~\bibnamefont {Yamamoto}}, \bibinfo {author} {\bibfnamefont
				{M.}~\bibnamefont {Tanaka}},\ and\ \bibinfo {author} {\bibfnamefont
				{Y.}~\bibnamefont {Takahashi}},\ }\href@noop {} {\bibinfo {title}
			{Observation of non-linearity of generalized {King} plot in the search for
				new boson}} (\bibinfo {year} {2021}),\ \Eprint
		{https://arxiv.org/abs/2110.13544} {arXiv:2110.13544 [physics.atom-ph]}
		\BibitemShut {NoStop}%
		\bibitem [{\citenamefont {Figueroa}\ \emph {et~al.}(2022)\citenamefont
			{Figueroa}, \citenamefont {Berengut}, \citenamefont {Dzuba}, \citenamefont
			{Flambaum}, \citenamefont {Budker},\ and\ \citenamefont
			{Antypas}}]{Figueroa2021}%
		\BibitemOpen
		\bibfield  {author} {\bibinfo {author} {\bibfnamefont {N.~L.}\ \bibnamefont
				{Figueroa}}, \bibinfo {author} {\bibfnamefont {J.~C.}\ \bibnamefont
				{Berengut}}, \bibinfo {author} {\bibfnamefont {V.~A.}\ \bibnamefont {Dzuba}},
			\bibinfo {author} {\bibfnamefont {V.~V.}\ \bibnamefont {Flambaum}}, \bibinfo
			{author} {\bibfnamefont {D.}~\bibnamefont {Budker}},\ and\ \bibinfo {author}
			{\bibfnamefont {D.}~\bibnamefont {Antypas}},\ }\bibfield  {title} {\bibinfo
			{title} {Precision determination of isotope shifts in ytterbium and
				implications for new physics},\ }\href
		{https://doi.org/10.1103/PhysRevLett.128.073001} {\bibfield  {journal}
			{\bibinfo  {journal} {Phys. Rev. Lett.}\ }\textbf {\bibinfo {volume} {128}},\
			\bibinfo {pages} {073001} (\bibinfo {year} {2022})}\BibitemShut {NoStop}%
		\bibitem [{\citenamefont {Reinhard}\ and\ \citenamefont
			{Nazarewicz}(2021)}]{Reinhard2021}%
		\BibitemOpen
		\bibfield  {author} {\bibinfo {author} {\bibfnamefont {P.-G.}\ \bibnamefont
				{Reinhard}}\ and\ \bibinfo {author} {\bibfnamefont {W.}~\bibnamefont
				{Nazarewicz}},\ }\bibfield  {title} {\bibinfo {title} {Nuclear charge
				densities in spherical and deformed nuclei: Toward precise calculations of
				charge radii},\ }\href {https://doi.org/10.1103/PhysRevC.103.054310}
		{\bibfield  {journal} {\bibinfo  {journal} {Phys. Rev. C}\ }\textbf {\bibinfo
				{volume} {103}},\ \bibinfo {pages} {054310} (\bibinfo {year}
			{2021})}\BibitemShut {NoStop}%
		\bibitem [{\citenamefont {Nesterenko}\ \emph {et~al.}(2020)\citenamefont
			{Nesterenko}, \citenamefont {{de Groote}}, \citenamefont {Eronen},
			\citenamefont {Ge}, \citenamefont {Hukkanen}, \citenamefont {Jokinen},\ and\
			\citenamefont {Kankainen}}]{Nesterenko2020}%
		\BibitemOpen
		\bibfield  {author} {\bibinfo {author} {\bibfnamefont {D.}~\bibnamefont
				{Nesterenko}}, \bibinfo {author} {\bibfnamefont {R.}~\bibnamefont {{de
						Groote}}}, \bibinfo {author} {\bibfnamefont {T.}~\bibnamefont {Eronen}},
			\bibinfo {author} {\bibfnamefont {Z.}~\bibnamefont {Ge}}, \bibinfo {author}
			{\bibfnamefont {M.}~\bibnamefont {Hukkanen}}, \bibinfo {author}
			{\bibfnamefont {A.}~\bibnamefont {Jokinen}},\ and\ \bibinfo {author}
			{\bibfnamefont {A.}~\bibnamefont {Kankainen}},\ }\bibfield  {title} {\bibinfo
			{title} {High-precision mass measurement of {${}^{168}$Yb} for verification
				of nonlinear isotope shift},\ }\href
		{https://doi.org/10.1016/j.ijms.2020.116435} {\bibfield  {journal} {\bibinfo
				{journal} {Int. J. Mass Spectrom.}\ }\textbf {\bibinfo {volume} {458}},\
			\bibinfo {pages} {116435} (\bibinfo {year} {2020})}\BibitemShut {NoStop}%
		\bibitem [{\citenamefont {Huang}\ \emph {et~al.}(2017)\citenamefont {Huang},
			\citenamefont {Audi}, \citenamefont {Wang}, \citenamefont {Kondev},
			\citenamefont {Naimi},\ and\ \citenamefont {Xu}}]{AME2016_1}%
		\BibitemOpen
		\bibfield  {author} {\bibinfo {author} {\bibfnamefont {W.}~\bibnamefont
				{Huang}}, \bibinfo {author} {\bibfnamefont {G.}~\bibnamefont {Audi}},
			\bibinfo {author} {\bibfnamefont {M.}~\bibnamefont {Wang}}, \bibinfo {author}
			{\bibfnamefont {F.~G.}\ \bibnamefont {Kondev}}, \bibinfo {author}
			{\bibfnamefont {S.}~\bibnamefont {Naimi}},\ and\ \bibinfo {author}
			{\bibfnamefont {X.}~\bibnamefont {Xu}},\ }\bibfield  {title} {\bibinfo
			{title} {The {AME2016} atomic mass evaluation ({I}). evaluation of input
				data; and adjustment procedures},\ }\href
		{https://doi.org/10.1088/1674-1137/41/3/030002} {\bibfield  {journal}
			{\bibinfo  {journal} {Chin. Phys. C}\ }\textbf {\bibinfo {volume} {41}},\
			\bibinfo {pages} {030002} (\bibinfo {year} {2017})}\BibitemShut {NoStop}%
		\bibitem [{\citenamefont {Wang}\ \emph {et~al.}(2017)\citenamefont {Wang},
			\citenamefont {Audi}, \citenamefont {Kondev}, \citenamefont {Huang},
			\citenamefont {Naimi},\ and\ \citenamefont {Xu}}]{AME2016_2}%
		\BibitemOpen
		\bibfield  {author} {\bibinfo {author} {\bibfnamefont {M.}~\bibnamefont
				{Wang}}, \bibinfo {author} {\bibfnamefont {G.}~\bibnamefont {Audi}}, \bibinfo
			{author} {\bibfnamefont {F.~G.}\ \bibnamefont {Kondev}}, \bibinfo {author}
			{\bibfnamefont {W.}~\bibnamefont {Huang}}, \bibinfo {author} {\bibfnamefont
				{S.}~\bibnamefont {Naimi}},\ and\ \bibinfo {author} {\bibfnamefont
				{X.}~\bibnamefont {Xu}},\ }\bibfield  {title} {\bibinfo {title} {The
				{AME2016} atomic mass evaluation ({II}). tables, graphs and references},\
		}\href {https://doi.org/10.1088/1674-1137/41/3/030003} {\bibfield  {journal}
			{\bibinfo  {journal} {Chin. Phys. C}\ }\textbf {\bibinfo {volume} {41}},\
			\bibinfo {pages} {030003} (\bibinfo {year} {2017})}\BibitemShut {NoStop}%
		\bibitem [{\citenamefont {Rana}\ \emph {et~al.}(2012)\citenamefont {Rana},
			\citenamefont {H\"ocker},\ and\ \citenamefont {Myers}}]{Rana2012}%
		\BibitemOpen
		\bibfield  {author} {\bibinfo {author} {\bibfnamefont {R.}~\bibnamefont
				{Rana}}, \bibinfo {author} {\bibfnamefont {M.}~\bibnamefont {H\"ocker}},\
			and\ \bibinfo {author} {\bibfnamefont {E.~G.}\ \bibnamefont {Myers}},\
		}\bibfield  {title} {\bibinfo {title} {Atomic masses of strontium and
				ytterbium},\ }\href {https://doi.org/10.1103/PhysRevA.86.050502} {\bibfield
			{journal} {\bibinfo  {journal} {Phys. Rev. A}\ }\textbf {\bibinfo {volume}
				{86}},\ \bibinfo {pages} {050502(R)} (\bibinfo {year} {2012})}\BibitemShut
		{NoStop}%
		\bibitem [{SM()}]{SM}%
		\BibitemOpen
		\href@noop {} {}\bibinfo {note} {See Supplemental Material
			for the details on experimental protocols, data analysis, estimation of
			systematic effects, and the calculation of electronic factors and nuclear
			parameters.}\BibitemShut
		{Stop}%
		\bibitem [{\citenamefont {F\"urst}\ \emph {et~al.}(2020)\citenamefont
			{F\"urst}, \citenamefont {Yeh}, \citenamefont {Kalincev}, \citenamefont
			{Kulosa}, \citenamefont {Dreissen}, \citenamefont {Lange}, \citenamefont
			{Benkler}, \citenamefont {Huntemann}, \citenamefont {Peik},\ and\
			\citenamefont {Mehlst\"aubler}}]{Furst2020}%
		\BibitemOpen
		\bibfield  {author} {\bibinfo {author} {\bibfnamefont {H.~A.}\ \bibnamefont
				{F\"urst}}, \bibinfo {author} {\bibfnamefont {C.-H.}\ \bibnamefont {Yeh}},
			\bibinfo {author} {\bibfnamefont {D.}~\bibnamefont {Kalincev}}, \bibinfo
			{author} {\bibfnamefont {A.~P.}\ \bibnamefont {Kulosa}}, \bibinfo {author}
			{\bibfnamefont {L.~S.}\ \bibnamefont {Dreissen}}, \bibinfo {author}
			{\bibfnamefont {R.}~\bibnamefont {Lange}}, \bibinfo {author} {\bibfnamefont
				{E.}~\bibnamefont {Benkler}}, \bibinfo {author} {\bibfnamefont
				{N.}~\bibnamefont {Huntemann}}, \bibinfo {author} {\bibfnamefont
				{E.}~\bibnamefont {Peik}},\ and\ \bibinfo {author} {\bibfnamefont {T.~E.}\
				\bibnamefont {Mehlst\"aubler}},\ }\bibfield  {title} {\bibinfo {title}
			{Coherent excitation of the highly forbidden electric octupole transition in
				{$^{172}{\mathrm{Yb}}^{+}$}},\ }\href
		{https://doi.org/10.1103/PhysRevLett.125.163001} {\bibfield  {journal}
			{\bibinfo  {journal} {Phys. Rev. Lett.}\ }\textbf {\bibinfo {volume} {125}},\
			\bibinfo {pages} {163001} (\bibinfo {year} {2020})}\BibitemShut {NoStop}%
		\bibitem [{Meh()}]{MehlstaublerPrivComm}%
		\BibitemOpen
		\href@noop {} {}\bibinfo {note} {T. E. Mehlst\"aubler (private communication)
			for the correction to the measured transition frequency reported in
			Ref.~\cite{Furst2020}.}\BibitemShut {Stop}%
		\bibitem [{\citenamefont {Counts}(2020)}]{CountsThesis}%
		\BibitemOpen
		\bibfield  {author} {\bibinfo {author} {\bibfnamefont {I.}~\bibnamefont
				{Counts}},\ }\emph {\bibinfo {title} {Surface Friction and Spectroscopic
				Probes of New Physics with Trapped Ions}},\ \href@noop {} {Ph.D. thesis},\
		\bibinfo  {school} {Massachusetts Institute of Technology} (\bibinfo {year}
		{2020})\BibitemShut {NoStop}%
		\bibitem [{\citenamefont {Reinhard}\ \emph {et~al.}(2021)\citenamefont
			{Reinhard}, \citenamefont {Schuetrumpf},\ and\ \citenamefont
			{Maruhn}}]{Reinhard2021c}%
		\BibitemOpen
		\bibfield  {author} {\bibinfo {author} {\bibfnamefont {P.-G.}\ \bibnamefont
				{Reinhard}}, \bibinfo {author} {\bibfnamefont {B.}~\bibnamefont
				{Schuetrumpf}},\ and\ \bibinfo {author} {\bibfnamefont {J.}~\bibnamefont
				{Maruhn}},\ }\bibfield  {title} {\bibinfo {title} {The axial {Hartree–Fock}
				+ {BCS} code {SkyAx}},\ }\href {https://doi.org/10.1016/j.cpc.2020.107603}
		{\bibfield  {journal} {\bibinfo  {journal} {Comput. Phys. Commun.}\ }\textbf
			{\bibinfo {volume} {258}},\ \bibinfo {pages} {107603} (\bibinfo {year}
			{2021})}\BibitemShut {NoStop}%
		\bibitem [{\citenamefont {Kl\"upfel}\ \emph {et~al.}(2009)\citenamefont
			{Kl\"upfel}, \citenamefont {Reinhard}, \citenamefont {B\"urvenich},\ and\
			\citenamefont {Maruhn}}]{Klupfel2009}%
		\BibitemOpen
		\bibfield  {author} {\bibinfo {author} {\bibfnamefont {P.}~\bibnamefont
				{Kl\"upfel}}, \bibinfo {author} {\bibfnamefont {P.-G.}\ \bibnamefont
				{Reinhard}}, \bibinfo {author} {\bibfnamefont {T.~J.}\ \bibnamefont
				{B\"urvenich}},\ and\ \bibinfo {author} {\bibfnamefont {J.~A.}\ \bibnamefont
				{Maruhn}},\ }\bibfield  {title} {\bibinfo {title} {Variations on a theme by
				{Skyrme}: A systematic study of adjustments of model parameters},\ }\href
		{https://doi.org/10.1103/PhysRevC.79.034310} {\bibfield  {journal} {\bibinfo
				{journal} {Phys. Rev. C}\ }\textbf {\bibinfo {volume} {79}},\ \bibinfo
			{pages} {034310} (\bibinfo {year} {2009})}\BibitemShut {NoStop}%
		\bibitem [{\citenamefont {Brown}\ \emph {et~al.}(1984)\citenamefont {Brown},
			\citenamefont {Bronk},\ and\ \citenamefont {Hodgson}}]{Brown1984}%
		\BibitemOpen
		\bibfield  {author} {\bibinfo {author} {\bibfnamefont {B.~A.}\ \bibnamefont
				{Brown}}, \bibinfo {author} {\bibfnamefont {C.~R.}\ \bibnamefont {Bronk}},\
			and\ \bibinfo {author} {\bibfnamefont {P.~E.}\ \bibnamefont {Hodgson}},\
		}\bibfield  {title} {\bibinfo {title} {Systematics of nuclear {RMS} charge
				radii},\ }\href {https://doi.org/10.1088/0305-4616/10/12/008} {\bibfield
			{journal} {\bibinfo  {journal} {J. Phys. G}\ }\textbf {\bibinfo {volume}
				{10}},\ \bibinfo {pages} {1683} (\bibinfo {year} {1984})}\BibitemShut
		{NoStop}%
		\bibitem [{\citenamefont {Otten}(1989)}]{Otten1989}%
		\BibitemOpen
		\bibfield  {author} {\bibinfo {author} {\bibfnamefont {E.~W.}\ \bibnamefont
				{Otten}},\ }\bibinfo {title} {Nuclear radii and moments of unstable
			isotopes},\ in\ \href {https://doi.org/10.1007/978-1-4613-0713-6_7} {\emph
			{\bibinfo {booktitle} {Treatise on Heavy Ion Science: Volume 8: Nuclei Far
					From Stability}}},\ \bibinfo {editor} {edited by\ \bibinfo {editor}
			{\bibfnamefont {D.~A.}\ \bibnamefont {Bromley}}}\ (\bibinfo  {publisher}
		{Springer US},\ \bibinfo {address} {Boston, MA},\ \bibinfo {year} {1989})\
		pp.\ \bibinfo {pages} {517--638}\BibitemShut {NoStop}%
		\bibitem [{\citenamefont {J\"onsson}\ \emph {et~al.}(1996)\citenamefont
			{J\"onsson}, \citenamefont {Ynnerman}, \citenamefont {Froese~Fischer},
			\citenamefont {Godefroid},\ and\ \citenamefont {Olsen}}]{Jonsson1996}%
		\BibitemOpen
		\bibfield  {author} {\bibinfo {author} {\bibfnamefont {P.}~\bibnamefont
				{J\"onsson}}, \bibinfo {author} {\bibfnamefont {A.}~\bibnamefont {Ynnerman}},
			\bibinfo {author} {\bibfnamefont {C.}~\bibnamefont {Froese~Fischer}},
			\bibinfo {author} {\bibfnamefont {M.~R.}\ \bibnamefont {Godefroid}},\ and\
			\bibinfo {author} {\bibfnamefont {J.}~\bibnamefont {Olsen}},\ }\bibfield
		{title} {\bibinfo {title} {Large-scale multiconfiguration hartree-fock and
				configuration-interaction calculations of the transition probability and
				hyperfine structures in the sodium resonance transition},\ }\href
		{https://doi.org/10.1103/PhysRevA.53.4021} {\bibfield  {journal} {\bibinfo
				{journal} {Phys. Rev. A}\ }\textbf {\bibinfo {volume} {53}},\ \bibinfo
			{pages} {4021} (\bibinfo {year} {1996})}\BibitemShut {NoStop}%
		\bibitem [{\citenamefont {Porsev}\ \emph {et~al.}(2009)\citenamefont {Porsev},
			\citenamefont {Kozlov},\ and\ \citenamefont {Reimers}}]{Porsev2009}%
		\BibitemOpen
		\bibfield  {author} {\bibinfo {author} {\bibfnamefont {S.~G.}\ \bibnamefont
				{Porsev}}, \bibinfo {author} {\bibfnamefont {M.~G.}\ \bibnamefont {Kozlov}},\
			and\ \bibinfo {author} {\bibfnamefont {D.}~\bibnamefont {Reimers}},\
		}\bibfield  {title} {\bibinfo {title} {Transition frequency shifts with
				fine-structure constant variation for {Fe I} and isotope-shift calculations
				in {Fe I} and {Fe II}},\ }\href {https://doi.org/10.1103/PhysRevA.79.032519}
		{\bibfield  {journal} {\bibinfo  {journal} {Phys. Rev. A}\ }\textbf {\bibinfo
				{volume} {79}},\ \bibinfo {pages} {032519} (\bibinfo {year}
			{2009})}\BibitemShut {NoStop}%
		\bibitem [{\citenamefont {Fawcett}\ and\ \citenamefont
			{Wilson}(1991)}]{Fawcett1991}%
		\BibitemOpen
		\bibfield  {author} {\bibinfo {author} {\bibfnamefont {B.}~\bibnamefont
				{Fawcett}}\ and\ \bibinfo {author} {\bibfnamefont {M.}~\bibnamefont
				{Wilson}},\ }\bibfield  {title} {\bibinfo {title} {Computed oscillator
				strengths, {L}and{\'{e}} {$g$} values, and lifetimes in {Yb II}},\ }\href
		{https://doi.org/10.1016/0092-640X(91)90003-M} {\bibfield  {journal}
			{\bibinfo  {journal} {At. Data Nucl. Data Tables}\ }\textbf {\bibinfo
				{volume} {47}},\ \bibinfo {pages} {241} (\bibinfo {year} {1991})}\BibitemShut
		{NoStop}%
		\bibitem [{\citenamefont {Bi{\'{e}}mont}\ \emph {et~al.}(1998)\citenamefont
			{Bi{\'{e}}mont}, \citenamefont {Dutrieux}, \citenamefont {Martin},\ and\
			\citenamefont {Quinet}}]{Biemont1998}%
		\BibitemOpen
		\bibfield  {author} {\bibinfo {author} {\bibfnamefont {E.}~\bibnamefont
				{Bi{\'{e}}mont}}, \bibinfo {author} {\bibfnamefont {J.-F.}\ \bibnamefont
				{Dutrieux}}, \bibinfo {author} {\bibfnamefont {I.}~\bibnamefont {Martin}},\
			and\ \bibinfo {author} {\bibfnamefont {P.}~\bibnamefont {Quinet}},\
		}\bibfield  {title} {\bibinfo {title} {Lifetime calculations in {Yb II}},\
		}\href {https://doi.org/10.1088/0953-4075/31/15/006} {\bibfield  {journal}
			{\bibinfo  {journal} {J. Phys. B}\ }\textbf {\bibinfo {volume} {31}},\
			\bibinfo {pages} {3321} (\bibinfo {year} {1998})}\BibitemShut {NoStop}%
		\bibitem [{\citenamefont {{Froese Fischer}}\ \emph {et~al.}(2019)\citenamefont
			{{Froese Fischer}}, \citenamefont {Gaigalas}, \citenamefont {J\"{o}nsson},\
			and\ \citenamefont {Biero\'{n}}}]{FroeseFischer2018}%
		\BibitemOpen
		\bibfield  {author} {\bibinfo {author} {\bibfnamefont {C.}~\bibnamefont
				{{Froese Fischer}}}, \bibinfo {author} {\bibfnamefont {G.}~\bibnamefont
				{Gaigalas}}, \bibinfo {author} {\bibfnamefont {P.}~\bibnamefont
				{J\"{o}nsson}},\ and\ \bibinfo {author} {\bibfnamefont {J.}~\bibnamefont
				{Biero\'{n}}},\ }\bibfield  {title} {\bibinfo {title} {{GRASP2018} - a
				{Fortran 95} version of the general relativistic atomic structure package},\
		}\href {https://doi.org/10.1016/j.cpc.2018.10.032} {\bibfield  {journal}
			{\bibinfo  {journal} {Comput. Phys. Commun.}\ }\textbf {\bibinfo {volume}
				{237}},\ \bibinfo {pages} {184} (\bibinfo {year} {2019})}\BibitemShut
		{NoStop}%
		\bibitem [{\citenamefont {Kahl}\ and\ \citenamefont
			{Berengut}(2019)}]{Kahl2019}%
		\BibitemOpen
		\bibfield  {author} {\bibinfo {author} {\bibfnamefont {E.}~\bibnamefont
				{Kahl}}\ and\ \bibinfo {author} {\bibfnamefont {J.}~\bibnamefont
				{Berengut}},\ }\bibfield  {title} {\bibinfo {title}
			{{\textsc{amb}{\footnotesize i}\textsc{t}}: A programme for high-precision
				relativistic atomic structure calculations},\ }\href
		{https://doi.org/10.1016/j.cpc.2018.12.014} {\bibfield  {journal} {\bibinfo
				{journal} {Comput. Phys. Commun.}\ }\textbf {\bibinfo {volume} {238}},\
			\bibinfo {pages} {232} (\bibinfo {year} {2019})}\BibitemShut {NoStop}%
		\bibitem [{\citenamefont {Dzuba}\ \emph {et~al.}(1996)\citenamefont {Dzuba},
			\citenamefont {Flambaum},\ and\ \citenamefont {Kozlov}}]{Dzuba1996}%
		\BibitemOpen
		\bibfield  {author} {\bibinfo {author} {\bibfnamefont {V.~A.}\ \bibnamefont
				{Dzuba}}, \bibinfo {author} {\bibfnamefont {V.~V.}\ \bibnamefont
				{Flambaum}},\ and\ \bibinfo {author} {\bibfnamefont {M.~G.}\ \bibnamefont
				{Kozlov}},\ }\bibfield  {title} {\bibinfo {title} {Combination of the
				many-body perturbation theory with the configuration-interaction method},\
		}\href {https://doi.org/10.1103/PhysRevA.54.3948} {\bibfield  {journal}
			{\bibinfo  {journal} {Phys. Rev. A}\ }\textbf {\bibinfo {volume} {54}},\
			\bibinfo {pages} {3948} (\bibinfo {year} {1996})}\BibitemShut {NoStop}%
		\bibitem [{\citenamefont {Hammen}\ \emph {et~al.}(2018)\citenamefont {Hammen},
			\citenamefont {N\"ortersh\"auser}, \citenamefont {Balabanski}, \citenamefont
			{Bissell}, \citenamefont {Blaum}, \citenamefont {Budin\v{c}evi\'c},
			\citenamefont {Cheal}, \citenamefont {Flanagan}, \citenamefont {Fr\"ommgen},
			\citenamefont {Georgiev}, \citenamefont {Geppert}, \citenamefont {Kowalska},
			\citenamefont {Kreim}, \citenamefont {Krieger}, \citenamefont {Nazarewicz},
			\citenamefont {Neugart}, \citenamefont {Neyens}, \citenamefont {Papuga},
			\citenamefont {Reinhard}, \citenamefont {Rajabali}, \citenamefont {Schmidt},\
			and\ \citenamefont {Yordanov}}]{Hammen2018}%
		\BibitemOpen
		\bibfield  {author} {\bibinfo {author} {\bibfnamefont {M.}~\bibnamefont
				{Hammen}}, \bibinfo {author} {\bibfnamefont {W.}~\bibnamefont
				{N\"ortersh\"auser}}, \bibinfo {author} {\bibfnamefont {D.~L.}\ \bibnamefont
				{Balabanski}}, \bibinfo {author} {\bibfnamefont {M.~L.}\ \bibnamefont
				{Bissell}}, \bibinfo {author} {\bibfnamefont {K.}~\bibnamefont {Blaum}},
			\bibinfo {author} {\bibfnamefont {I.}~\bibnamefont {Budin\v{c}evi\'c}},
			\bibinfo {author} {\bibfnamefont {B.}~\bibnamefont {Cheal}}, \bibinfo
			{author} {\bibfnamefont {K.~T.}\ \bibnamefont {Flanagan}}, \bibinfo {author}
			{\bibfnamefont {N.}~\bibnamefont {Fr\"ommgen}}, \bibinfo {author}
			{\bibfnamefont {G.}~\bibnamefont {Georgiev}}, \bibinfo {author}
			{\bibfnamefont {C.}~\bibnamefont {Geppert}}, \bibinfo {author} {\bibfnamefont
				{M.}~\bibnamefont {Kowalska}}, \bibinfo {author} {\bibfnamefont
				{K.}~\bibnamefont {Kreim}}, \bibinfo {author} {\bibfnamefont
				{A.}~\bibnamefont {Krieger}}, \bibinfo {author} {\bibfnamefont
				{W.}~\bibnamefont {Nazarewicz}}, \bibinfo {author} {\bibfnamefont
				{R.}~\bibnamefont {Neugart}}, \bibinfo {author} {\bibfnamefont
				{G.}~\bibnamefont {Neyens}}, \bibinfo {author} {\bibfnamefont
				{J.}~\bibnamefont {Papuga}}, \bibinfo {author} {\bibfnamefont {P.-G.}\
				\bibnamefont {Reinhard}}, \bibinfo {author} {\bibfnamefont {M.~M.}\
				\bibnamefont {Rajabali}}, \bibinfo {author} {\bibfnamefont {S.}~\bibnamefont
				{Schmidt}},\ and\ \bibinfo {author} {\bibfnamefont {D.~T.}\ \bibnamefont
				{Yordanov}},\ }\bibfield  {title} {\bibinfo {title} {From calcium to cadmium:
				Testing the pairing functional through charge radii measurements of
				$^{100-130}${Cd}},\ }\href {https://doi.org/10.1103/PhysRevLett.121.102501}
		{\bibfield  {journal} {\bibinfo  {journal} {Phys. Rev. Lett.}\ }\textbf
			{\bibinfo {volume} {121}},\ \bibinfo {pages} {102501} (\bibinfo {year}
			{2018})}\BibitemShut {NoStop}%
		\bibitem [{\citenamefont {Gorges}\ \emph {et~al.}(2019)\citenamefont {Gorges},
			\citenamefont {Rodr\'{\i}guez}, \citenamefont {Balabanski}, \citenamefont
			{Bissell}, \citenamefont {Blaum}, \citenamefont {Cheal}, \citenamefont
			{Garcia~Ruiz}, \citenamefont {Georgiev}, \citenamefont {Gins}, \citenamefont
			{Heylen}, \citenamefont {Kanellakopoulos}, \citenamefont {Kaufmann},
			\citenamefont {Kowalska}, \citenamefont {Lagaki}, \citenamefont {Lechner},
			\citenamefont {Maa\ss{}}, \citenamefont {Malbrunot-Ettenauer}, \citenamefont
			{Nazarewicz}, \citenamefont {Neugart}, \citenamefont {Neyens}, \citenamefont
			{N\"ortersh\"auser}, \citenamefont {Reinhard}, \citenamefont {Sailer},
			\citenamefont {S\'anchez}, \citenamefont {Schmidt}, \citenamefont {Wehner},
			\citenamefont {Wraith}, \citenamefont {Xie}, \citenamefont {Xu},
			\citenamefont {Yang},\ and\ \citenamefont {Yordanov}}]{Gorges2019}%
		\BibitemOpen
		\bibfield  {author} {\bibinfo {author} {\bibfnamefont {C.}~\bibnamefont
				{Gorges}}, \bibinfo {author} {\bibfnamefont {L.~V.}\ \bibnamefont
				{Rodr\'{\i}guez}}, \bibinfo {author} {\bibfnamefont {D.~L.}\ \bibnamefont
				{Balabanski}}, \bibinfo {author} {\bibfnamefont {M.~L.}\ \bibnamefont
				{Bissell}}, \bibinfo {author} {\bibfnamefont {K.}~\bibnamefont {Blaum}},
			\bibinfo {author} {\bibfnamefont {B.}~\bibnamefont {Cheal}}, \bibinfo
			{author} {\bibfnamefont {R.~F.}\ \bibnamefont {Garcia~Ruiz}}, \bibinfo
			{author} {\bibfnamefont {G.}~\bibnamefont {Georgiev}}, \bibinfo {author}
			{\bibfnamefont {W.}~\bibnamefont {Gins}}, \bibinfo {author} {\bibfnamefont
				{H.}~\bibnamefont {Heylen}}, \bibinfo {author} {\bibfnamefont
				{A.}~\bibnamefont {Kanellakopoulos}}, \bibinfo {author} {\bibfnamefont
				{S.}~\bibnamefont {Kaufmann}}, \bibinfo {author} {\bibfnamefont
				{M.}~\bibnamefont {Kowalska}}, \bibinfo {author} {\bibfnamefont
				{V.}~\bibnamefont {Lagaki}}, \bibinfo {author} {\bibfnamefont
				{S.}~\bibnamefont {Lechner}}, \bibinfo {author} {\bibfnamefont
				{B.}~\bibnamefont {Maa\ss{}}}, \bibinfo {author} {\bibfnamefont
				{S.}~\bibnamefont {Malbrunot-Ettenauer}}, \bibinfo {author} {\bibfnamefont
				{W.}~\bibnamefont {Nazarewicz}}, \bibinfo {author} {\bibfnamefont
				{R.}~\bibnamefont {Neugart}}, \bibinfo {author} {\bibfnamefont
				{G.}~\bibnamefont {Neyens}}, \bibinfo {author} {\bibfnamefont
				{W.}~\bibnamefont {N\"ortersh\"auser}}, \bibinfo {author} {\bibfnamefont
				{P.-G.}\ \bibnamefont {Reinhard}}, \bibinfo {author} {\bibfnamefont
				{S.}~\bibnamefont {Sailer}}, \bibinfo {author} {\bibfnamefont
				{R.}~\bibnamefont {S\'anchez}}, \bibinfo {author} {\bibfnamefont
				{S.}~\bibnamefont {Schmidt}}, \bibinfo {author} {\bibfnamefont
				{L.}~\bibnamefont {Wehner}}, \bibinfo {author} {\bibfnamefont
				{C.}~\bibnamefont {Wraith}}, \bibinfo {author} {\bibfnamefont
				{L.}~\bibnamefont {Xie}}, \bibinfo {author} {\bibfnamefont {Z.~Y.}\
				\bibnamefont {Xu}}, \bibinfo {author} {\bibfnamefont {X.~F.}\ \bibnamefont
				{Yang}},\ and\ \bibinfo {author} {\bibfnamefont {D.~T.}\ \bibnamefont
				{Yordanov}},\ }\bibfield  {title} {\bibinfo {title} {Laser spectroscopy of
				neutron-rich tin isotopes: A discontinuity in charge radii across the
				{$N=82$} shell closure},\ }\href
		{https://doi.org/10.1103/PhysRevLett.122.192502} {\bibfield  {journal}
			{\bibinfo  {journal} {Phys. Rev. Lett.}\ }\textbf {\bibinfo {volume} {122}},\
			\bibinfo {pages} {192502} (\bibinfo {year} {2019})}\BibitemShut {NoStop}%
		\bibitem [{\citenamefont {Miller}\ \emph {et~al.}(2019)\citenamefont {Miller},
			\citenamefont {Minamisono}, \citenamefont {Klose}, \citenamefont {Garand},
			\citenamefont {Kujawa}, \citenamefont {Lantis}, \citenamefont {Liu},
			\citenamefont {Maaß}, \citenamefont {Mantica}, \citenamefont {Nazarewicz},
			\citenamefont {N\"ortersh\"auser}, \citenamefont {Pineda}, \citenamefont
			{Reinhard}, \citenamefont {Rossi}, \citenamefont {Sommer}, \citenamefont
			{Sumithrarachchi}, \citenamefont {Teigelh\"ofer},\ and\ \citenamefont
			{Watkins}}]{Miller2019}%
		\BibitemOpen
		\bibfield  {author} {\bibinfo {author} {\bibfnamefont {A.~J.}\ \bibnamefont
				{Miller}}, \bibinfo {author} {\bibfnamefont {K.}~\bibnamefont {Minamisono}},
			\bibinfo {author} {\bibfnamefont {A.}~\bibnamefont {Klose}}, \bibinfo
			{author} {\bibfnamefont {D.}~\bibnamefont {Garand}}, \bibinfo {author}
			{\bibfnamefont {C.}~\bibnamefont {Kujawa}}, \bibinfo {author} {\bibfnamefont
				{J.~D.}\ \bibnamefont {Lantis}}, \bibinfo {author} {\bibfnamefont
				{Y.}~\bibnamefont {Liu}}, \bibinfo {author} {\bibfnamefont {B.}~\bibnamefont
				{Maaß}}, \bibinfo {author} {\bibfnamefont {P.~F.}\ \bibnamefont {Mantica}},
			\bibinfo {author} {\bibfnamefont {W.}~\bibnamefont {Nazarewicz}}, \bibinfo
			{author} {\bibfnamefont {W.}~\bibnamefont {N\"ortersh\"auser}}, \bibinfo
			{author} {\bibfnamefont {S.~V.}\ \bibnamefont {Pineda}}, \bibinfo {author}
			{\bibfnamefont {P.-G.}\ \bibnamefont {Reinhard}}, \bibinfo {author}
			{\bibfnamefont {D.~M.}\ \bibnamefont {Rossi}}, \bibinfo {author}
			{\bibfnamefont {F.}~\bibnamefont {Sommer}}, \bibinfo {author} {\bibfnamefont
				{C.}~\bibnamefont {Sumithrarachchi}}, \bibinfo {author} {\bibfnamefont
				{A.}~\bibnamefont {Teigelh\"ofer}},\ and\ \bibinfo {author} {\bibfnamefont
				{J.}~\bibnamefont {Watkins}},\ }\bibfield  {title} {\bibinfo {title} {Proton
				superfluidity and charge radii in proton-rich calcium isotopes},\ }\href
		{https://doi.org/10.1038/s41567-019-0416-9} {\bibfield  {journal} {\bibinfo
				{journal} {Nat. Phys.}\ }\textbf {\bibinfo {volume} {15}},\ \bibinfo {pages}
			{1} (\bibinfo {year} {2019})}\BibitemShut {NoStop}%
		\bibitem [{\citenamefont {Berengut}\ \emph {et~al.}(2020)\citenamefont
			{Berengut}, \citenamefont {Delaunay}, \citenamefont {Geddes},\ and\
			\citenamefont {Soreq}}]{Berengut2020}%
		\BibitemOpen
		\bibfield  {author} {\bibinfo {author} {\bibfnamefont {J.~C.}\ \bibnamefont
				{Berengut}}, \bibinfo {author} {\bibfnamefont {C.}~\bibnamefont {Delaunay}},
			\bibinfo {author} {\bibfnamefont {A.}~\bibnamefont {Geddes}},\ and\ \bibinfo
			{author} {\bibfnamefont {Y.}~\bibnamefont {Soreq}},\ }\bibfield  {title}
		{\bibinfo {title} {Generalized {King} linearity and new physics searches with
				isotope shifts},\ }\href {https://doi.org/10.1103/PhysRevResearch.2.043444}
		{\bibfield  {journal} {\bibinfo  {journal} {Phys. Rev. Res.}\ }\textbf
			{\bibinfo {volume} {2}},\ \bibinfo {pages} {043444} (\bibinfo {year}
			{2020})}\BibitemShut {NoStop}%
		\bibitem [{\citenamefont {Hanneke}\ \emph {et~al.}(2008)\citenamefont
			{Hanneke}, \citenamefont {Fogwell},\ and\ \citenamefont
			{Gabrielse}}]{Hanneke2008}%
		\BibitemOpen
		\bibfield  {author} {\bibinfo {author} {\bibfnamefont {D.}~\bibnamefont
				{Hanneke}}, \bibinfo {author} {\bibfnamefont {S.}~\bibnamefont {Fogwell}},\
			and\ \bibinfo {author} {\bibfnamefont {G.}~\bibnamefont {Gabrielse}},\
		}\bibfield  {title} {\bibinfo {title} {New measurement of the electron
				magnetic moment and the fine structure constant},\ }\href
		{https://doi.org/10.1103/PhysRevLett.100.120801} {\bibfield  {journal}
			{\bibinfo  {journal} {Phys. Rev. Lett.}\ }\textbf {\bibinfo {volume} {100}},\
			\bibinfo {pages} {120801} (\bibinfo {year} {2008})}\BibitemShut {NoStop}%
		\bibitem [{\citenamefont {Aoyama}\ \emph {et~al.}(2012)\citenamefont {Aoyama},
			\citenamefont {Hayakawa}, \citenamefont {Kinoshita},\ and\ \citenamefont
			{Nio}}]{Aoyama2012}%
		\BibitemOpen
		\bibfield  {author} {\bibinfo {author} {\bibfnamefont {T.}~\bibnamefont
				{Aoyama}}, \bibinfo {author} {\bibfnamefont {M.}~\bibnamefont {Hayakawa}},
			\bibinfo {author} {\bibfnamefont {T.}~\bibnamefont {Kinoshita}},\ and\
			\bibinfo {author} {\bibfnamefont {M.}~\bibnamefont {Nio}},\ }\bibfield
		{title} {\bibinfo {title} {Tenth-order qed contribution to the electron
				{$g\ensuremath{-}2$} and an improved value of the fine structure constant},\
		}\href {https://doi.org/10.1103/PhysRevLett.109.111807} {\bibfield  {journal}
			{\bibinfo  {journal} {Phys. Rev. Lett.}\ }\textbf {\bibinfo {volume} {109}},\
			\bibinfo {pages} {111807} (\bibinfo {year} {2012})}\BibitemShut {NoStop}%
		\bibitem [{\citenamefont {Bouchendira}\ \emph {et~al.}(2011)\citenamefont
			{Bouchendira}, \citenamefont {Clad\'e}, \citenamefont {Guellati-Kh\'elifa},
			\citenamefont {Nez},\ and\ \citenamefont {Biraben}}]{Bouchendira2011}%
		\BibitemOpen
		\bibfield  {author} {\bibinfo {author} {\bibfnamefont {R.}~\bibnamefont
				{Bouchendira}}, \bibinfo {author} {\bibfnamefont {P.}~\bibnamefont
				{Clad\'e}}, \bibinfo {author} {\bibfnamefont {S.}~\bibnamefont
				{Guellati-Kh\'elifa}}, \bibinfo {author} {\bibfnamefont {F.}~\bibnamefont
				{Nez}},\ and\ \bibinfo {author} {\bibfnamefont {F.}~\bibnamefont {Biraben}},\
		}\bibfield  {title} {\bibinfo {title} {New determination of the fine
				structure constant and test of the quantum electrodynamics},\ }\href
		{https://doi.org/10.1103/PhysRevLett.106.080801} {\bibfield  {journal}
			{\bibinfo  {journal} {Phys. Rev. Lett.}\ }\textbf {\bibinfo {volume} {106}},\
			\bibinfo {pages} {080801} (\bibinfo {year} {2011})}\BibitemShut {NoStop}%
		\bibitem [{\citenamefont {Davoudiasl}\ \emph {et~al.}(2014)\citenamefont
			{Davoudiasl}, \citenamefont {Lee},\ and\ \citenamefont
			{Marciano}}]{Davoudiasl2014}%
		\BibitemOpen
		\bibfield  {author} {\bibinfo {author} {\bibfnamefont {H.}~\bibnamefont
				{Davoudiasl}}, \bibinfo {author} {\bibfnamefont {H.-S.}\ \bibnamefont
				{Lee}},\ and\ \bibinfo {author} {\bibfnamefont {W.~J.}\ \bibnamefont
				{Marciano}},\ }\bibfield  {title} {\bibinfo {title} {Muon $g\ensuremath{-}2$,
				rare kaon decays, and parity violation from dark bosons},\ }\href
		{https://doi.org/10.1103/PhysRevD.89.095006} {\bibfield  {journal} {\bibinfo
				{journal} {Phys. Rev. D}\ }\textbf {\bibinfo {volume} {89}},\ \bibinfo
			{pages} {095006} (\bibinfo {year} {2014})}\BibitemShut {NoStop}%
		\bibitem [{\citenamefont {Barbieri}\ and\ \citenamefont
			{Ericson}(1975)}]{Barbieri1975}%
		\BibitemOpen
		\bibfield  {author} {\bibinfo {author} {\bibfnamefont {R.}~\bibnamefont
				{Barbieri}}\ and\ \bibinfo {author} {\bibfnamefont {T.}~\bibnamefont
				{Ericson}},\ }\bibfield  {title} {\bibinfo {title} {Evidence against the
				existence of a low mass scalar boson from neutron-nucleus scattering},\
		}\href {https://doi.org/10.1016/0370-2693(75)90073-8} {\bibfield  {journal}
			{\bibinfo  {journal} {Phys. Lett. B}\ }\textbf {\bibinfo {volume} {57}},\
			\bibinfo {pages} {270} (\bibinfo {year} {1975})}\BibitemShut {NoStop}%
		\bibitem [{\citenamefont {Leeb}\ and\ \citenamefont
			{Schmiedmayer}(1992)}]{Leeb1992}%
		\BibitemOpen
		\bibfield  {author} {\bibinfo {author} {\bibfnamefont {H.}~\bibnamefont
				{Leeb}}\ and\ \bibinfo {author} {\bibfnamefont {J.}~\bibnamefont
				{Schmiedmayer}},\ }\bibfield  {title} {\bibinfo {title} {Constraint on
				hypothetical light interacting bosons from low-energy neutron experiments},\
		}\href {https://doi.org/10.1103/PhysRevLett.68.1472} {\bibfield  {journal}
			{\bibinfo  {journal} {Phys. Rev. Lett.}\ }\textbf {\bibinfo {volume} {68}},\
			\bibinfo {pages} {1472} (\bibinfo {year} {1992})}\BibitemShut {NoStop}%
		\bibitem [{\citenamefont {Pokotilovski}(2006)}]{Pokotilovski2006}%
		\BibitemOpen
		\bibfield  {author} {\bibinfo {author} {\bibfnamefont {Y.~N.}\ \bibnamefont
				{Pokotilovski}},\ }\bibfield  {title} {\bibinfo {title} {Constraints on new
				interactions from neutron scattering experiments},\ }\href
		{https://doi.org/10.1134/S1063778806060020} {\bibfield  {journal} {\bibinfo
				{journal} {Phys. At. Nucl.}\ }\textbf {\bibinfo {volume} {69}},\ \bibinfo
			{pages} {924} (\bibinfo {year} {2006})}\BibitemShut {NoStop}%
		\bibitem [{\citenamefont {Nesvizhevsky}\ \emph {et~al.}(2008)\citenamefont
			{Nesvizhevsky}, \citenamefont {Pignol},\ and\ \citenamefont
			{Protasov}}]{Nesvizhevsky2008}%
		\BibitemOpen
		\bibfield  {author} {\bibinfo {author} {\bibfnamefont {V.~V.}\ \bibnamefont
				{Nesvizhevsky}}, \bibinfo {author} {\bibfnamefont {G.}~\bibnamefont
				{Pignol}},\ and\ \bibinfo {author} {\bibfnamefont {K.~V.}\ \bibnamefont
				{Protasov}},\ }\bibfield  {title} {\bibinfo {title} {Neutron scattering and
				extra-short-range interactions},\ }\href
		{https://doi.org/10.1103/PhysRevD.77.034020} {\bibfield  {journal} {\bibinfo
				{journal} {Phys. Rev. D}\ }\textbf {\bibinfo {volume} {77}},\ \bibinfo
			{pages} {034020} (\bibinfo {year} {2008})}\BibitemShut {NoStop}%
		\bibitem [{\citenamefont {Manovitz}\ \emph {et~al.}(2019)\citenamefont
			{Manovitz}, \citenamefont {Shaniv}, \citenamefont {Shapira}, \citenamefont
			{Ozeri},\ and\ \citenamefont {Akerman}}]{Manovitz2019}%
		\BibitemOpen
		\bibfield  {author} {\bibinfo {author} {\bibfnamefont {T.}~\bibnamefont
				{Manovitz}}, \bibinfo {author} {\bibfnamefont {R.}~\bibnamefont {Shaniv}},
			\bibinfo {author} {\bibfnamefont {Y.}~\bibnamefont {Shapira}}, \bibinfo
			{author} {\bibfnamefont {R.}~\bibnamefont {Ozeri}},\ and\ \bibinfo {author}
			{\bibfnamefont {N.}~\bibnamefont {Akerman}},\ }\bibfield  {title} {\bibinfo
			{title} {Precision measurement of atomic isotope shifts using a two-isotope
				entangled state},\ }\href {https://doi.org/10.1103/PhysRevLett.123.203001}
		{\bibfield  {journal} {\bibinfo  {journal} {Phys. Rev. Lett.}\ }\textbf
			{\bibinfo {volume} {123}},\ \bibinfo {pages} {203001} (\bibinfo {year}
			{2019})}\BibitemShut {NoStop}%
		\bibitem [{\citenamefont {Aymar}\ \emph {et~al.}(1980)\citenamefont {Aymar},
			\citenamefont {Debarre},\ and\ \citenamefont {Robaux}}]{Aymar1980}%
		\BibitemOpen
		\bibfield  {author} {\bibinfo {author} {\bibfnamefont {M.}~\bibnamefont
				{Aymar}}, \bibinfo {author} {\bibfnamefont {A.}~\bibnamefont {Debarre}},\
			and\ \bibinfo {author} {\bibfnamefont {O.}~\bibnamefont {Robaux}},\
		}\bibfield  {title} {\bibinfo {title} {Highly excited levels of neutral
				ytterbium. {II}. multichannel quantum defect analysis of odd- and even-parity
				spectra},\ }\href {https://doi.org/10.1088/0022-3700/13/6/016} {\bibfield
			{journal} {\bibinfo  {journal} {J. Phys. B}\ }\textbf {\bibinfo {volume}
				{13}},\ \bibinfo {pages} {1089} (\bibinfo {year} {1980})}\BibitemShut
		{NoStop}%
		\bibitem [{\citenamefont {Bender}\ \emph {et~al.}(2003)\citenamefont {Bender},
			\citenamefont {Heenen},\ and\ \citenamefont {Reinhard}}]{Bender2003}%
		\BibitemOpen
		\bibfield  {author} {\bibinfo {author} {\bibfnamefont {M.}~\bibnamefont
				{Bender}}, \bibinfo {author} {\bibfnamefont {P.-H.}\ \bibnamefont {Heenen}},\
			and\ \bibinfo {author} {\bibfnamefont {P.-G.}\ \bibnamefont {Reinhard}},\
		}\bibfield  {title} {\bibinfo {title} {Self-consistent mean-field models for
				nuclear structure},\ }\href {https://doi.org/10.1103/RevModPhys.75.121}
		{\bibfield  {journal} {\bibinfo  {journal} {Rev. Mod. Phys.}\ }\textbf
			{\bibinfo {volume} {75}},\ \bibinfo {pages} {121} (\bibinfo {year}
			{2003})}\BibitemShut {NoStop}%
		\bibitem [{\citenamefont {Berengut}\ \emph {et~al.}(2003)\citenamefont
			{Berengut}, \citenamefont {Dzuba},\ and\ \citenamefont
			{Flambaum}}]{Berengut2003}%
		\BibitemOpen
		\bibfield  {author} {\bibinfo {author} {\bibfnamefont {J.~C.}\ \bibnamefont
				{Berengut}}, \bibinfo {author} {\bibfnamefont {V.~A.}\ \bibnamefont
				{Dzuba}},\ and\ \bibinfo {author} {\bibfnamefont {V.~V.}\ \bibnamefont
				{Flambaum}},\ }\bibfield  {title} {\bibinfo {title} {Isotope-shift
				calculations for atoms with one valence electron},\ }\href
		{https://doi.org/10.1103/PhysRevA.68.022502} {\bibfield  {journal} {\bibinfo
				{journal} {Phys. Rev. A}\ }\textbf {\bibinfo {volume} {68}},\ \bibinfo
			{pages} {022502} (\bibinfo {year} {2003})}\BibitemShut {NoStop}%
		\bibitem [{\citenamefont {Bogdanovich}\ and\ \citenamefont
			{Zukauskas}(1983)}]{Bogdanovich1983}%
		\BibitemOpen
		\bibfield  {author} {\bibinfo {author} {\bibfnamefont {P.}~\bibnamefont
				{Bogdanovich}}\ and\ \bibinfo {author} {\bibfnamefont {G.}~\bibnamefont
				{Zukauskas}},\ }\bibfield  {title} {\bibinfo {title} {Approximate allowance
				for superposition of configurations in atomic spectra},\ }\href@noop {}
		{\bibfield  {journal} {\bibinfo  {journal} {Sov. Phys. Collect.}\ }\textbf
			{\bibinfo {volume} {23}},\ \bibinfo {pages} {13} (\bibinfo {year}
			{1983})}\BibitemShut {NoStop}%
		\bibitem [{\citenamefont {Dyall}\ \emph {et~al.}(1989)\citenamefont {Dyall},
			\citenamefont {Grant}, \citenamefont {Johnson}, \citenamefont {Parpia},\ and\
			\citenamefont {Plummer}}]{Dyall1989}%
		\BibitemOpen
		\bibfield  {author} {\bibinfo {author} {\bibfnamefont {K.}~\bibnamefont
				{Dyall}}, \bibinfo {author} {\bibfnamefont {I.}~\bibnamefont {Grant}},
			\bibinfo {author} {\bibfnamefont {C.}~\bibnamefont {Johnson}}, \bibinfo
			{author} {\bibfnamefont {F.}~\bibnamefont {Parpia}},\ and\ \bibinfo {author}
			{\bibfnamefont {E.}~\bibnamefont {Plummer}},\ }\bibfield  {title} {\bibinfo
			{title} {{GRASP}: A general-purpose relativistic atomic structure program},\
		}\href {https://doi.org/10.1016/0010-4655(89)90136-7} {\bibfield  {journal}
			{\bibinfo  {journal} {Comput. Phys. Commun.}\ }\textbf {\bibinfo {volume}
				{55}},\ \bibinfo {pages} {425} (\bibinfo {year} {1989})}\BibitemShut
		{NoStop}%
		\bibitem [{\citenamefont {Edmunds}\ \emph {et~al.}(2021)\citenamefont
			{Edmunds}, \citenamefont {Tan}, \citenamefont {Milne}, \citenamefont {Singh},
			\citenamefont {Biercuk},\ and\ \citenamefont {Hempel}}]{Edmunds2021}%
		\BibitemOpen
		\bibfield  {author} {\bibinfo {author} {\bibfnamefont {C.~L.}\ \bibnamefont
				{Edmunds}}, \bibinfo {author} {\bibfnamefont {T.~R.}\ \bibnamefont {Tan}},
			\bibinfo {author} {\bibfnamefont {A.~R.}\ \bibnamefont {Milne}}, \bibinfo
			{author} {\bibfnamefont {A.}~\bibnamefont {Singh}}, \bibinfo {author}
			{\bibfnamefont {M.~J.}\ \bibnamefont {Biercuk}},\ and\ \bibinfo {author}
			{\bibfnamefont {C.}~\bibnamefont {Hempel}},\ }\bibfield  {title} {\bibinfo
			{title} {Scalable hyperfine qubit state detection via electron shelving in
				the {${}^{2}{D}_{5/2}$} and {${}^{2}{F}_{7/2}$} manifolds in
				{${}^{171}$Yb${}^+$}},\ }\href {https://doi.org/10.1103/PhysRevA.104.012606}
		{\bibfield  {journal} {\bibinfo  {journal} {Phys. Rev. A}\ }\textbf {\bibinfo
				{volume} {104}},\ \bibinfo {pages} {012606} (\bibinfo {year}
			{2021})}\BibitemShut {NoStop}%
		\bibitem [{\citenamefont {Ekman}\ \emph {et~al.}(2019)\citenamefont {Ekman},
			\citenamefont {Jönsson}, \citenamefont {Godefroid}, \citenamefont {Nazé},
			\citenamefont {Gaigalas},\ and\ \citenamefont {Bieroń}}]{Ekman2019}%
		\BibitemOpen
		\bibfield  {author} {\bibinfo {author} {\bibfnamefont {J.}~\bibnamefont
				{Ekman}}, \bibinfo {author} {\bibfnamefont {P.}~\bibnamefont {Jönsson}},
			\bibinfo {author} {\bibfnamefont {M.}~\bibnamefont {Godefroid}}, \bibinfo
			{author} {\bibfnamefont {C.}~\bibnamefont {Nazé}}, \bibinfo {author}
			{\bibfnamefont {G.}~\bibnamefont {Gaigalas}},\ and\ \bibinfo {author}
			{\bibfnamefont {J.}~\bibnamefont {Bieroń}},\ }\bibfield  {title} {\bibinfo
			{title} {{RIS4}: A program for relativistic isotope shift calculations},\
		}\href {https://doi.org/10.1016/j.cpc.2018.08.017} {\bibfield  {journal}
			{\bibinfo  {journal} {Comput. Phys. Commun.}\ }\textbf {\bibinfo {volume}
				{235}},\ \bibinfo {pages} {433} (\bibinfo {year} {2019})}\BibitemShut
		{NoStop}%
		\bibitem [{\citenamefont {Erler}\ \emph {et~al.}(2010)\citenamefont {Erler},
			\citenamefont {Kl\"upfel},\ and\ \citenamefont {Reinhard}}]{Erler2010}%
		\BibitemOpen
		\bibfield  {author} {\bibinfo {author} {\bibfnamefont {J.}~\bibnamefont
				{Erler}}, \bibinfo {author} {\bibfnamefont {P.}~\bibnamefont {Kl\"upfel}},\
			and\ \bibinfo {author} {\bibfnamefont {P.-G.}\ \bibnamefont {Reinhard}},\
		}\bibfield  {title} {\bibinfo {title} {Exploration of a modified density
				dependence in the {Skyrme} functional},\ }\href
		{https://doi.org/10.1103/PhysRevC.82.044307} {\bibfield  {journal} {\bibinfo
				{journal} {Phys. Rev. C}\ }\textbf {\bibinfo {volume} {82}},\ \bibinfo
			{pages} {044307} (\bibinfo {year} {2010})}\BibitemShut {NoStop}%
		\bibitem [{\citenamefont {Grant}\ \emph {et~al.}(1980)\citenamefont {Grant},
			\citenamefont {McKenzie}, \citenamefont {Norrington}, \citenamefont
			{Mayers},\ and\ \citenamefont {Pyper}}]{Grant1980}%
		\BibitemOpen
		\bibfield  {author} {\bibinfo {author} {\bibfnamefont {I.}~\bibnamefont
				{Grant}}, \bibinfo {author} {\bibfnamefont {B.}~\bibnamefont {McKenzie}},
			\bibinfo {author} {\bibfnamefont {P.}~\bibnamefont {Norrington}}, \bibinfo
			{author} {\bibfnamefont {D.}~\bibnamefont {Mayers}},\ and\ \bibinfo {author}
			{\bibfnamefont {N.}~\bibnamefont {Pyper}},\ }\bibfield  {title} {\bibinfo
			{title} {An atomic multiconfigurational {Dirac}-{Fock} package},\ }\href
		{https://doi.org/10.1016/0010-4655(80)90041-7} {\bibfield  {journal}
			{\bibinfo  {journal} {Comput. Phys. Commun.}\ }\textbf {\bibinfo {volume}
				{21}},\ \bibinfo {pages} {207} (\bibinfo {year} {1980})}\BibitemShut
		{NoStop}%
		\bibitem [{\citenamefont {Huntemann}\ \emph {et~al.}(2012)\citenamefont
			{Huntemann}, \citenamefont {Okhapkin}, \citenamefont {Lipphardt},
			\citenamefont {Weyers}, \citenamefont {Tamm},\ and\ \citenamefont
			{Peik}}]{Huntemann2012}%
		\BibitemOpen
		\bibfield  {author} {\bibinfo {author} {\bibfnamefont {N.}~\bibnamefont
				{Huntemann}}, \bibinfo {author} {\bibfnamefont {M.}~\bibnamefont {Okhapkin}},
			\bibinfo {author} {\bibfnamefont {B.}~\bibnamefont {Lipphardt}}, \bibinfo
			{author} {\bibfnamefont {S.}~\bibnamefont {Weyers}}, \bibinfo {author}
			{\bibfnamefont {C.}~\bibnamefont {Tamm}},\ and\ \bibinfo {author}
			{\bibfnamefont {E.}~\bibnamefont {Peik}},\ }\bibfield  {title} {\bibinfo
			{title} {High-accuracy optical clock based on the octupole transition in
				$^{171}\mathrm{Yb}^{+}$},\ }\href
		{https://doi.org/10.1103/PhysRevLett.108.090801} {\bibfield  {journal}
			{\bibinfo  {journal} {Phys. Rev. Lett.}\ }\textbf {\bibinfo {volume} {108}},\
			\bibinfo {pages} {090801} (\bibinfo {year} {2012})}\BibitemShut {NoStop}%
		\bibitem [{\citenamefont {Jau}\ \emph {et~al.}(2015)\citenamefont {Jau},
			\citenamefont {Hunker},\ and\ \citenamefont {Schwindt}}]{Jau2015}%
		\BibitemOpen
		\bibfield  {author} {\bibinfo {author} {\bibfnamefont {Y.-Y.}\ \bibnamefont
				{Jau}}, \bibinfo {author} {\bibfnamefont {J.~D.}\ \bibnamefont {Hunker}},\
			and\ \bibinfo {author} {\bibfnamefont {P.~D.~D.}\ \bibnamefont {Schwindt}},\
		}\bibfield  {title} {\bibinfo {title} {{F}-state quenching with {CH4} for
				buffer-gas cooled {${}^{171}$Yb$^+$} frequency standard},\ }\href
		{https://doi.org/10.1063/1.4935562} {\bibfield  {journal} {\bibinfo
				{journal} {AIP Adv.}\ }\textbf {\bibinfo {volume} {5}},\ \bibinfo {pages}
			{117209} (\bibinfo {year} {2015})}\BibitemShut {NoStop}%
		\bibitem [{\citenamefont {King}\ \emph {et~al.}(2012)\citenamefont {King},
			\citenamefont {Godun}, \citenamefont {Webster}, \citenamefont {Margolis},
			\citenamefont {Johnson}, \citenamefont {Szymaniec}, \citenamefont {Baird},\
			and\ \citenamefont {Gill}}]{King2012}%
		\BibitemOpen
		\bibfield  {author} {\bibinfo {author} {\bibfnamefont {S.~A.}\ \bibnamefont
				{King}}, \bibinfo {author} {\bibfnamefont {R.~M.}\ \bibnamefont {Godun}},
			\bibinfo {author} {\bibfnamefont {S.~A.}\ \bibnamefont {Webster}}, \bibinfo
			{author} {\bibfnamefont {H.~S.}\ \bibnamefont {Margolis}}, \bibinfo {author}
			{\bibfnamefont {L.~A.~M.}\ \bibnamefont {Johnson}}, \bibinfo {author}
			{\bibfnamefont {K.}~\bibnamefont {Szymaniec}}, \bibinfo {author}
			{\bibfnamefont {P.~E.~G.}\ \bibnamefont {Baird}},\ and\ \bibinfo {author}
			{\bibfnamefont {P.}~\bibnamefont {Gill}},\ }\bibfield  {title} {\bibinfo
			{title} {Absolute frequency measurement of the {${}^2S_{1/2} \textendash
					{}^2F_{7/2}$} electric octupole transition in a single ion of
				{${}^{171}$Yb${}^+$} with {$10^{-15}$} fractional uncertainty},\ }\href
		{https://doi.org/10.1088/1367-2630/14/1/013045} {\bibfield  {journal}
			{\bibinfo  {journal} {New J. Phys.}\ }\textbf {\bibinfo {volume} {14}},\
			\bibinfo {pages} {013045} (\bibinfo {year} {2012})}\BibitemShut {NoStop}%
		\bibitem [{\citenamefont {Kortelainen}\ \emph {et~al.}(2012)\citenamefont
			{Kortelainen}, \citenamefont {McDonnell}, \citenamefont {Nazarewicz},
			\citenamefont {Reinhard}, \citenamefont {Sarich}, \citenamefont {Schunck},
			\citenamefont {Stoitsov},\ and\ \citenamefont {Wild}}]{Kortelainen2012}%
		\BibitemOpen
		\bibfield  {author} {\bibinfo {author} {\bibfnamefont {M.}~\bibnamefont
				{Kortelainen}}, \bibinfo {author} {\bibfnamefont {J.}~\bibnamefont
				{McDonnell}}, \bibinfo {author} {\bibfnamefont {W.}~\bibnamefont
				{Nazarewicz}}, \bibinfo {author} {\bibfnamefont {P.-G.}\ \bibnamefont
				{Reinhard}}, \bibinfo {author} {\bibfnamefont {J.}~\bibnamefont {Sarich}},
			\bibinfo {author} {\bibfnamefont {N.}~\bibnamefont {Schunck}}, \bibinfo
			{author} {\bibfnamefont {M.~V.}\ \bibnamefont {Stoitsov}},\ and\ \bibinfo
			{author} {\bibfnamefont {S.~M.}\ \bibnamefont {Wild}},\ }\bibfield  {title}
		{\bibinfo {title} {Nuclear energy density optimization: Large deformations},\
		}\href {https://doi.org/10.1103/PhysRevC.85.024304} {\bibfield  {journal}
			{\bibinfo  {journal} {Phys. Rev. C}\ }\textbf {\bibinfo {volume} {85}},\
			\bibinfo {pages} {024304} (\bibinfo {year} {2012})}\BibitemShut {NoStop}%
		\bibitem [{\citenamefont {McLoughlin}\ \emph {et~al.}(2011)\citenamefont
			{McLoughlin}, \citenamefont {Nizamani}, \citenamefont {Siverns},
			\citenamefont {Sterling}, \citenamefont {Hughes}, \citenamefont {Lekitsch},
			\citenamefont {Stein}, \citenamefont {Weidt},\ and\ \citenamefont
			{Hensinger}}]{McLoughlin2011}%
		\BibitemOpen
		\bibfield  {author} {\bibinfo {author} {\bibfnamefont {J.~J.}\ \bibnamefont
				{McLoughlin}}, \bibinfo {author} {\bibfnamefont {A.~H.}\ \bibnamefont
				{Nizamani}}, \bibinfo {author} {\bibfnamefont {J.~D.}\ \bibnamefont
				{Siverns}}, \bibinfo {author} {\bibfnamefont {R.~C.}\ \bibnamefont
				{Sterling}}, \bibinfo {author} {\bibfnamefont {M.~D.}\ \bibnamefont
				{Hughes}}, \bibinfo {author} {\bibfnamefont {B.}~\bibnamefont {Lekitsch}},
			\bibinfo {author} {\bibfnamefont {B.}~\bibnamefont {Stein}}, \bibinfo
			{author} {\bibfnamefont {S.}~\bibnamefont {Weidt}},\ and\ \bibinfo {author}
			{\bibfnamefont {W.~K.}\ \bibnamefont {Hensinger}},\ }\bibfield  {title}
		{\bibinfo {title} {Versatile ytterbium ion trap experiment for operation of
				scalable ion-trap chips with motional heating and transition-frequency
				measurements},\ }\href {https://doi.org/10.1103/PhysRevA.83.013406}
		{\bibfield  {journal} {\bibinfo  {journal} {Phys. Rev. A}\ }\textbf {\bibinfo
				{volume} {83}},\ \bibinfo {pages} {013406} (\bibinfo {year}
			{2011})}\BibitemShut {NoStop}%
		\bibitem [{\citenamefont {Mulholland}\ \emph {et~al.}(2019)\citenamefont
			{Mulholland}, \citenamefont {Klein}, \citenamefont {Barwood}, \citenamefont
			{Donnellan}, \citenamefont {Nisbet-Jones}, \citenamefont {Huang},
			\citenamefont {Walsh}, \citenamefont {Baird},\ and\ \citenamefont
			{Gill}}]{Mulholland2019}%
		\BibitemOpen
		\bibfield  {author} {\bibinfo {author} {\bibfnamefont {S.}~\bibnamefont
				{Mulholland}}, \bibinfo {author} {\bibfnamefont {H.~A.}\ \bibnamefont
				{Klein}}, \bibinfo {author} {\bibfnamefont {G.~P.}\ \bibnamefont {Barwood}},
			\bibinfo {author} {\bibfnamefont {S.}~\bibnamefont {Donnellan}}, \bibinfo
			{author} {\bibfnamefont {P.~B.~R.}\ \bibnamefont {Nisbet-Jones}}, \bibinfo
			{author} {\bibfnamefont {G.}~\bibnamefont {Huang}}, \bibinfo {author}
			{\bibfnamefont {G.}~\bibnamefont {Walsh}}, \bibinfo {author} {\bibfnamefont
				{P.~E.~G.}\ \bibnamefont {Baird}},\ and\ \bibinfo {author} {\bibfnamefont
				{P.}~\bibnamefont {Gill}},\ }\bibfield  {title} {\bibinfo {title} {Compact
				laser system for a laser-cooled ytterbium ion microwave frequency standard},\
		}\href {https://doi.org/10.1063/1.5082703} {\bibfield  {journal} {\bibinfo
				{journal} {Rev. Sci. Instrum.}\ }\textbf {\bibinfo {volume} {90}},\ \bibinfo
			{pages} {033105} (\bibinfo {year} {2019})}\BibitemShut {NoStop}%
		\bibitem [{\citenamefont {Kramida}\ \emph {et~al.}(2021)\citenamefont
			{Kramida}, \citenamefont {{Yu.~Ralchenko}}, \citenamefont {Reader},\ and\
			\citenamefont {{and NIST ASD Team}}}]{NIST_ASD}%
		\BibitemOpen
		\bibfield  {author} {\bibinfo {author} {\bibfnamefont {A.}~\bibnamefont
				{Kramida}}, \bibinfo {author} {\bibnamefont {{Yu.~Ralchenko}}}, \bibinfo
			{author} {\bibfnamefont {J.}~\bibnamefont {Reader}},\ and\ \bibinfo {author}
			{\bibnamefont {{and NIST ASD Team}}},\ }\href@noop {} {}\bibinfo
		{howpublished} {{NIST Atomic Spectra Database (ver. 5.9), [Online].
				Available: {\tt{https://physics.nist.gov/asd}} [2021, November 2]. National
				Institute of Standards and Technology, Gaithersburg, MD.}} (\bibinfo {year}
		{2021})\BibitemShut {NoStop}%
		\bibitem [{\citenamefont {Olmschenk}\ \emph {et~al.}(2007)\citenamefont
			{Olmschenk}, \citenamefont {Younge}, \citenamefont {Moehring}, \citenamefont
			{Matsukevich}, \citenamefont {Maunz},\ and\ \citenamefont
			{Monroe}}]{Olmschenk2007}%
		\BibitemOpen
		\bibfield  {author} {\bibinfo {author} {\bibfnamefont {S.}~\bibnamefont
				{Olmschenk}}, \bibinfo {author} {\bibfnamefont {K.~C.}\ \bibnamefont
				{Younge}}, \bibinfo {author} {\bibfnamefont {D.~L.}\ \bibnamefont
				{Moehring}}, \bibinfo {author} {\bibfnamefont {D.~N.}\ \bibnamefont
				{Matsukevich}}, \bibinfo {author} {\bibfnamefont {P.}~\bibnamefont {Maunz}},\
			and\ \bibinfo {author} {\bibfnamefont {C.}~\bibnamefont {Monroe}},\
		}\bibfield  {title} {\bibinfo {title} {Manipulation and detection of a
				trapped {$\mathrm{Yb}^{+}$} hyperfine qubit},\ }\href
		{https://doi.org/10.1103/PhysRevA.76.052314} {\bibfield  {journal} {\bibinfo
				{journal} {Phys. Rev. A}\ }\textbf {\bibinfo {volume} {76}},\ \bibinfo
			{pages} {052314} (\bibinfo {year} {2007})}\BibitemShut {NoStop}%
		\bibitem [{\citenamefont {Papoulia}\ \emph {et~al.}(2016)\citenamefont
			{Papoulia}, \citenamefont {Carlsson},\ and\ \citenamefont
			{Ekman}}]{Papoulia2016}%
		\BibitemOpen
		\bibfield  {author} {\bibinfo {author} {\bibfnamefont {A.}~\bibnamefont
				{Papoulia}}, \bibinfo {author} {\bibfnamefont {B.~G.}\ \bibnamefont
				{Carlsson}},\ and\ \bibinfo {author} {\bibfnamefont {J.}~\bibnamefont
				{Ekman}},\ }\bibfield  {title} {\bibinfo {title} {Effect of realistic nuclear
				charge distributions on isotope shifts and progress towards the extraction of
				higher-order nuclear radial moments},\ }\href
		{https://doi.org/10.1103/PhysRevA.94.042502} {\bibfield  {journal} {\bibinfo
				{journal} {Phys. Rev. A}\ }\textbf {\bibinfo {volume} {94}},\ \bibinfo
			{pages} {042502} (\bibinfo {year} {2016})}\BibitemShut {NoStop}%
		\bibitem [{\citenamefont {Pizzocaro}\ \emph {et~al.}(2020)\citenamefont
			{Pizzocaro}, \citenamefont {Bregolin}, \citenamefont {Barbieri},
			\citenamefont {Rauf}, \citenamefont {Levi},\ and\ \citenamefont
			{Calonico}}]{Pizzocaro2020}%
		\BibitemOpen
		\bibfield  {author} {\bibinfo {author} {\bibfnamefont {M.}~\bibnamefont
				{Pizzocaro}}, \bibinfo {author} {\bibfnamefont {F.}~\bibnamefont {Bregolin}},
			\bibinfo {author} {\bibfnamefont {P.}~\bibnamefont {Barbieri}}, \bibinfo
			{author} {\bibfnamefont {B.}~\bibnamefont {Rauf}}, \bibinfo {author}
			{\bibfnamefont {F.}~\bibnamefont {Levi}},\ and\ \bibinfo {author}
			{\bibfnamefont {D.}~\bibnamefont {Calonico}},\ }\bibfield  {title} {\bibinfo
			{title} {Absolute frequency measurement of the {${}^1S_{0} \textendash
					{}^3P_{0}$} transition of {${}^{171}$Yb} with a link to international atomic
				time},\ }\href {https://doi.org/10.1088/1681-7575/ab50e8} {\bibfield
			{journal} {\bibinfo  {journal} {Metrologia}\ }\textbf {\bibinfo {volume}
				{57}},\ \bibinfo {pages} {035007} (\bibinfo {year} {2020})}\BibitemShut
		{NoStop}%
		\bibitem [{\citenamefont {Pritychenko}\ \emph {et~al.}(2016)\citenamefont
			{Pritychenko}, \citenamefont {Birch}, \citenamefont {Singh},\ and\
			\citenamefont {Horoi}}]{Pritychenko2016}%
		\BibitemOpen
		\bibfield  {author} {\bibinfo {author} {\bibfnamefont {B.}~\bibnamefont
				{Pritychenko}}, \bibinfo {author} {\bibfnamefont {M.}~\bibnamefont {Birch}},
			\bibinfo {author} {\bibfnamefont {B.}~\bibnamefont {Singh}},\ and\ \bibinfo
			{author} {\bibfnamefont {M.}~\bibnamefont {Horoi}},\ }\bibfield  {title}
		{\bibinfo {title} {Tables of {E2} transition probabilities from the first 2+
				states in even–even nuclei},\ }\href
		{https://doi.org/10.1016/j.adt.2015.10.001} {\bibfield  {journal} {\bibinfo
				{journal} {At. Data Nucl. Data Tables}\ }\textbf {\bibinfo {volume} {107}},\
			\bibinfo {pages} {1} (\bibinfo {year} {2016})}\BibitemShut {NoStop}%
		\bibitem [{\citenamefont {Puchalski}\ and\ \citenamefont
			{Pachucki}(2010)}]{Puchalski2010}%
		\BibitemOpen
		\bibfield  {author} {\bibinfo {author} {\bibfnamefont {M.}~\bibnamefont
				{Puchalski}}\ and\ \bibinfo {author} {\bibfnamefont {K.}~\bibnamefont
				{Pachucki}},\ }\bibfield  {title} {\bibinfo {title} {Nuclear structure
				effects in the isotope shift with halo nuclei},\ }\href
		{https://doi.org/10.1007/s10751-009-0137-z} {\bibfield  {journal} {\bibinfo
				{journal} {Hyperfine Interact.}\ }\textbf {\bibinfo {volume} {196}},\
			\bibinfo {pages} {35} (\bibinfo {year} {2010})}\BibitemShut {NoStop}%
		\bibitem [{\citenamefont {Hur}()}]{REDF}%
		\BibitemOpen
		\bibfield  {author} {\bibinfo {author} {\bibfnamefont {J.}~\bibnamefont
				{Hur}},\ }\href {https://doi.org/10.5281/zenodo.5818081} {\bibinfo {title}
			{{REDF} (v1.0.0)}},\ \bibinfo {note} {{Zenodo}. doi:10.5281/zenodo.5818081
			(2022)}\BibitemShut {NoStop}%
		\bibitem [{\citenamefont {Ransford}(2020)}]{Ransford2020}%
		\BibitemOpen
		\bibfield  {author} {\bibinfo {author} {\bibfnamefont {A.~M.}\ \bibnamefont
				{Ransford}},\ }\emph {\bibinfo {title} {Old Dog, New Trick: High Fidelity,
				Background-free State Detection of an Ytterbium Ion Qubit}},\ \href
		{https://www.proquest.com/dissertations-theses/old-dog-new-trick-high-fidelity-background-free/docview/2395237565/se-2?accountid=12492}
		{Ph.D. thesis},\ \bibinfo  {school} {The University of California, Los
			Angeles} (\bibinfo {year} {2020})\BibitemShut {NoStop}%
		\bibitem [{\citenamefont {Reinhard}\ and\ \citenamefont
			{Nazarewicz}(2017)}]{Reinhard2017a}%
		\BibitemOpen
		\bibfield  {author} {\bibinfo {author} {\bibfnamefont {P.-G.}\ \bibnamefont
				{Reinhard}}\ and\ \bibinfo {author} {\bibfnamefont {W.}~\bibnamefont
				{Nazarewicz}},\ }\bibfield  {title} {\bibinfo {title} {Toward a global
				description of nuclear charge radii: {Exploring the Fayans} energy density
				functional},\ }\href {https://doi.org/10.1103/PhysRevC.95.064328} {\bibfield
			{journal} {\bibinfo  {journal} {Phys. Rev. C}\ }\textbf {\bibinfo {volume}
				{95}},\ \bibinfo {pages} {064328} (\bibinfo {year} {2017})}\BibitemShut
		{NoStop}%
		\bibitem [{\citenamefont {Roberts}\ \emph {et~al.}(1999)\citenamefont
			{Roberts}, \citenamefont {Taylor}, \citenamefont {Gateva-Kostova},
			\citenamefont {Clarke}, \citenamefont {Rowley},\ and\ \citenamefont
			{Gill}}]{Roberts1999}%
		\BibitemOpen
		\bibfield  {author} {\bibinfo {author} {\bibfnamefont {M.}~\bibnamefont
				{Roberts}}, \bibinfo {author} {\bibfnamefont {P.}~\bibnamefont {Taylor}},
			\bibinfo {author} {\bibfnamefont {S.~V.}\ \bibnamefont {Gateva-Kostova}},
			\bibinfo {author} {\bibfnamefont {R.~B.~M.}\ \bibnamefont {Clarke}}, \bibinfo
			{author} {\bibfnamefont {W.~R.~C.}\ \bibnamefont {Rowley}},\ and\ \bibinfo
			{author} {\bibfnamefont {P.}~\bibnamefont {Gill}},\ }\bibfield  {title}
		{\bibinfo {title} {Measurement of the
				${}^{2}{S}_{1/2}{\ensuremath{-}}^{2}{D}_{5/2}$ clock transition in a single
				{${}^{171}{\mathrm{Yb}}^{+}$} ion},\ }\href
		{https://doi.org/10.1103/PhysRevA.60.2867} {\bibfield  {journal} {\bibinfo
				{journal} {Phys. Rev. A}\ }\textbf {\bibinfo {volume} {60}},\ \bibinfo
			{pages} {2867} (\bibinfo {year} {1999})}\BibitemShut {NoStop}%
		\bibitem [{\citenamefont {Safronova}\ \emph {et~al.}(2012)\citenamefont
			{Safronova}, \citenamefont {Kozlov},\ and\ \citenamefont
			{Clark}}]{Safronova2012}%
		\BibitemOpen
		\bibfield  {author} {\bibinfo {author} {\bibfnamefont {M.~S.}\ \bibnamefont
				{Safronova}}, \bibinfo {author} {\bibfnamefont {M.~G.}\ \bibnamefont
				{Kozlov}},\ and\ \bibinfo {author} {\bibfnamefont {C.~W.}\ \bibnamefont
				{Clark}},\ }\bibfield  {title} {\bibinfo {title} {Blackbody radiation shifts
				in optical atomic clocks},\ }\href {https://doi.org/10.1109/TUFFC.2012.2213}
		{\bibfield  {journal} {\bibinfo  {journal} {IEEE Trans. Ultrason.,
					Ferroelectr., Freq. Control}\ }\textbf {\bibinfo {volume} {59}},\ \bibinfo
			{pages} {439} (\bibinfo {year} {2012})}\BibitemShut {NoStop}%
		\bibitem [{\citenamefont {Sugiyama}\ \emph {et~al.}(2000)\citenamefont
			{Sugiyama}, \citenamefont {Wakita},\ and\ \citenamefont
			{Nakata}}]{Sugiyama2000}%
		\BibitemOpen
		\bibfield  {author} {\bibinfo {author} {\bibfnamefont {K.}~\bibnamefont
				{Sugiyama}}, \bibinfo {author} {\bibfnamefont {A.}~\bibnamefont {Wakita}},\
			and\ \bibinfo {author} {\bibfnamefont {A.}~\bibnamefont {Nakata}},\
		}\bibfield  {title} {\bibinfo {title} {Diode-laser-based light sources for
				laser cooling of trapped {Yb$^+$} ions},\ }in\ \href
		{https://doi.org/10.1109/CPEM.2000.851105} {\emph {\bibinfo {booktitle}
				{Conference on Precision Electromagnetic Measurements. Conference Digest.
					CPEM 2000 (Cat. No.00CH37031)}}}\ (\bibinfo {year} {2000})\ pp.\ \bibinfo
		{pages} {509--510}\BibitemShut {NoStop}%
		\bibitem [{\citenamefont {Tamm}\ \emph {et~al.}(2009)\citenamefont {Tamm},
			\citenamefont {Weyers}, \citenamefont {Lipphardt},\ and\ \citenamefont
			{Peik}}]{Tamm2009}%
		\BibitemOpen
		\bibfield  {author} {\bibinfo {author} {\bibfnamefont {C.}~\bibnamefont
				{Tamm}}, \bibinfo {author} {\bibfnamefont {S.}~\bibnamefont {Weyers}},
			\bibinfo {author} {\bibfnamefont {B.}~\bibnamefont {Lipphardt}},\ and\
			\bibinfo {author} {\bibfnamefont {E.}~\bibnamefont {Peik}},\ }\bibfield
		{title} {\bibinfo {title} {Stray-field-induced quadrupole shift and absolute
				frequency of the 688-{THz} ${^{171}\text{Y}\text{b}}^{+}$ single-ion optical
				frequency standard},\ }\href {https://doi.org/10.1103/PhysRevA.80.043403}
		{\bibfield  {journal} {\bibinfo  {journal} {Phys. Rev. A}\ }\textbf {\bibinfo
				{volume} {80}},\ \bibinfo {pages} {043403} (\bibinfo {year}
			{2009})}\BibitemShut {NoStop}%
		\bibitem [{\citenamefont {Taylor}\ \emph {et~al.}(1997)\citenamefont {Taylor},
			\citenamefont {Roberts}, \citenamefont {Gateva-Kostova}, \citenamefont
			{Clarke}, \citenamefont {Barwood}, \citenamefont {Rowley},\ and\
			\citenamefont {Gill}}]{Taylor1997}%
		\BibitemOpen
		\bibfield  {author} {\bibinfo {author} {\bibfnamefont {P.}~\bibnamefont
				{Taylor}}, \bibinfo {author} {\bibfnamefont {M.}~\bibnamefont {Roberts}},
			\bibinfo {author} {\bibfnamefont {S.~V.}\ \bibnamefont {Gateva-Kostova}},
			\bibinfo {author} {\bibfnamefont {R.~B.~M.}\ \bibnamefont {Clarke}}, \bibinfo
			{author} {\bibfnamefont {G.~P.}\ \bibnamefont {Barwood}}, \bibinfo {author}
			{\bibfnamefont {W.~R.~C.}\ \bibnamefont {Rowley}},\ and\ \bibinfo {author}
			{\bibfnamefont {P.}~\bibnamefont {Gill}},\ }\bibfield  {title} {\bibinfo
			{title} {Investigation of the ${}^{2}{S}_{1/2}{\ensuremath{-}}^{2}{D}_{5/2}$
				clock transition in a single ytterbium ion},\ }\href
		{https://doi.org/10.1103/PhysRevA.56.2699} {\bibfield  {journal} {\bibinfo
				{journal} {Phys. Rev. A}\ }\textbf {\bibinfo {volume} {56}},\ \bibinfo
			{pages} {2699} (\bibinfo {year} {1997})}\BibitemShut {NoStop}%
		\bibitem [{\citenamefont {Webster}\ \emph {et~al.}(2010)\citenamefont
			{Webster}, \citenamefont {Godun}, \citenamefont {King}, \citenamefont
			{Huang}, \citenamefont {Walton}, \citenamefont {Tsatourian}, \citenamefont
			{Margolis}, \citenamefont {Lea},\ and\ \citenamefont {Gill}}]{Webster2010}%
		\BibitemOpen
		\bibfield  {author} {\bibinfo {author} {\bibfnamefont {S.}~\bibnamefont
				{Webster}}, \bibinfo {author} {\bibfnamefont {R.}~\bibnamefont {Godun}},
			\bibinfo {author} {\bibfnamefont {S.}~\bibnamefont {King}}, \bibinfo {author}
			{\bibfnamefont {G.}~\bibnamefont {Huang}}, \bibinfo {author} {\bibfnamefont
				{B.}~\bibnamefont {Walton}}, \bibinfo {author} {\bibfnamefont
				{V.}~\bibnamefont {Tsatourian}}, \bibinfo {author} {\bibfnamefont
				{H.}~\bibnamefont {Margolis}}, \bibinfo {author} {\bibfnamefont
				{S.}~\bibnamefont {Lea}},\ and\ \bibinfo {author} {\bibfnamefont
				{P.}~\bibnamefont {Gill}},\ }\bibfield  {title} {\bibinfo {title} {Frequency
				measurement of the {$^2$S$_{1/2}$--$^2$D$_{3/2}$} electric quadrupole
				transition in a single {$^{171}$Yb$^+$} ion},\ }\href
		{https://doi.org/10.1109/TUFFC.2010.1452} {\bibfield  {journal} {\bibinfo
				{journal} {IEEE Trans. Ultrason., Ferroelectr., Freq. Control}\ }\textbf
			{\bibinfo {volume} {57}},\ \bibinfo {pages} {592} (\bibinfo {year}
			{2010})}\BibitemShut {NoStop}%
		\bibitem [{\citenamefont {Zhang}\ \emph {et~al.}(2014)\citenamefont {Zhang},
			\citenamefont {Martin}, \citenamefont {Benko}, \citenamefont {Hall},
			\citenamefont {Ye}, \citenamefont {Hagemann}, \citenamefont {Legero},
			\citenamefont {Sterr}, \citenamefont {Riehle}, \citenamefont {Cole},\ and\
			\citenamefont {Aspelmeyer}}]{Zhang2014}%
		\BibitemOpen
		\bibfield  {author} {\bibinfo {author} {\bibfnamefont {W.}~\bibnamefont
				{Zhang}}, \bibinfo {author} {\bibfnamefont {M.~J.}\ \bibnamefont {Martin}},
			\bibinfo {author} {\bibfnamefont {C.}~\bibnamefont {Benko}}, \bibinfo
			{author} {\bibfnamefont {J.~L.}\ \bibnamefont {Hall}}, \bibinfo {author}
			{\bibfnamefont {J.}~\bibnamefont {Ye}}, \bibinfo {author} {\bibfnamefont
				{C.}~\bibnamefont {Hagemann}}, \bibinfo {author} {\bibfnamefont
				{T.}~\bibnamefont {Legero}}, \bibinfo {author} {\bibfnamefont
				{U.}~\bibnamefont {Sterr}}, \bibinfo {author} {\bibfnamefont
				{F.}~\bibnamefont {Riehle}}, \bibinfo {author} {\bibfnamefont {G.~D.}\
				\bibnamefont {Cole}},\ and\ \bibinfo {author} {\bibfnamefont
				{M.}~\bibnamefont {Aspelmeyer}},\ }\bibfield  {title} {\bibinfo {title}
			{Reduction of residual amplitude modulation to {$1\times10^{-6}$} for
				frequency modulation and laser stabilization},\ }\href
		{https://doi.org/10.1364/OL.39.001980} {\bibfield  {journal} {\bibinfo
				{journal} {Opt. Lett.}\ }\textbf {\bibinfo {volume} {39}},\ \bibinfo {pages}
			{1980} (\bibinfo {year} {2014})}\BibitemShut {NoStop}%
	\end{thebibliography}
\end{document}